\documentclass{jfm}

\usepackage{graphicx}
\usepackage{newtxtext}
\usepackage{newtxmath}
\usepackage{natbib}
\usepackage{hyperref}
\hypersetup{
    colorlinks = true,
    urlcolor   = blue,
    citecolor  = black,
}

\usepackage{empheq}
\usepackage{tikz}
\usepackage{bm}
\usepackage{float}
\usepackage[title]{appendix}
\usepackage{mathrsfs}

\newcommand{\sv}{\xi}

\renewcommand{\vec}{\mathbf}

\newtheorem{defin}{Definition}
\usepackage{subcaption}
\allowdisplaybreaks  
\usepackage{algorithm}
\usepackage{algpseudocode}
\usepackage{caption}
\usepackage{varwidth}
\usepackage[symbol]{footmisc}

\title{Structured Input-Output Modeling and\\ Robust Stability Analysis of Compressible Flows}    
\author{Diganta Bhattacharjee\aff{1}
  \corresp{\email{dbhattac@umn.edu}},
  Talha Mushtaq\aff{1}, Peter Seiler\aff{2},
 \and Maziar S. Hemati\aff{1}}
\affiliation{\aff{1}Department of Aerospace Engineering and Mechanics, University of Minnesota, Minneapolis, MN 55455, USA
\aff{2}Department of Electrical Engineering and Computer Science, University of Michigan, Ann Arbor, MI 48109, USA}

\begin{document}

\maketitle
             
\begin{abstract}
The recently introduced structured input-output analysis 
is a powerful 
method for capturing nonlinear phenomena associated with incompressible flows, and this paper extends that method to the compressible regime. 
The proposed method relies upon a reformulation of the compressible Navier-Stokes equations, which allows for an exact quadratic formulation of the dynamics of perturbations about a steady base flow. 
To facilitate the structured input-output analysis, a pseudo-linear model for the quadratic nonlinearity is proposed and the structural information of the nonlinearity is embedded into a structured uncertainty comprising unknown `perturbations'.
The structured singular value framework is employed to compute the input-output gain, which provides an estimate of the robust stability margin of the flow perturbations, as well as the forcing and response modes that are consistent with the nonlinearity structure.
The analysis is then carried out on a plane, laminar compressible Couette flow over a range of Mach numbers.   
The structured input-output gains identify an instability mechanism, characterized by a spanwise elongated structure in the streamwise-spanwise wavenumber space at a subsonic Mach number, that 
{takes the form of} an oblique structure at sonic and supersonic Mach numbers.
In addition, the structured input-output forcing and response modes provide insight into the thermodynamic and momentum characteristics associated with a source of instability.
Comparisons with a resolvent/unstructured analysis reveal discrepancies in the distribution of input-output gains over the wavenumber space as well as in the modal behavior of an instability, thus highlighting the strong correlation between the structural information of the nonlinearity and the underlying flow physics.
\end{abstract}



\section{Introduction}
Compressible flows arise in most aerospace applications, and these flows are governed by the compressible Navier-Stokes equations~(NSE), which can result in highly complex flow physics with a rich array of nonlinear flow interactions. 
Modal analysis techniques have proven to be invaluable for unraveling these complex flow physics to arrive at an improved understanding of key instability mechanisms and coherent structures that drive the associated fluid dynamics~\citep{tairaAIAAJ2020,tairaAIAAJ2017,tairaAIAAJ2020b}.
In that regard, input-output~(I/O) and resolvent-based techniques play a key role in extracting information related to modal analysis~\citep{jovanovicARFM2021}.
%
%
These techniques were primarily adopted and developed in the context of incompressible flows~\citep{jovanovicJFM2005,mckeonJFM2010,mckeon2017engine}, but have since been adopted and developed for studies of compressible fluid dynamics~\citep{nicholsAIAA2019,dawsonAIAA2019,baeJFM2019,baeAIAA2020,sunAIAAJ2020,chen2023linear,dwivedi2019reattachment,dawson2020prediction}.
I/O methods are inherently physics-based and work by decomposing the governing flow equations into a feedback interconnection between the linear dynamics and the nonlinear terms.
Traditionally, the outputs of the nonlinear terms are treated as an implicit (unstructured) forcing on the linear dynamics, which results in an optimization problem that is relatively straightforward to solve using linear systems analysis techniques.
Despite their successes, linear I/O analysis neglects any known structure regarding the nonlinear terms.

Most existing analysis methods that account for the nonlinearities in NSE result in computationally expensive---or even intractable---solution algorithms~\citep{Kerswell_2018,kalurAIAA2020,kalur2021nonlinear,kalur2021estimating,liu20, goulart2012global}.
These include recently proposed frameworks for nonlinear stability and transient energy growth analysis of fluid flows based on variational approaches \citep{Kerswell_2018} and linear matrix inequalities (LMIs) \citep{kalurAIAA2020, goulart2012global}.
%
%
%
The recent methods/tools on estimating the regions of attraction of transitional fluid flows using LMIs and quadratic constraints \citep{kalur2021nonlinear,kalur2021estimating, liu20} also fall under this category. 
These are typically restricted to reduced-order models of transitional flows due to the high computational cost associated with solving optimization problems involving LMIs.
%

Recently, structured I/O analysis was proposed as a computationally tractable method for analyzing incompressible flows~\citep{liu21}.
Unlike traditional unstructured I/O techniques like the resolvent analysis~\citep{mckeonJFM2010, mckeon2017engine}, structured I/O analysis retains the essence of the nonlinear interactions by imposing the structure of the convective nonlinear terms within a linear I/O analysis framework~\citep{liu21}.
%
%
The structured I/O framework involves a linear system which is a feedback interconnection between the linearized fluid flow dynamics and 
a static map that comprises constraints enforcing the structure of the quadratic nonlinearity.
The corresponding I/O analysis can be carried out using established ideas from the robust controls literature~\citep{packard1993, zhou1996robust}. 
Thus, structured I/O analysis results in physically-consistent I/O gains and mode shapes that depict important underlying features in the fluid flow system. 
This is evidenced by the structured I/O analysis of incompressible flows in \cite{liu21}, which led to results that were in agreement with both experimental observations and direct numerical simulation results.
%
%
It was further shown in~\cite{mushtaq2024algorithms, mushtaq2023exact} that additional repeated structure in the convective nonlinearity in the incompressible NSE can be exploited to further refine the structured I/O analysis.
Moreover, the structured I/O framework can be utilized to conduct modal analysis, as shown in \cite{mushtaq2023structured}, and has been used to elucidate flow physics in stratified flow \citep{liu2022structured} and turbulent flows \citep{mushtaq2023exact,mushtaq2023riblets}.
It is noteworthy that the structured I/O modal analysis is able to capture nonlinear flow behaviors like the dampening of the near-wall cycle and creation of Kelvin-Helmholtz-type instabilities for turbulent flows over riblets that are consistent with direct numerical simulation results \citep{mushtaq2023riblets}.
These indicate the usefulness of structured I/O analysis and serve as a motivation for this work.
%

%
The main contribution of this work, 
an earlier version of which has been reported in \cite{bhattacharjee2023structured}, lies in extending the incompressible structured I/O framework proposed in \cite{liu21} to compressible flows where the nonlinearities are substantially more complicated than their incompressible counterparts. 
%
%
First, we reformulate the compressible NSE such that the nonlinearity in the governing equations is quadratic, which allows for an efficient structured I/O modeling of these nonlinear terms.
The modeling essentially involves pseudo-linearization of the quadratic nonlinearity and constructing a `structured uncertainty' that embeds structural information and properties of the nonlinearity. 
Next, we utilize the structured singular value formalism (or simply `$\mu$') from robust control theory to perform robust stability analysis of the perturbed fluid flow system.
As part of the analysis, the upper and lower bounds on $\mu$ are computed. 
While the upper bound provides a sufficient condition for robust stability of the flow perturbations, the lower bound serves as a sufficient condition for instability \citep{zhou1996robust,packard1993}.
The lower bound algorithm (i.e., power iteration) also provides the forcing and response modes consistent with the structure of the nonlinearity, which can be utilized for modal analysis of the flow \citep{mushtaq2023structured}.
{Note that, compared to \cite{bhattacharjee2023structured}, the structured I/O framework here contains a refined uncertainty structure and we have incorporated the Chu energy \citep{chu1965energy} norm into the analysis. 
Furthermore, we have employed the lower bound algorithm to compute the essential ingredients of structured modal analysis---structured I/O modes and structured uncertainty.}

The proposed method is implemented on a compressible laminar Couette flow over a range of Mach numbers.
This flow has been investigated in numerous other works on flow stability and I/O analysis due to its simplicity~\citep{duckJFM1994,malikPOF2006,dawsonAIAA2019,huPOF1998}. 
Results of the structured I/O analysis are compared with those obtained through a resolvent analysis which does not incorporate the nonlinearity structure.
Our results illustrate that accounting for the structure of the nonlinearity reduces the overall conservatism in the I/O gains at subsonic Mach numbers, which translates to a larger estimate of stability margin of the perturbed flow. 
Despite some qualitative similarities, there are distinct differences in the I/O gain distributions over the grid of spatial wavenumbers across the two sets of results, which indicate that resolvent analysis might predict potentially redundant/non-physical flow instability mechanisms that are not consistent with the admissible input-output behavior of the nonlinearity in the governing flow equations.   
In addition, the amplified flow features are better highlighted locally in the streamwise and spanwise wavenumber space in terms of the structured I/O gains, making it easier to identify and isolate different sources of instability.
%
%
%
%
%
While both the analyses are able to identify the same mechanism of instability---characterized by local maxima in the corresponding I/O gains at a particular value of the streamwise and spanwise wavenumber pair---the associated forcing and response modes point towards disparate mechanisms causing the instability as well as different flow perturbations getting amplified as a result of the instability. 
For example, the structured forcing and response modes indicate that the instability associated with the maximum structured I/O gain at a subsonic Mach number primarily affects the thermodynamic properties of the flow; however, the resolvent modes indicate that the instability is related to both the momentum and thermodynamic variables of the flow.
Note that although the outcome of an I/O analysis generally depends on the variables used to describe the fluid flow \citep{karban2020ambiguity, bhattacharjee2024formulation}, all the discrepancies reported in this paper are primarily due to the fact that structured I/O utilizes the structure of the nonlinearity, whereas resolvent analysis does not.
%



The remainder of the paper proceeds as follows. 
The reformulation of compressible NSE, perturbation dynamics setup for a generic, steady base flow, and structured I/O modeling and analysis are detailed in Section \ref{Sec: Compressible Structured I/O}. 
The compressible plane Couette flow is described in Section \ref{Sec: Compressible Plane Couette Flow}, which includes details on the base flow calculations, linear operators required for the analysis and validation of our numerical implementation.
Furthermore, detailed numerical results and the corresponding discussion are included in Section \ref{Sec: Results}.
Finally, the conclusions and future directions of this work are provided in Section \ref{Sec: Conclusion}. 
 
We use symbols $\mathbb{C}^n$, $\mathbb{C}^{n \times m}$ and $\mathbb{R}^{n \times m}$ to denote the sets of $n$-dimensional complex vectors, complex matrices of dimension $n \times m$, real matrices of dimension $n \times m$, respectively.
%
%
The transpose and conjugate (Hermitian) transpose are denoted by $(\cdot)^\text{T}$ and $(\cdot)^\dagger$, respectively, and $(\cdot)^\ddagger$ stands for the pseudo-inverse of a matrix.
%
%
The symbol $\| \cdot \|_2$ means the Euclidean norm for a vector and the spectral norm for a matrix, and $\| \cdot \|_F$ denotes the Frobenius norm of a matrix. 
Also, an $n \times n$ identity matrix is denoted by $\vec{I}_n$ and we use $\vec{i} = \sqrt{-1} $ as the imaginary unit. 
The notation $\text{diag}(\cdot)$ stands for a block-diagonal operation and/or a block-diagonal matrix.

 
\section{Structured I/O Analysis of Compressible Flows} \label{Sec: Compressible Structured I/O}
Consider a compressible fluid in the domain~${\Gamma\subset\mathbb{R}^3}$.
The state of the fluid at any instant in time can be characterized solely based on the primitive variables of density $\check{\varrho}(\vec{x},t)$, velocity $\check{\vec{u}}(\vec{x},t)=(\check{u}(\vec{x},t),\check{v}(\vec{x},t),\check{w}(\vec{x},t))$, and pressure~$\check{p}(\vec{x},t)$.
Here, $\vec{x}\in\Gamma$ is the spatial coordinate and $t\in\mathbb{R}_{\geq0}$ is time.
The equations of motion governing the dynamics of the flow in $\Gamma$ are derived from the conservation laws for mass, momentum, and energy. These equations can be expressed in terms of the primitive variables $\check{\vec{q}} = (\check{\sv}:=1/\check{\varrho},\check{\vec{u}},\check{p})$.
All variables are non-dimensionalized in the usual way using $L$, $u_r$, $T_r$, $\xi_r=1/\rho_r$, and $\eta_r$ as the reference length, velocity, temperature, specific volume (density), and viscosity, respectively.
Using the dimensional equation of state, the reference pressure is chosen to be $p_r=T_rR/\sv_r$, where $R$ denotes the  gas constant.
Denoting all dimensional quantities with a superscript~$(\cdot)^d$, we define the following non-dimensional quantities:
\begin{align*}
    \check{\sv}&=\frac{\check{\sv}^d}{\sv_r},\enspace 
    \check{\vec{u}} = \frac{\check{\vec{u}}^d}{u_r},\enspace
    t=\frac{t^d}{L/u_r},\enspace
    \check{\eta}=\frac{\check{\eta}^d}{\eta_r},\enspace
    \check{T}=\frac{\check{T}^d}{T_r},\enspace
    \check{p}=\frac{\check{p}^d}{p_r}, \\ 
    Re&=\frac{u_r L}{\eta_r \sv_r},\enspace
    M_r=\frac{u_r}{a_r}=\frac{u_r}{\sqrt{\gamma R T_r}},
\end{align*}
where $\eta$ is the coefficient of shear viscosity and $\gamma$ is the specific heat ratio, while $Re$ and $M_r$ denote the Reynolds number and Mach number, respectively.
The resulting non-dimensional compressible NSE can be expressed as
\begin{align}
\partial_t \check{\sv} + \check{\vec{u}} \cdot \nabla\check{\sv} - \check{\sv} \nabla \cdot \check{\vec{u}} &= 0\label{eq:nondim_sv}\\
\partial_t \check{\vec{u}} + \check{\vec{u}} \cdot\nabla \check{\vec{u}} + \frac{1}{\gamma M_r^2} \check{\sv}\nabla \check{p} - \frac{1}{Re} \check{\sv}\nabla\cdot\Pi(\check{\vec{u}}, \check{\eta}) &= 0 \label{eq:nondim_u}\\
\partial_t \check{p} + \check{\vec{u}} \cdot \nabla \check{p} + \gamma \check{p}\nabla\cdot\check{\vec{u}} - \frac{\gamma(\gamma-1)M_r^2}{Re} \Phi(\check{\vec{u}}, \check{\eta}) - \frac{\gamma}{Re Pr}\nabla\cdot (\check{\eta} \nabla (\check{p} \check{\sv})) &= 0 \label{eq:nondim_p}
\end{align} 
with the associated non-dimensional equation of state for a perfect polytropic gas $\check{p} \check{\xi} = \check{T}$.
Here, $Pr$ is the Prandtl number and $\Pi(\check{\vec{u}}, \check{\eta})$ denotes the viscous stress tensor, which for a Newtonian fluid takes the form
\begin{equation*}
\Pi(\check{\vec{u}},\check{\eta})=\begin{bmatrix}\tau_{xx}&\tau_{xy}&\tau_{xz}\\\tau_{yx}&\tau_{yy}&\tau_{yz}\\\tau_{zx}&\tau_{zy}&\tau_{zz}\end{bmatrix}=2 \check{\eta} \mathcal{D} + (\check{\eta}_b - \frac23 \check{\eta})(\nabla\cdot\check{\vec{u}})\vec{I}_3,
\end{equation*}
where $\check{\eta}_b$ is the coefficient of bulk viscosity, and $\mathcal{D}$ is the deformation tensor given by
\begin{equation*}
    \mathcal{D}=\begin{bmatrix}
    \partial_x \check{u} & \frac12 (\partial_x \check{v} + \partial_y \check{u}) & \frac12 (\partial_x \check{w} + \partial_z \check{u})\\
    \frac12 (\partial_y \check{u} + \partial_x \check{v}) & \partial_y \check{v} & \frac12 (\partial_y \check{w} + \partial_z \check{v})\\
    \frac12 (\partial_z \check{u} + \partial_x \check{w}) & \frac12 (\partial_z \check{v} +\partial_y \check{w}) & \partial_z \check{w}
    \end{bmatrix}.
\end{equation*}
%
In this work, we will apply Stokes' hypothesis, so that $\check{\eta}_b=0$. Also, the term $\Phi(\check{\vec{u}}, \check{\eta})$ in \eqref{eq:nondim_p} is the viscous dissipation term given by 
\begin{align*}
\Phi (\check{\vec{u}}, \check{\eta}) &= \check{\eta} \left( 2 \left( (\partial_x \check{u})^2 + (\partial_y \check{v})^2 + (\partial_z \check{w})^2 \right) + (\partial_y \check{u} + \partial_x \check{v})^2 + (\partial_z \check{v} + \partial_y \check{w})^2 + (\partial_z \check{u} + \partial_x \check{w})^2  \right) \\
& \quad -\frac23 \check{\eta} \left( \nabla \cdot \check{\vec{u}} \right)^2 \\
&= \frac{\check{\eta}}{2} \left[ \nabla \check{\vec{u}}  + (\nabla \check{\vec{u}})^\text{T} \right]^2 -\frac23 \check{\eta} \left( \nabla \cdot \check{\vec{u}} \right)^2 .
\end{align*}

\subsection{Perturbation Dynamics about Steady Base Flows: A Quadratic Formulation} \label{Sec:Quadratic Formulation}
In this section, the dynamics of flow perturbations about a steady base flow will be discussed.
{In particular, we seek a quadratic formulation of the perturbation dynamics---similar, in principle, to some of the existing quadratic reformulations of NSEs in the literature \citep{vigo1998proper, iollo2000stability, qian2019transform}---that will facilitate the subsequent structured I/O modeling.}
To this end, we will assume that the base flow viscosity $\eta_0$ depends on the base flow temperature $T_0$, i.e., $\eta_0=\eta_0(T_0)$; however, the temperature dependence of viscosity in the perturbation dynamics---and therefore perturbations to the base viscosity $\eta_0$---will be neglected in the ensuing analysis. 
The total field $\check{\vec{q}}$ is decomposed as $\check{\vec{q}} = \vec{q}_0 + \vec{q}$, where $\vec{q}_0$ denotes the base flow and $\vec{q}$ are the corresponding perturbations about that base flow.
Then, the equations governing flow perturbations $\vec{q} = (\xi,\vec{u},p) = (\xi,u,v,w,p)$ about the base flow $\vec{q}_0 = (\xi_0,\vec{u}_0,p_0) = (\xi_0, u_0,v_0,w_0,p_0)$ are obtained using the compressible NSE in \eqref{eq:nondim_sv}, \eqref{eq:nondim_u}, \eqref{eq:nondim_p}.
%
%
Isolating the linear dynamics on the left-hand side and the nonlinear terms on the right-hand side, the dynamics of perturbations can be expressed as 
\begin{align}
    \partial_t \sv - L_\sv(\vec{q}) &= f_\sv(\vec{q}) \notag \\
    \partial_t \vec{u} - L_\vec{u}(\vec{q}) &= f_\vec{u}(\vec{q}) \label{eq:perturbation_dynamics_1} \\
    \partial_t p -L_p(\vec{q}) &= f_p(\vec{q}) \notag 
\end{align}
where 
\begin{align}
    L_\sv(\vec{q}) &= -\vec{u}_0\cdot\nabla\sv -\vec{u}\cdot\nabla\sv_0 + \sv\nabla\cdot\vec{u}_0 + \sv_0\nabla\cdot\vec{u} \notag \\
    L_\vec{u}(\vec{q}) &= -\vec{u}_0\cdot\nabla\vec{u} - \vec{u}\cdot\nabla\vec{u}_0 - \frac{1}{\gamma M_r^2} \sv_0\nabla p - \frac{1}{\gamma M_r^2} \sv\nabla p_0 \notag \\
    & \quad + \frac{1}{Re}\left(\sv\nabla\cdot\Pi(\vec{u}_0, \eta_0) + \sv_0\nabla\cdot\Pi(\vec{u}, \eta_0) \right) \notag \\
   L_p(\vec{q}) &= -\vec{u}_0\cdot\nabla p -\vec{u}\cdot\nabla p_0 -\gamma\left(p\nabla\cdot\vec{u}_0+p_0\nabla\cdot\vec{u}\right) \notag \\
   & \quad + \frac{\gamma(\gamma-1)M_r^2}{Re} \Bigg[ \eta_0 \Big( 4 \partial_x u \partial_x u_0 + 4 \partial_y v \partial_y v_0 + 4 \partial_z w \partial_z w_0 \notag \\
   & \quad + 2 (\partial_y u + \partial_x v) (\partial_y u_0 + \partial_x v_0) + 2 (\partial_z v + \partial_y w) (\partial_z v_0 + \partial_y w_0)  \label{eq:linear_and_nonlinear_terms_1} \\
   & \quad + 2 (\partial_z u + \partial_x w) (\partial_z u_0 + \partial_x w_0) \Big) - \frac{4}{3} \eta_0 (\nabla \cdot \vec{u}_0) (\nabla \cdot \vec{u}) \Bigg] \notag \\
   & \quad + \frac{\gamma}{Re Pr}\nabla\cdot\eta_0\left(\nabla(p_0\sv +p\sv_0)\right)\notag \\
    f_\sv(\vec{q}) &= \sv\nabla\cdot\vec{u} - \vec{u}\cdot\nabla\sv \notag \\
    f_\vec{u}(\vec{q}) &= - \vec{u}\cdot\nabla\vec{u} - \frac{1}{\gamma M_r^2} \sv\nabla p+\frac{1}{Re}\sv\nabla\cdot\Pi(\vec{u}, \eta_0) \notag \\
    f_p(\vec{q}) &= -\vec{u}\cdot\nabla p - \gamma p\nabla\vec{u} + \frac{\gamma(\gamma-1)M_r^2}{Re} \Phi(\vec{u}, \eta_0) +\frac{\gamma}{Re Pr}\nabla\cdot\left(\eta_0\nabla(p\sv)\right). \notag 
\end{align}
Also, the equation of state can be expressed as $T - L_T(\vec{q}) = f_T(\vec{q})$ where $L_T(\vec{q}) = p_0 \xi + p \xi_0, f_T(\vec{q}) = p \xi$.
This description of the perturbation dynamics in \eqref{eq:perturbation_dynamics_1} provides a feedback
interpretation of the equations, which is 
shown in Fig. \ref{subfig:original_system_schematic} where the perturbed quantities ($\xi, \vec{u}, p$) denote the state of a linear system (i.e., the linear perturbation dynamics). 
The states are also the outputs of the linear system and the inputs forcing the linear system are associated with the nonlinear feedback of the outputs (see Fig. \ref{subfig:original_system_schematic}).
Furthermore, the term $\Phi (\vec{u}, \eta_0)$ can be expressed as
\begin{equation*}
\Phi (\vec{u}, \eta_0) = \frac{{\eta_0}}{2} \underbrace{\left[ \nabla \vec{{u}}  + (\nabla \vec{{u}})^\text{T} \right]^2}_{\Psi_1(\vec{u})} -\frac23 \eta_0 \underbrace{(\nabla \cdot \vec{u})^2}_{\Psi_2(\vec{u})} 
\end{equation*}
where
\begin{align*}
\Psi_1 (\vec{u}) &= \begin{bmatrix}
(\nabla u)^\text{T}  & (\nabla v)^\text{T} & (\nabla w)^\text{T} & (\partial_x \vec{u})^\text{T} & (\partial_y \vec{u})^\text{T} & ( \partial_z \vec{u})^\text{T} 
\end{bmatrix} \begin{bmatrix}
\nabla u \\ \nabla v \\ \nabla w \\ (2 \nabla u + \partial_x \vec{u}) \\ (2 \nabla v + \partial_y \vec{u}) \\ (2 \nabla w + \partial_z \vec{u})
\end{bmatrix}, \\
\Psi_2 (\vec{u}) &= (\nabla \cdot \vec{u})^2.
\end{align*}
Also, we have the following identity
\begin{equation*}
\nabla^2 (p \xi) = \xi \nabla^2 p + p \nabla^2 \xi + 2 \nabla p \cdot \nabla \xi .
\end{equation*}
After substituting the above expressions into \eqref{eq:linear_and_nonlinear_terms_1}, we derive
\begin{align}
f_\xi (\vec{q}) &=  - \vec{u} \cdot \nabla \xi + \xi \nabla \cdot \vec{u}, \notag \\
f_{\vec{u}} (\vec{q}) &= - \vec{u} \cdot \nabla \vec{u} - \frac{1}{\gamma M_r^2} \xi \nabla p  + \frac{\xi}{Re} \nabla \cdot \Pi (\vec{u}, \eta_0), \notag \\
f_p (\vec{q}) &=  - \vec{u} \cdot \nabla p - \gamma p \nabla \cdot \vec{u} + \frac{\gamma (\gamma - 1) M_r^2}{Re} \frac{{\eta_0}}{2} \Bigg[ (\nabla u)^\text{T} \nabla u + (\nabla v)^\text{T} \nabla v + (\nabla w)^\text{T} \nabla w \label{Nonlinear terms in perturbation dynamics-2} \\
& \quad + (\partial_x \vec{u})^\text{T} (2 \nabla u + \partial_x \vec{u}) + (\partial_y \vec{u})^\text{T} (2 \nabla v + \partial_y \vec{u}) + ( \partial_z \vec{u})^\text{T} (2 \nabla w + \partial_z \vec{u})  \Bigg] \notag \\
& \quad - \frac{\gamma (\gamma - 1) M_r^2}{Re} \frac23 \eta_0 (\nabla \cdot \vec{u})^2 \notag \\
& \quad + \frac{\gamma}{Re Pr} \left(\eta_0 \xi \nabla^2 p + \eta_0 p \nabla^2 \xi + 2 \eta_0 \nabla p \cdot \nabla \xi + \nabla \eta_0 \cdot \xi \nabla p + \nabla \eta_0 \cdot p \nabla \xi \right). \notag
\end{align}
In most systems-theoretic formulations, the nonlinearity in the equations governing flow perturbations is cubic~\citep{rowleyPD2004}, which creates non-trivial challenges with regards to the necessary modeling of the nonlinear terms for subsequent structured I/O analysis.
However, the reformulation of the compressible NSE here, combined with the the above-mentioned assumption on the viscosity perturbations, makes the resulting nonlinearity in \eqref{Nonlinear terms in perturbation dynamics-2} quadratic in the perturbed flow states $\vec{q}$. 
The quadratic nonlinearity in our current formulation makes the application of structured I/O analysis more tractable through pseudo-linearization, as described in Section \ref{Sec:Modeling the Nonlinear Terms}.
%

 \begin{figure}
 \captionsetup[subfigure]{justification=centering}
 \centering
 \begin{subfigure}{0.48\textwidth}
 \scalebox{0.9}{
 \begin{picture} (100,100) (-60,0)
  \put(0,0){\framebox(60,40){\parbox{40\unitlength}{Linear \\ dynamics \\ of $\vec{q}$}}}%
  \put(0,60){\framebox(60,40){\parbox{40\unitlength}{Nonlinear \\ maps $f_{(\cdot)}$}}}\
  \put(60,20){\line(1,0){20}}
   \put(80,20){\line(0,1){55}} 
  \put(80,75){\vector(-1,0){20}}  
  \put(0,75){\line(-1,0){20}} 
  \put(-20,75){\line(0,-1){55}} 
  \put(-20,20){\vector(1,0){20}}  
  \put(-60,45){$\begin{bmatrix} f_\xi(\vec{q}) \\ f_{\vec{u}}(\vec{q}) \\ f_p(\vec{q}) \end{bmatrix}$}
  \put(85,45){$\vec{q} = \begin{bmatrix} \xi \\ \vec{u} \\ p \end{bmatrix}$}
 \end{picture}
 }
 \caption{Quadratic nonlinear system} \label{subfig:original_system_schematic}
 \end{subfigure}
 \hfill
 \centering
\begin{subfigure}{0.48\textwidth}
 \scalebox{0.9}{
\begin{picture} (100,100) (-60,15)
  \put(0,0){\framebox(60,40){\parbox{40\unitlength}{Linear \\ dynamics \\ of $\vec{q}$}}}
  \put(0,70){\framebox(60,40){\parbox{45\unitlength}{Structured \\ uncertainty $\Delta$ }}}
  \put(70,30){\framebox(60,30){\parbox{40\unitlength}{Linear \\ map $\vec{C}_\chi$}}}
  \put(-70,30){\framebox(60,30){\parbox{40\unitlength}{Linear \\ map $\vec{B}_\chi$ }}}
  \put(60,20){\line(1,0){40}}
 \put(100,20){\vector(0,1){10}}
 \put(100,60){\line(0,1){25}} 
 \put(100,85){\vector(-1,0){40}} 
 \put(0,85){\line(-1,0){45}}
 \put(-45,85){\vector(0,-1){25}}
  \put(-45,30){\line(0,-1){10}}
  \put(-45,20){\vector(1,0){45}} 
  \put(-70,5){$\begin{bmatrix} f_\xi & f_{\vec{u}}^\text{T} & f_p \end{bmatrix}_\chi^\text{T}$} 
  \put(70,5){$\vec{q} = \begin{bmatrix} \xi & \vec{u}^\text{T} & p \end{bmatrix}^\text{T}$}
  \put(80,90){$\vec{y}_\chi$}  
  \put(-30,90){$\vec{f}_\chi$}  
  \put(-78,-4){\dashbox(218,70)}
  \end{picture} }
  \vspace{0.6cm}
 \caption{Modeled system (before discretization)} \label{subfig:modeled_system_schematic}
 \end{subfigure}
 \centering
 \begin{subfigure}{0.495\textwidth}
\scalebox{0.9}{
\begin{picture}(120,110)(-70,-5)
 \thicklines
 \put(0,0){\framebox(70,40){$
     \mathcal{H}(k_x, k_z, \omega)
$}}
 \put(15,50){\framebox(40,40){$\hat{\Delta}$}}
 \put(-35,45){$\hat{\vec{f}}_\chi$}
 \put(70,20){\line(1,0){20}}
 \put(90,20){\line(0,1){50}}
 \put(90,70){\vector(-1,0){35}}
 \put(15,70){\line(-1,0){35}}
 \put(-20,70){\line(0,-1){50}}
 \put(-20,20){\vector(1,0){20}}
 \put(94,45) {$\hat{\vec{y}}_\chi$}
\end{picture}
} 
\caption{Modeled system (after discretization)} \label{subfig:modeled_system_after_discretization}
 \end{subfigure}
\caption{The perturbation dynamics expressed in feedback forms: (a) the nonlinear system in \eqref{eq:perturbation_dynamics_1}; (b) the system in \eqref{eq:structuredI/O_continuous_form} obtained after the structured I/O modeling; (c) the system in \eqref{eq:IO-operator_discretized_perturbation_dynamics} resulting from spectral discretization of the structured I/O system in  \eqref{eq:structuredI/O_continuous_form}.
%
%
Note that the dashed box in (b), which represents the dynamic (linear) map that takes the modeled forcing/inputs $\vec{f}_\chi$ to the corresponding modeled outputs $\vec{y}_\chi$, becomes the frequency response operator $\mathcal{H}(k_x, k_z, \omega)$ in (c) after discretization.}
\label{fig:modeling_approximation}
\end{figure}

\subsection{Modeling the Nonlinear Terms: Structured Uncertainty} \label{Sec:Modeling the Nonlinear Terms}
We now describe a modeling of the quadratic nonlinearity 
in \eqref{Nonlinear terms in perturbation dynamics-2} 
that enables the structured I/O analysis using the structured singular value formalism \citep{packard1993}. 
This involves 
decomposing each quadratic nonlinearity into its constituent linear parts, one of which gets modeled as an unknown/uncertain perturbation or gain, while the other is treated as a known quantity.
%
%
Following the terminology used in the robust control literature, we refer to the resulting matrix of uncertain gains as a structured uncertainty which typically takes a block-diagonal structure\footnote[4]{In fact, the modeling requires two linear transformations/operators to make the structured uncertainty block-diagonal. 
{These are the operators $\vec{B}_\chi$ and $\vec{C}_\chi$ in Fig. \ref{subfig:modeled_system_schematic}}.}.
%
%
Also, the known quantities in the above decomposition are treated as measured outputs from the linear dynamics governing the flow perturbations. 
Results of the system modeling are schematically shown in Fig. \ref{subfig:modeled_system_schematic}, where the subscript $(\cdot)_\chi$ denotes
the modeled quantities. 
Note that the uncertain gains are
independent of the flow perturbations due to the pseudo-linearization of the quadratic nonlinearity in the structured I/O modeling.
{In other words, the map that takes the modeled outputs $\vec{y}_\chi$ to the modeled forcing/inputs $\vec{f}_\chi$ is linear} and does not depend on the flow perturbations, which would not have been the case for the true inputs-outputs $\vec{f}$, $\vec{y}$ and the associated nonlinear map (compare Figs. \ref{subfig:original_system_schematic}, \ref{subfig:modeled_system_schematic}).
However, the modeled system retains the essence of the true nonlinear system through the structured uncertainty which embeds structural information of the nonlinearity.
%
%
The details of our model are provided next.

First, we separate out the nonlinear forcing in \eqref{Nonlinear terms in perturbation dynamics-2} into three different vectors as 
\begin{equation} \label{eq:true_nonlinearity_separated}
\begin{split}
\vec{f}_{{1}} = \begin{bmatrix}
\xi \nabla^2 p \\ \xi \nabla p \\ \xi (\nabla \cdot \vec{u}) \\ \xi \nabla \cdot  \Pi(\vec{u}, \eta_0) \\ p \nabla^2 \xi \\ p \nabla \xi \\ p \nabla \cdot \vec{u}
\end{bmatrix}, \quad 
\vec{f}_{{2}} = \begin{bmatrix}
\vec{u} \cdot \nabla \xi \\ \vec{u} \cdot \nabla u \\ \vec{u} \cdot \nabla v \\ \vec{u} \cdot \nabla w \\ \vec{u} \cdot \nabla p  
\end{bmatrix}, \quad 
\vec{f}_{{3}} = \begin{bmatrix}
\nabla u \cdot \nabla u \\ \nabla v \cdot \nabla v \\ \nabla w \cdot \nabla w \\ \partial_x \vec{u} \cdot (2 \nabla u + \partial_x \vec{u}) \\ \partial_y \vec{u} \cdot (2 \nabla v + \partial_y \vec{u}) \\ \partial_z \vec{u} \cdot (2 \nabla w + \partial_z \vec{u}) \\ \nabla p \cdot \nabla \xi \\ (\nabla \cdot \vec{u})^2
\end{bmatrix},
\end{split}
\end{equation}
which are related to the nonlinear forcing in \eqref{Nonlinear terms in perturbation dynamics-2} as
\begin{equation} \label{eq:truth_to_modeled_nonlinearity}
\begin{bmatrix}
f_\xi \\ f_{\vec{u}} \\ f_p
\end{bmatrix} = \vec{B}_{1_{\chi}} \vec{f}_{{1}} + \vec{B}_{2_{\chi}} \vec{f}_{{2}} + \vec{B}_{3_{\chi}} \vec{f}_{{3}} = \begin{bmatrix}
\vec{B}_{1_{\chi}} & \vec{B}_{2_{\chi}} & \vec{B}_{3_{\chi}}
\end{bmatrix} \begin{bmatrix}
\vec{f}_{{1}} \\ \vec{f}_{{2}} \\ \vec{f}_{{3}}
\end{bmatrix} = \vec{B}_{\chi} \vec{f}
\end{equation}
where $\vec{B}_{i_{\chi}}, i=1,2,3,$ are as shown in Appendix \ref{app:operators-1}.
Next, we introduce a pseudo-linear model for all the nonlinear entries in the vectors $\vec{f}_i$ in \eqref{eq:true_nonlinearity_separated} as
\begin{align*}
\vec{f}_{1_{\chi}} &= \begin{bmatrix}
\xi_\chi \nabla^2 p \\ \xi_\chi \nabla p \\ \xi_\chi (\nabla \cdot \vec{u}) \\ \xi_\chi \nabla \cdot  \Pi(\vec{u}, \eta_0) \\ p_\chi \nabla^2 \xi \\ p_\chi \nabla \xi \\ p_\chi \nabla \cdot \vec{u}
\end{bmatrix} = \begin{bmatrix}
\xi_\chi & 0 & 0 & 0 & 0 & 0 & 0 \\
0   & \xi_\chi \vec{I}_3 & 0 & 0 & 0 & 0 & 0 \\
0 & 0 & \xi_\chi & 0 & 0 & 0 & 0 \\
0 & 0 & 0 & \xi_\chi \vec{I}_3 & 0 & 0 & 0 \\
0 & 0 & 0 & 0 & p_\chi & 0 & 0 \\
0 & 0 & 0 & 0 & 0 & p_\chi \vec{I}_3 & 0 \\
0 & 0 & 0 & 0 & 0 & 0 & p_\chi \\
\end{bmatrix} \begin{bmatrix}
\nabla^2 p \\ \nabla p \\ \nabla \cdot \vec{u} \\ \nabla \cdot  \Pi(\vec{u}, \eta_0) \\ \nabla^2 \xi \\ \nabla \xi \\ \nabla \cdot \vec{u}
\end{bmatrix} = \bar{\Delta}_1 \vec{y}_{1_{\chi}}, \\
\vec{f}_{2_{\chi}} &= \begin{bmatrix}
\vec{u}_\chi \cdot \nabla \xi \\ \vec{u}_\chi \cdot \nabla u \\ \vec{u}_\chi \cdot \nabla v \\ \vec{u}_\chi \cdot \nabla w \\ \vec{u}_\chi \cdot \nabla p 
\end{bmatrix} = \begin{bmatrix}
\vec{u}_\chi^\text{T} & 0 & 0 & 0 & 0 \\
0 & \vec{u}_\chi^\text{T} & 0 & 0 & 0 \\
0 & 0 & \vec{u}_\chi^\text{T} & 0 & 0 \\
0 & 0 &  0 & \vec{u}_\chi^\text{T} & 0 \\
0 & 0 & 0 & 0 & \vec{u}_\chi^\text{T} \\
\end{bmatrix} \begin{bmatrix}
\nabla \xi  \\ \nabla u \\ \nabla v \\ \nabla w \\ \nabla p
\end{bmatrix} = \bar{\Delta}_2 \vec{y}_{2_{\chi}}, \\
\vec{f}_{3_{\chi}} &=  \begin{bmatrix}
(\nabla u)_\chi \cdot \nabla u \\ (\nabla v)_\chi \cdot \nabla v \\ (\nabla w)_\chi \cdot \nabla w \\ (\partial_x \vec{u})_\chi \cdot (2 \nabla u + \partial_x \vec{u}) \\ (\partial_y \vec{u})_\chi \cdot (2 \nabla v + \partial_y \vec{u}) \\ (\partial_z \vec{u})_\chi \cdot (2 \nabla w + \partial_z \vec{u}) \\ (\nabla p)_\chi \cdot \nabla \xi \\ (\nabla \cdot \vec{u})_\chi (\nabla \cdot \vec{u})
\end{bmatrix} \\
&= \begin{bmatrix}
(\nabla u)_\chi^\text{T} & 0 & 0 & 0 & 0 & 0 & 0 & 0 \\
0  &  (\nabla v)_\chi^\text{T} & 0 & 0 & 0 & 0 & 0 & 0 \\
0  &  0  &  (\nabla w)_\chi^\text{T} & 0 & 0 & 0 & 0 & 0 \\
0 &  0 & 0 & \left(\partial_x \vec{u}\right)_\chi^\text{T} & 0 & 0 & 0 & 0 \\
0 &  0 & 0 & 0 & \left(\partial_y \vec{u}\right)_\chi^\text{T} & 0 & 0 & 0 \\
0 &  0 & 0 & 0 & 0 & \left(\partial_z \vec{u}\right)_\chi^\text{T} & 0 & 0 \\
0 &  0 & 0 & 0 & 0 & 0 & (\nabla p)_\chi^\text{T} & 0 \\
0 &  0 & 0 & 0 & 0 & 0 & 0 & \left(\nabla \cdot \vec{u}\right)_\chi \\
\end{bmatrix} \begin{bmatrix}
 \nabla u \\ \nabla v  \\ \nabla w  \\ (2 \nabla u + \partial_x \vec{u})  \\ (2 \nabla v + \partial_y \vec{u})  \\ (2 \nabla w + \partial_z \vec{u})  \\ \nabla \xi \\ (\nabla \cdot \vec{u})
\end{bmatrix} \\ 
&= \bar{\Delta}_3 \vec{y}_{3_{\chi}},
\end{align*}
where the approximated/unknown variables are labeled using the subscript $\chi$.
Therefore, the resulting model for the entire vector $\vec{f} = \begin{bmatrix}
    \vec{f}_1^\text{T} & \vec{f}_2^\text{T} & \vec{f}_3^\text{T}
\end{bmatrix}^\text{T}$ can be expressed as 
\begin{equation}
\vec{f}_\chi = \begin{bmatrix}
\vec{f}_{1_{\chi}} \\ \vec{f}_{2_{\chi}} \\ \vec{f}_{3_{\chi}}
\end{bmatrix} = \text{diag} \left( \bar{\Delta}_1, \bar{\Delta}_2, \bar{\Delta}_3 \right) \begin{bmatrix}
\vec{y}_{1_{\chi}} \\ \vec{y}_{2_{\chi}} \\ \vec{y}_{3_{\chi}}
\end{bmatrix} = \text{diag}\left( \Delta_1, \Delta_2, \dots, \Delta_{11} \right) \vec{y}_\chi = \Delta \vec{y}_\chi
\end{equation}
where $\vec{y}_\chi = \begin{bmatrix}
    \vec{y}_{1_{\chi}}^\text{T} & \vec{y}_{2_{\chi}}^\text{T} & \vec{y}_{3_{\chi}}^\text{T}
\end{bmatrix}^\text{T}$ denote the modeled outputs and
\begin{equation} \label{eq:Delta sub-blocks}
\begin{split}
\Delta_1 &= \xi_\chi \vec{I}_8, \enspace \Delta_2 =  p_\chi \vec{I}_5, \enspace \Delta_3 = \vec{I}_5 \otimes \vec{u}_\chi^\text{T}, \enspace \Delta_4 = (\nabla u)_\chi^\text{T}, \enspace \Delta_5 = (\nabla v)_\chi^\text{T}, \enspace \Delta_6 = (\nabla w)_\chi^\text{T}, \\
\Delta_7 &= (\partial_x \vec{u})_\chi^\text{T}, \enspace \Delta_8 = (\partial_y \vec{u})_\chi^\text{T}, \enspace \Delta_9 = (\partial_z \vec{u})_\chi^\text{T}, \enspace \Delta_{10} = (\nabla p)_\chi^\text{T}, \enspace \Delta_{11} = (\nabla \cdot \vec{u})_\chi. \\
\end{split}
\end{equation}
The modeled output vectors $\vec{y}_{i_{\chi}}$ can be expressed in terms of the perturbed flow quantities $\vec{q}$ as 
\begin{align*}
\vec{y}_{1_{\chi}} &= \begin{bmatrix}
\nabla^2 p \\ \nabla p \\ \nabla \cdot \vec{u} \\ \nabla \cdot  \Pi(\vec{u}, \eta_0) \\ \nabla^2 \xi \\ \nabla \xi \\ \nabla \cdot \vec{u}
\end{bmatrix} = \vec{C}_{1_{\chi}} \vec{q}, \enspace
\vec{y}_{2_{\chi}} = \begin{bmatrix}
\nabla \xi  \\ \nabla u \\ \nabla v \\ \nabla w \\ \nabla p
\end{bmatrix} = \vec{C}_{2_{\chi}} \vec{q}, \notag \\
\vec{y}_{3_{\chi}} &= \begin{bmatrix}
 \nabla u \\ \nabla v  \\ \nabla w  \\ (2 \nabla u + \partial_x \vec{u})  \\ (2 \nabla v + \partial_y \vec{u})  \\ (2 \nabla w + \partial_z \vec{u})  \\ \nabla \xi \\ (\nabla \cdot \vec{u})
\end{bmatrix} = \vec{C}_{{3_{1}}_\chi} \begin{bmatrix}
 \nabla u \\ \nabla v  \\ \nabla w  \\  \partial_x \vec{u}  \\ \partial_y \vec{u}  \\ \partial_z \vec{u}  \\ \nabla \xi 
\end{bmatrix} = \vec{C}_{{3_{1}}_\chi} \vec{C}_{{3_{2}}_\chi} \vec{q}.
\end{align*}
The expressions of the above operators are provided in Appendix~\ref{app:operators-1}.
Hence, the mapping from perturbed flow states $\vec{q}$ to the modeled outputs $\vec{y}_\chi$ 
is given by 
\begin{equation}
\vec{y}_\chi = \begin{bmatrix}
\vec{y}_{1_{\chi}} \\ \vec{y}_{2_{\chi}} \\ \vec{y}_{3_{\chi}}
\end{bmatrix} = \begin{bmatrix}
\vec{C}_{1_{\chi}} \\ \vec{C}_{2_{\chi}} \\ \vec{C}_{{3_{1}}_\chi}  \vec{C}_{{3_{2}}_\chi} 
\end{bmatrix} \vec{q} = \vec{C}_\chi \vec{q}.
\end{equation}
Finally, the overall perturbation dynamics obtained through the structured I/O modeling is given by 
\begin{align}
\begin{bmatrix} 
\partial_t \xi \\ \partial_t \vec{u} \\ \partial_t p 
\end{bmatrix} &= \begin{bmatrix} L_\xi (\vec{q}) \\ L_{\vec{u}} (\vec{q}) \\ L_p (\vec{q}) \end{bmatrix} + \vec{B}_\chi \vec{f}_\chi, \notag \\
\vec{y}_\chi &= \vec{C}_\chi \vec{q}, \label{eq:structuredI/O_continuous_form} \\
\vec{f}_\chi &= \Delta \vec{y}_\chi, \notag
\end{align}
a schematic of which is provided in Fig. \ref{subfig:modeled_system_schematic}.

\subsection{Spectral Discretization}  \label{Sec: Spectral Discretization}
Now, we employ a Fourier transform under the assumption that the flow is homogeneous in the streamwise and spanwise directions and in time. 
This leads to the following triple Fourier transform that relates the true flow state $\vec{q}(x,y,z,t)$ with the transformed state $\tilde{\vec{q}}(y;k_x,k_z,\omega)$ as
\begin{equation}
    \vec{q}(x,y,z,t) = \int_{-\infty}^{\infty} \int_{-\infty}^{\infty} \int_{-\infty}^{\infty} \tilde{\vec{q}}(y;k_x,k_z,\omega) \exp \left(\vec{i}\left(\omega t + k_x x + k_z z \right) \right) d\omega dk_x dk_z 
\end{equation}
where $k_x$ and $k_z$ are the wavenumbers along the streamwise ($x$) and spanwise ($z$) directions, respectively, and $\omega$ is the temporal frequency.
A similar transform is carried out on the modeled nonlinear terms $\vec{f}_\chi$ and the modeled outputs $\vec{y}_\chi$.
These Fourier transformed quantities are then numerically discretized using a Chebyshev spectral collocation method \citep{trefethen2000spectral} in the wall-normal ($y$) direction, and the discretized quantities are denoted using a $\hat{(\cdot)}$.
Therefore, the discretized form of the equations in \eqref{eq:structuredI/O_continuous_form} can be expressed in the following form:
\begin{equation} \label{eq:discretized system}
\begin{split}
\vec{i} \omega \hat{\vec{q}} &= \hat{\vec{L}}(k_x, k_z) \hat{\vec{q}} + \hat{\vec{B}}_\chi \hat{\vec{f}}_\chi  \\
\hat{\vec{y}}_\chi &= \hat{\vec{C}}_\chi (k_x, k_z) \hat{\vec{q}} \\
\hat{\vec{f}}_\chi &= 
\text{diag}\left( \hat{\Delta}_1, \hat{\Delta}_2, \dots, \hat{\Delta}_{11} \right) \hat{\vec{y}}_\chi = \hat{\Delta} \hat{\vec{y}}_\chi
\end{split}
\end{equation}
where $\hat{\vec{L}}(k_x, k_z) \in \mathbb{C}^{n_q \times n_q}, \hat{\vec{C}}_\chi (k_x, k_z) \in \mathbb{C}^{n_y \times n_q}$ are the discretized operators, 
and $\hat{\vec{B}}_\chi \in \mathbb{R}^{n_q \times n_f}$ is the dimensionally consistent form of $\vec{B}_\chi$ for the discretized variables.
Thus, we have $n_q = 5 N_y$ where $N_y$ denotes the number of Chebyshev collocation points in the wall-normal direction. 
Also in the above, each $\hat{\Delta}_i$ is a complex block matrix---referred to as a \emph{complex full-block}---of appropriate dimensions and represents the discrete counterpart of the associated $\Delta_i$ shown in \eqref{eq:Delta sub-blocks}. 
For example, $\hat{\Delta}_1$ and $\hat{\Delta}_2$, which are the discretized versions of the uncertainties $\Delta_1 = \xi_\chi \vec{I}_8$ and $\Delta_2 =  p_\chi \vec{I}_5$ respectively, consist of eight and five $N_y \times N_y$ full-blocks, respectively.  
Similarly,  $\hat{\Delta}_3$ contains five $N_y \times 3N_y$ complex full-blocks, with each full-block representing a discretization of $\vec{u}^\text{T}_\chi$ (see \eqref{eq:Delta sub-blocks} for comparison).
%
%
Therefore, $\hat{\Delta}$ here contains twenty six complex full-blocks and the expanded $\hat{\Delta}$ can be expressed as 
\begin{equation} \label{eq:uncertainty_discretized_nonrep_fullblocks}
    \hat{\Delta}=\text{diag} \left(\hat{\Delta}_{F_{1}}, \hat{\Delta}_{F_{2}}, \dots, \hat{\Delta}_{F_{26}} \right)
\end{equation}
where each $\hat{\Delta}_{F_{i}} \in \mathbb{C}^{m_i \times n_i}$ such that $\sum_{i=1}^{26} m_i = n_f$ and $\sum_{i=1}^{26} n_i = n_y$ for consistent dimensions. 
Furthermore, the dimensions of each of these full-blocks are as follows:
\begin{align}
    m_i &= n_i = N_y, i=1,2,\dots,13, \notag \\
    m_i &= N_y, n_i = 3N_y, i=14,15,\dots,25, \label{eq:structured_uncertainty_dimensions} \\
    m_i &= n_i = N_y, i=26. \notag
\end{align}
Note that the full-blocks in \eqref{eq:uncertainty_discretized_nonrep_fullblocks} should ideally be a mixture of repeated and non-repeated ones due to the structured I/O modeling.
But, we introduce a simplifying assumption that all $\Delta_{F_{i}}$ in \eqref{eq:uncertainty_discretized_nonrep_fullblocks} are non-repeated, which enables efficient computations for the structured I/O analysis (see Section \ref{Sec:Structured I/O Analysis}). 
Exploiting additional structure in the uncertainty model will be part of our future work.

The system of equations in \eqref{eq:discretized system} can be interpreted as a feedback interconnection between a linear time-invariant~(LTI) system and a structured uncertainty $\hat{\Delta}$. 
In this interpretation, the inputs and outputs of the LTI system are $\hat{\vec{f}}_\chi \in \mathbb{C}^{n_f}$ and $\hat{\vec{y}}_\chi \in \mathbb{C}^{n_y}$, respectively.
Furthermore, the I/O relationship can be written as 
\begin{equation} \label{eq:IO-operator_discretized_perturbation_dynamics}
\hat{\vec{y}}_\chi = \mathcal{H}(k_x, k_z, \omega) \hat{\vec{f}}_\chi
\end{equation}
where $\mathcal{H}(k_x, k_z, \omega) = \hat{\vec{C}}_\chi (k_x, k_z) (\vec{i} \omega \vec{I}_{n_q} - \hat{\vec{L}}(k_x, k_z))^{-1} \hat{\vec{B}}_\chi$ is the frequency response operator that maps the inputs $\hat{\vec{f}}_\chi$ to the corresponding outputs $\hat{\vec{y}}_\chi$ at a given tuple $(k_x, k_z, \omega)$.
%
%
{Figure \ref{fig:modeling_approximation} schematically outlines the two forms of the structured I/O perturbation dynamics: before discretization (Fig. \ref{subfig:modeled_system_schematic}) and after discretization (Fig. \ref{subfig:modeled_system_after_discretization}).}

\subsection{Chu Energy Expression for Quadratic Formulation}
The I/O analysis requires an inner product or norm of the form
\begin{equation*}
    \langle \vec{q}_1, \vec{q}_2 \rangle =\int_\Gamma \left(\vec{q}_1\right)^\dagger W_E \vec{q}_2 \mathrm{d}\vec{x}
\end{equation*}
where the operator $W_{E}$ can be used to specify a particular choice for the inner product.
{In particular, Chu energy \citep{chu1965energy} has been extensively utilized for the inner product in 
I/O analysis of compressible flows \citep{baeJFM2019,baeAIAA2020, dawsonAIAA2019, nicholsAIAA2019, chen2023linear, dwivedi2019reattachment}.}
The expression of Chu energy in terms of the variables $\vec{q}^c=(\rho,u,v,w,T)$ is well known and can be derived from first principles by enforcing the principles of energy conservation and eliminating pressure-related compression work, as shown in \cite{hanifi1996transient}. 

Here, we derive an expression for the Chu energy in terms of the variables used in the quadratic formulation, $\vec{q} = (\xi, u,v,w,p)$,
based on the principles outlined in \cite{karban2020ambiguity}.
We start by defining the nonlinear transformation between $\vec{q}^c$ and $\vec{q}$ as
\begin{equation*}
\vec{q}^c = (\rho,u,v,w,T) = \left(\frac{1}{\xi},u,v,w,p\xi \right) = g(\xi, u,v,w,p) = g(\vec{q}).
\end{equation*}
We need the expression of the Jacobian of $g(\cdot)$ with respect to $\vec{q}$, which is given by
\begin{equation*}
\partial_{\vec{q}} g = \begin{bmatrix}
-\frac{1}{\xi^2} & 0 & 0 & 0 & 0 \\
0 & 1 & 0 & 0 & 0 \\
0 & 0 & 1 & 0 & 0 \\
0 & 0 & 0 & 1 & 0 \\
p & 0 & 0 & 0 & \xi \\
\end{bmatrix}.
\end{equation*}
Let the Chu energy expressions in terms of $\vec{q}^c$ and $\vec{q}$ be denoted by $W_{E}^c$ and $W_{E}$, respectively, with $W_{E}^c$ given by \citep{hanifi1996transient}
\begin{equation*}
W_{E}^c = \text{diag}\left(\frac{T_0}{\rho_0 \gamma M_r^2}, \rho_0, \rho_0, \rho_0, \frac{\rho_0}{\gamma(\gamma-1)M_r^2T_0} \right).
\end{equation*} 
%
The equivalent expression for $W_{E}$ is then given by
\begin{equation*}
W_{E} = \left( \partial_{\vec{q}} g \big\rvert_{\vec{q} = \vec{q}_{{0}}} \right)^\dagger W_{E}^c \left( \partial_{\vec{q}} g \big\rvert_{\vec{q} = \vec{q}_{{0}}} \right)
\end{equation*}
where $\vec{q}_{{0}}$ denotes the base flow described in terms of the flow variables $\vec{q}$.
Carrying out these calculations leads to the following expression: 
\begin{align*}
W_{E} = \begin{bmatrix}
\frac{T_0}{\xi_0^3 \gamma M_r^2} + \frac{p_0^2 \rho_0}{\gamma(\gamma-1)M_r^2T_0} & 0 & 0 & 0 & \frac{p_0}{\gamma(\gamma-1)M_r^2T_0} \\
0 & \rho_0 & 0 & 0 & 0 \\
0 & 0 & \rho_0 & 0 & 0 \\
0 & 0 & 0 & \rho_0 & 0 \\
\frac{p_0}{\gamma(\gamma-1)M_r^2T_0} & 0 & 0 & 0 & \frac{\xi_0}{\gamma(\gamma-1)M_r^2T_0} \\
\end{bmatrix}
\end{align*}
which subsequently will be used as the inner product for the compressible Couette flow results in Section \ref{Sec: Results}. 

\subsection{Structured I/O Analysis: Structured Singular Value} \label{Sec:Structured I/O Analysis}
The structured I/O analysis utilizes the concept of structured singular value (often simply referred to as `$\mu$') which is a robust analysis tool for LTI systems subject to structured uncertainty. 
We will start the discussion with the matrix case by recalling the definition of $\mu$ for a given matrix $\vec{H} \in \mathbb{C}^{n \times m}$ and a set of structured matrices $\vec{\hat{\Delta}} \subset \mathbb{C}^{m \times n}$.  
\begin{defin}[\cite{packard1993, zhou1996robust}] \label{def:mu_definition}
For a given matrix $\vec{H} \in \mathbb{C}^{n \times m}$ and a set of structured matrices $\vec{\hat{\Delta}} \subset \mathbb{C}^{m \times n}$, the structured singular value is defined as 
\begin{equation}
    \mu_{\vec{\hat{\Delta}}}(\vec{H}) = \frac{1}{\min(\|\hat{\Delta}\|_2: \hat{\Delta} \in \vec{\hat{\Delta}}, \det(\mathbf{I}_n - \vec{H} \hat{\Delta}) = 0)} .
    \label{eq:ssv}
\end{equation}
If there does not exist any $\hat{\Delta} \in \vec{\hat{\Delta}}$ such that $\det(\mathbf{I}_n - \vec{H} \hat{\Delta}) = 0$, then $\mu_{\vec{\hat{\Delta}}}(\vec{H}) = 0$.
\end{defin}
Note that $\mu_{\vec{\hat{\Delta}}}(\vec{H})$ depends both on the matrix $\vec{H}$ and the set $\vec{\hat{\Delta}}$. 
However, for simplicity, we will omit the subscript $\hat{\mathbf{\Delta}}$ when the uncertainty structure is clear from the context of the discussion. 
{Also, $\mu$ is inversely related to the smallest structured uncertainty $\hat{\Delta}$ (in the sense of $\|\cdot\|_2$) that can make the feedback interconnection of the form shown in Fig.~\ref{subfig:modeled_system_after_discretization} unstable (see Remark 3.4 in \cite{packard1993} for more details).}
Thus, $\mu$ is closely related to flow stability, i.e., a large value indicates that the system is sensitive to small perturbations that can cause instability and vice versa, which is a consequence of a variation of the small-gain condition for structured uncertainties \citep{zhou1996robust}. 
%

%
For the dynamics of perturbations, computing $\mu$ reduces to carrying out computations on the frequency response operator $\mathcal{H}(k_x, k_z, \omega)$ in \eqref{eq:IO-operator_discretized_perturbation_dynamics} at a given wavenumber pair $(k_x, k_z)$ on a grid of temporal frequencies $\omega$ for a set of structured uncertainties comprising non-repeated complex full-blocks as in \eqref{eq:uncertainty_discretized_nonrep_fullblocks}.
This approach essentially provides information about wavenumber pairs $(k_x, k_z)$ where $\mu$ is higher, 
thereby indicating an instability mechanism. 
However, exactly computing the $\mu$ is NP-hard for a general uncertainty structure \citep{braatz1994computational, poljak1993checking, coxson1994computational}. 
Thus, it is a common practice to compute upper and lower bounds on the $\mu$ instead. 
Here, we will outline computationally efficient methods for computing the upper and lower bounds on $\mu$ for the particular uncertainty structure/set of interest (i.e., non-repeated complex full-blocks).
Note that while the upper bound provides a sufficient condition for robust stability, the lower bound serves as a sufficient condition for instability \citep{zhou1996robust,dullerud13,packard1993,young90,young1992}.
Details on the bounds and the accompanying computations are provided next.

\subsubsection{Upper Bound Calculation} \label{sec:upper_bound_calculation}
This section provides an overview on computing the $\mu$ upper bound for a set of non-repeated complex full-block uncertainties.
First, note that we have $\mu_{\vec{\hat{\Delta}}} (\vec{H}) \leq \| \vec{H} \|_2$ 
for any given matrix $\vec{H} \in \mathbb{C}^{n \times m}$ \citep{packard1993}. 
Furthermore, for each set of uncertainties $\vec{\hat{\Delta}} \subset \mathbb{C}^{m \times n}$, there are sets of non-singular, commuting matrices $\mathbb{D}_1 \subset \mathbb{C}^{n \times n}, \ \mathbb{D}_2 \subset \mathbb{C}^{m \times m}$ such that $ \hat{\Delta} \vec{D}_1 = \vec{D}_2 \hat{\Delta}$ for any $\vec{D}_1 \in \mathbb{D}_1, \vec{D}_2 \in \mathbb{D}_2, \hat{\Delta} \in \vec{\hat{\Delta}}$. 
Therefore, using this relationship between the matrices and the structured uncertainty and Sylvester's determinant identity\footnote[4]{$\det(\mathbf{I}_n - \vec{A} \vec{B}) = \det(\mathbf{I}_m - \vec{B} \vec{A})$ for any $\vec{A} \in \mathbb{C}^{n \times m}$, $\vec{B} \in \mathbb{C}^{m \times n}$}, we have 
\begin{equation*}
    \det(\mathbf{I}_n - \vec{H} \hat{\Delta}) = \det(\mathbf{I}_n - \vec{H} \vec{D}_2^{-1} \hat{\Delta} \vec{D}_1) = \det(\mathbf{I}_m - \hat{\Delta} \vec{D}_1 \vec{H} \vec{D}_2^{-1}) = \det(\mathbf{I}_n - \vec{D}_1 \vec{H} \vec{D}_2^{-1} \hat{\Delta})
\end{equation*}
which means $\mu_{\vec{\hat{\Delta}}}(\vec{H}) = \mu_{\vec{\hat{\Delta}}} (\vec{D}_1 \vec{H} \vec{D}_2^{-1}) \leq \| \vec{D}_1 \vec{H} \vec{D}_2^{-1} \|_2$.
Therefore, it is possible to  tighten the upper bound by computing optimal values of $\vec{D}_1 \in \mathbb{D}_1, \vec{D}_2 \in \mathbb{D}_2$.
Since the focus here is 
on uncertainties of the form in \eqref{eq:uncertainty_discretized_nonrep_fullblocks}, 
we define a generic set of non-repeating full-block uncertainties as 
\begin{equation*}
    \vec{\hat{\Delta}}_{F} := \{\text{diag}(\hat{\Delta}_{F_{1}}, \hat{\Delta}_{F_{2}}, \dots, \hat{\Delta}_{F_{N_{\Delta}}}): \hat{\Delta}_{F_{i}} \in \mathbb{C}^{m_i \times n_i} \} \subset \mathbb{C}^{m \times n}
\end{equation*}
where $\sum_{i=1}^{N_\Delta} m_i = m$ and $\sum_{i=1}^{N_\Delta} n_i = n$ for consistent dimensions.
Due to the above structure of $\vec{\hat{\Delta}}_{F}$, the corresponding sets of the scaling matrices take the following block-diagonal forms:
\begin{equation} \label{eq:D-scales_non-repeated_fullblocks}
\begin{split}
\mathbb{D}_{1_{F}} & := \{ \text{diag}(d_1 \vec{I}_{n_{1}}, d_2 \vec{I}_{n_{2}}, \dots, d_{N_{\Delta}} \vec{I}_{n_{N_{\Delta}}}): d_i >0 , \ i=1, 2, \dots, N_{\Delta} \} \subset \mathbb{R}^{n \times n}, \\
\mathbb{D}_{2_{F}} & := \{ \text{diag}(d_1 \vec{I}_{m_{1}}, d_2 \vec{I}_{m_{2}},\dots, d_{N_{\Delta}} \vec{I}_{m_{N_{\Delta}}}): d_i >0 , \ i=1, 2, \dots, N_{\Delta} \} \subset \mathbb{R}^{m \times m}.
\end{split}
\end{equation} 
Let $\vec{d} = \begin{bmatrix}
    d_1 & d_2 & \cdots & d_{N_{\Delta}}
\end{bmatrix}^\text{T} \in \mathbb{R}^{N_{\Delta}}$, and we utilize $\vec{D}_1 (\vec{d})$, $\vec{D}_2 (\vec{d})$ to denote scaling matrices belonging to the sets $\mathbb{D}_{1_{F}}$, $\mathbb{D}_{2_{F}}$, respectively.
Thus, the upper bound can be tightened by computing the optimal $\vec{d} \in \mathbb{R}^{N_{\Delta}}$ such that
\begin{equation} \label{eq:D-scale upper bound}
 \mu_{\vec{\hat{\Delta}}_F} (\vec{H}) \leq \min_{\vec{d} \in \mathbb{R}^{N_{\Delta}} } \| \vec{D}_1 (\vec{d}) \vec{H} \left(\vec{D}_2(\vec{d})\right)^{-1} \|_2. 
\end{equation}
This upper bound is called the \emph{$D$-scale upper bound}.

The optimization problem 
in \eqref{eq:D-scale upper bound} 
can be posed as a generalized eigenvalue problem \citep{packard1993, zhou1996robust}. 
However, well-established methods for solving a generalized eigenvalue problem, 
such as the method of centers \citep{boyd1993method}, can be computationally expensive due to the large-dimensionality of compressible fluid flow analysis problems (see, for example, the discussion in \cite{mushtaq2024algorithms}).
Instead, a Frobenius norm-based relaxation of the right-hand side of \eqref{eq:D-scale upper bound} leads to a computationally cheaper alternative and is often sufficient for practical purposes \citep{beck1992mixed}.
In this case, the $D$-scale upper bound for a given matrix $\vec{H}$ becomes
\begin{equation}
\label{eq:osbopt}
    \mu_{\vec{\hat{\Delta}}_F}(\vec{H}) \leq \min_{\vec{d} \in \mathbb{R}^{N_{\Delta}} } \| \vec{D}_1 (\vec{d}) \vec{H} \left(\vec{D}_2(\vec{d})\right)^{-1} \|_F. 
\end{equation}
The optimization problem on the right-hand side of \eqref{eq:osbopt} can be efficiently solved using a variation/generalization of the standard Osborne's iteration \citep{osborne1960}, and Algorithm \ref{alg:Osborne} summarizes the process of computing the optimal $D$-scales using one such variant/generalization of Osborne's iteration (see Section 3.1 in \cite{mushtaq2024algorithms} for details).
It should be noted that although the algorithm here is stated for a fixed number ($k_m$) of iterations, additional stopping criteria (such as terminating when the updates in the objective function value or cost falls below a pre-specified threshold) 
can be easily incorporated as needed.
\begin{algorithm}[H]
\caption{Upper Bound: Osborne's Iteration}
\begin{algorithmic}[1]
  \State (Initialization) Choose the maximum number of iterations $k_m$. Set $k=0$ and $\vec{H}^{[0]}=\vec{H}$.
  \While{$k < k_m$}
  \State \begin{varwidth}[t]{0.95\linewidth} Partition $\vec{H}^{[k]}$ into $n_i\times m_j$ sub-blocks $\vec{H}_{ij}$ according to the dimensions of the full-blocks in the set $\vec{\hat{\Delta}}_{F}$. \end{varwidth}
  \State \begin{varwidth}[t]{0.95\linewidth} Compute each optimal scale using $d_{i}^\star = \left( \frac{\sum_{r = 1, r\neq i}^{N}\|\vec{H}_{ri}\|^2_F}{\sum_{r = 1, r\neq i}^{N}\|\vec{H}_{ir}\|^2_F} \right)^{1/4}, i=1,2,\dots,N_\Delta.$ \end{varwidth}
  \State \begin{varwidth}[t]{0.95\linewidth} Get the optimal $D$-scale matrices $\vec{D}_1^\star= \vec{D}_1(\vec{d}^\star) = \text{diag}\left(d_1^\star \vec{I}_{n_{1}}, d_2^\star \vec{I}_{n_{2}}, \dots, d_{N_{\Delta}}^\star \vec{I}_{n_{N_{\Delta}}}\right)$ and $\vec{D}_2^\star= \vec{D}_2(\vec{d}^\star) = \text{diag}\left(d_1^\star \vec{I}_{m_{1}}, d_2^\star \vec{I}_{m_{2}}, \dots, d_{N_{\Delta}}^\star \vec{I}_{m_{N_{\Delta}}}\right)$. \end{varwidth}
  \State Set $\vec{H}^{[k+1]} = \vec{D}_1^\star \vec{H}^{[k]} \left(\vec{D}_2^\star\right)^{-1}$.
  \State Set $k=k+1$
  \EndWhile
\State (Output) Compute the upper bound $\alpha = \|\vec{H}^{[k_m]}\|_2$ and the optimal $D$-scale matrices $\vec{D}_1^\star, \vec{D}_2^\star$.
\end{algorithmic}
\label{alg:Osborne}
\end{algorithm}

The uncertainty in \eqref{eq:uncertainty_discretized_nonrep_fullblocks} belongs to the set $\hat{\vec{\Delta}}_F$ with $N_\Delta = 26$ and dimensions of each full-block as shown in \eqref{eq:structured_uncertainty_dimensions}.
In terms of the frequency response operator $\mathcal{H}(k_x, k_z, \omega)$ at a given wavenumber pair $(k_x, k_z)$, 
we choose the `best' upper bound, denoted by $\alpha_\mu (k_x, k_z)$, as the maximum of the upper bounds computed on a temporal frequency ($\omega$) grid. 
This is given by 
\begin{equation} \label{eq:muub_operator}
    \alpha_\mu (k_x, k_z) = \max_{\omega \in \Omega} \left[ \min_{\vec{d} \in \mathbb{R}^{26}} \| \vec{D}_1(\vec{d}) \mathcal{H}(k_x, k_z, \omega) \left(\vec{D}_2(\vec{d})\right)^{-1} \|_F \right]
\end{equation}
where $\Omega \subset \mathbb{R}$ is the $\omega$ grid. 
It should also be noted that we consider both positive and negative temporal frequencies (i.e., $\omega \in \mathbb{R}$) since the system matrices involved here are complex-valued and the corresponding frequency response is not necessarily symmetric about $\omega=0$.

\subsubsection{Lower Bound Calculation and Structured I/O Modes} \label{Sec:Lower bound and mode shapes}
This section briefly summarizes the procedure for computing a lower bound on $\mu$ for structured uncertainties comprising non-repeated complex full-blocks. 
To that end, we 
utilize the power iteration method (or, simply the power method) for complex uncertainties in \cite{packard1988power, packard1993}, which is an efficient method to find uncertainties $\hat{\Delta} \in \hat{\vec{\Delta}}$ that satisfy the determinant condition in Definition \ref{def:mu_definition}.
Note that any particular $\hat{\Delta} \in \hat{\vec{\Delta}}$ such that
$\det(\vec{I}_n-\vec{H} \hat{\Delta})=0$ yields a lower bound $\mu(\vec{H}) \geq \frac{1}{\|\hat{\Delta}\|_2}$.
The exact value of $\mu(\vec{H})$ corresponds to the ``smallest'' $\hat{\Delta} \in \hat{\vec{\Delta}}$ such that $\det(\vec{I}_n-\vec{H} \hat{\Delta})=0$ (see Definition \ref{def:mu_definition}).
The determinant condition is equivalent to finding $\hat{\Delta} \in \hat{\mathbf{\Delta}}$ and non-zero vectors $\vec{p} \in \mathbb{C}^{n}$ and $\vec{q} \in \mathbb{C}^{m}$ such that $\vec{p}=\vec{H} \vec{q}$ and $\vec{q}= \hat{\Delta} \vec{p}$, and the power iteration computes these quantities by exploring optimality conditions associated with $\mu$ (see \cite{packard1988power} for details).
Thus, the power iteration computes complex-valued vectors $\vec{p}$ and $\vec{q}$ 
that are consistent with the input-output relationships of the feedback interconnection, meaning that these vectors abide by 
the structural constraints originating from the uncertainty---an aspect not captured in traditional resolvent-based modal analysis of fluid flows (see, for example, \cite{dawsonAIAA2019}). 
Therefore, power iteration provides a pathway for structured modal analysis.\footnote[4]{See \cite{mushtaq2023structured,mushtaq2023riblets} for such modal analysis of incompressible channel flows and turbulent flows over riblets.} 
Although the power iteration usually yields a lower bound on $\mu$, these lower bounds are often accurate in practice \citep{packard1988power, packard1993}.
Moreover, the particular uncertainty returned by the power iteration can be studied further for insight.
In the context of analyzing fluid flow perturbations, the uncertainty returned by power iteration can be interpreted as a collection of time-invariant gains that describe the spatial structures associated with the quantities modeled as the uncertainty. 
For example, in case of incompressible flows, the structured uncertainty models the velocity field related to the quadratic convective nonlinearity in the incompressible NSE \citep{liu21, mushtaq2024algorithms}, and the gains in that case represent spatial structures of the velocity field in the wall-normal coordinate \citep{LiuACC2023}. 
These gains are also closely linked with `optimal' perturbations responsible for maximum amplification of the flow perturbations \citep{LiuACC2023}.
We are currently working on extending these arguments for the compressible flows \citep{bhattacharjeehemati2025}.

Consider a given matrix $\vec{H} \in \mathbb{C}^{n \times m}$ and a set of structured uncertainties with two non-repeated full blocks as 
\begin{equation}
    \vec{\hat{\Delta}} = \{\text{diag}(\hat{\Delta}_{F_{1}}, \hat{\Delta}_{F_{2}}): \hat{\Delta}_{F_{1}} \in \mathbb{C}^{m_1 \times n_1}, \hat{\Delta}_{F_{2}} \in \mathbb{C}^{m_2 \times n_2} \} 
\end{equation}
where $m_1+m_2=m$ and $n_1+n_2=n$ for consistent dimensions.
We will describe the power iteration for the above set of structured uncertainties, which can be generalized for any number of full-blocks by simply duplicating the formulae provided here.
The power iteration is described in terms of vectors $\vec{a}, \vec{z}\in \mathbb{C}^{n}$ and $\vec{b} ,\vec{w} \in \mathbb{C}^{m}$.
Now, these vectors are partitioned according to the dimensions of the full-blocks as 
\begin{equation*}
    \vec{a}=\begin{bmatrix}
        \vec{a}_1 \\ \vec{a}_2
    \end{bmatrix}, \
    \vec{z}=\begin{bmatrix}
        \vec{z}_1 \\ \vec{z}_2
    \end{bmatrix}, \
    \vec{b}=\begin{bmatrix}
        \vec{b}_1 \\ \vec{b}_2
    \end{bmatrix}, \
    \vec{w}=\begin{bmatrix}
        \vec{w}_1 \\ \vec{w}_2
    \end{bmatrix},
\end{equation*}
where, for example, $\vec{a}_1 \in \mathbb{C}^{n_1}$, $\vec{a}_2 \in \mathbb{C}^{n_2}$, $\vec{b}_1 \in \mathbb{C}^{m_1}$, $\vec{b}_2 \in \mathbb{C}^{m_2}$. 
The power iteration is then defined based on the following set of equations for some $\beta>0$:
\begin{subequations}
\label{eq:SSValign}
\begin{align}
\label{eq:SSValignA}
& \beta \vec{a} = \vec{H} \vec{b} \\
\label{eq:SSValignB}
& \vec{z}_1= \frac{\|\vec{w}_1\|_2}{\|\vec{a}_1\|_2} \vec{a}_1, \,\,
\vec{z}_2 = \frac{\|\vec{w}_2\|_2}{\|\vec{a}_2\|_2} \vec{a}_2\\
\label{eq:SSValignC}
& \beta \vec{w} = \vec{H}^\dagger \vec{z} \\
\label{eq:SSValignD}
& \vec{b}_1 = \frac{\|\vec{a}_1\|_2}{\|\vec{w}_1\|_2} \vec{w}_1, \,\,
\vec{b}_2 = \frac{\|\vec{a}_2\|_2}{\|\vec{w}_2\|_2} \vec{w}_2
\end{align}
\end{subequations}
where $\vec{a}$ and $\vec{w}$ are chosen to be unit norm.
The details for the derivation of these equations can be found in \cite{packard1988power,packard1993}.
Note that \eqref{eq:SSValignD} always implies $\|\vec{b}_1\|_2 = \|\vec{a}_1\|_2$ and $\|\vec{b}_2\|_2 = \|\vec{a}_2\|_2$. 
Hence, there are matrices
$\vec{Q}_i \in \mathbb{C}^{m_i \times n_i}$ 
with $\|\vec{Q}_i\|_2 = 1, i=1,2$ such that $\vec{b}_i = \vec{Q}_i \vec{a}_i, i=1,2$. 
Finally, define $\vec{q}:= \vec{b}$,
$\vec{p} := \beta \vec{a}$ and
$\hat{\Delta} := \frac{1}{\beta} \text{diag}(\vec{Q}_1, \, \vec{Q}_2)$.
It can be verified from \eqref{eq:SSValignA} that $\vec{p} = \vec{H} \vec{q}$.
Moreover, by construction, we have $\vec{q} = \hat{\Delta} \vec{p}$ and $\|\hat{\Delta}\|_2
=\frac{1}{\beta}$ and therefore, $\hat{\Delta} \in \hat{\vec{\Delta}}$ satisfies the
determinant condition and yields the lower bound $\mu(\vec{H}) \geq \frac{1}{\|\hat{\Delta}\|_2}=\beta$.

The power iteration is summarized in Algorithm \ref{alg:piter}, which works by iterating through the expressions in \eqref{eq:SSValign}.
For simplicity of presentation here, the algorithm is stated for a fixed number of iterations $k_m$ (similar to Algorithm \ref{alg:Osborne}), but more sophisticated stopping criteria (e.g., terminating when the various vectors have small updates as measured in the Euclidean norm) can be easily incorporated as needed.
The iterations for the vectors $b^{[k]},w^{[k]}$ can be initialized with some unit-norm random vectors in $\mathbb{C}^{m}$. 
However, as suggested in \cite{packard1988power}, the initial values $b^{[0]},w^{[0]}$ can be specifically chosen as the right singular vector associated with the largest singular value of $\vec{D}_1^\star \vec{H} \left(\vec{D}_2^\star\right)^{-1}$, where $\vec{D}_1^\star$, $\vec{D}_2^\star$ are obtained using a variant/generalization of Osborne's iteration (e.g., refer to Algorithm \ref{alg:Osborne}). 
Note that this initialization works well in practice and we utilize it for all the numerical results in this paper.
\begin{algorithm}[H]
\caption{Lower Bound: Power Iteration}
\begin{algorithmic}[1]
\State (Initialization) Choose the maximum number of iterations $k_m$ and set $k=0$.
Select $\vec{b}^{[0]}, \vec{w}^{[0]} \in \mathbb{C}^{m}$ as some unit-norm vectors, and 
set $\vec{a}^{[0]} = \vec{z}^{[0]} = 0\in \mathbb{C}^{n}$.
\While{$k < k_m$}
  \State{\eqref{eq:SSValignA}: $\beta:= \|\vec{H} \vec{b}^{[k]}\|_2$ and $\vec{a}^{[k+1]}:= \vec{H} \vec{b}^{[k]}/\beta$.}
  \State \eqref{eq:SSValignB}: Use $(\vec{a}^{[k+1]},\vec{w}^{[k]})$ to compute $\vec{z}^{[k+1]}$.
  \State \eqref{eq:SSValignC}: 
  $\beta:= \|\vec{H}^\dagger \vec{z}^{[k+1]}\|_2$ and $\vec{w}^{[k+1]}:= \vec{H}^\dagger \vec{z}^{[k+1]}/\beta$.
  \State \eqref{eq:SSValignD}: Use $(\vec{a}^{[k+1]},\vec{w}^{[k+1]})$ to compute $\vec{b}^{[k+1]}$.
  \State Set $k = k + 1$.
\EndWhile
\State (Output) Use $\vec{a}^{[k_m]}$, $\vec{b}^{[k_m]}$ and $\beta$ to compute $\vec{p}$, $\vec{q}$ and $\hat{\Delta}$.
\end{algorithmic}
\label{alg:piter}
\end{algorithm}

For the structured I/O analysis, we compute vectors of the modeled nonlinearity $\hat{\vec{f}}_\chi \in \mathbb{C}^{n_f}$ and modeled outputs $\hat{\vec{y}}_\chi \in \mathbb{C}^{n_y}$ that satisfy the input-output relations, i.e., $\hat{\vec{f}}_\chi = \hat{\Delta} \hat{\vec{y}}_\chi$ and $\hat{\vec{y}}_\chi = \mathcal{H}(k_x, k_z, \omega) \hat{\vec{f}}_\chi$, by executing the power iteration at a given value of the tuple $(k_x, k_z, \omega)$. 
The vectors $\hat{\vec{f}}_\chi$ and $\hat{\vec{y}}_\chi$ returned by the power iteration are subsequently used to compute the forcing and response modes, respectively. 
The forcing modes are computed using $\begin{bmatrix} f_\xi & f_{\vec{u}}^\text{T} & f_p \end{bmatrix}_\chi^\text{T} = \hat{\vec{B}}_\chi \hat{\vec{f}}_\chi$, where $\hat{\vec{B}}_\chi$ is the discretized operator related to the structured I/O modeling.
For the response modes, we first isolate the vector $\hat{\vec{y}}_{2_{\chi}}$ from $\hat{\vec{y}}_\chi$ returned by power iteration. 
The response modes are then computed using $\hat{\vec{q}}=\hat{\vec{C}}_{2_{\chi}}^\ddagger \hat{\vec{y}}_2$, where $\hat{\vec{C}}_{2_{\chi}}$ denotes the discretized operator corresponding to ${\vec{C}}_{2_{\chi}}$ in \eqref{eq:output_operators_continuous_form_1}.
Also, similar to the computation of $\alpha_\mu (k_x,k_z)$ in \eqref{eq:muub_operator}, the `best' lower bound for the frequency response operator $\mathcal{H}(k_x, k_z, \omega)$ at a given wavenumber pair $(k_x, k_z)$, denoted by $\beta_\mu (k_x, k_z)$, is selected as the maximum of the lower bounds computed by executing the power iteration on a grid of temporal frequencies $\omega \in \Omega$.


\section{Structured I/O Analysis of Compressible Plane Couette Flow} \label{Sec: Compressible Plane Couette Flow}
We choose the compressible plane Couette flow in this study, which is a convenient canonical setup for investigation that has been considered in many prior works---see, e.g.,~\cite{duckJFM1994,malikPOF2006,dawsonAIAA2019,huPOF1998}. 
In this section, we will discuss the base flow calculations, the perturbation dynamics and boundary conditions imposed on the perturbations, aspects of the structured I/O modeling specific to this flow as well as validation of our numerical implementation. 
\begin{figure}
\captionsetup[subfigure]{justification=centering}
    \centering
\begin{subfigure}[t]{0.495\linewidth}
\includegraphics[width=1\textwidth]{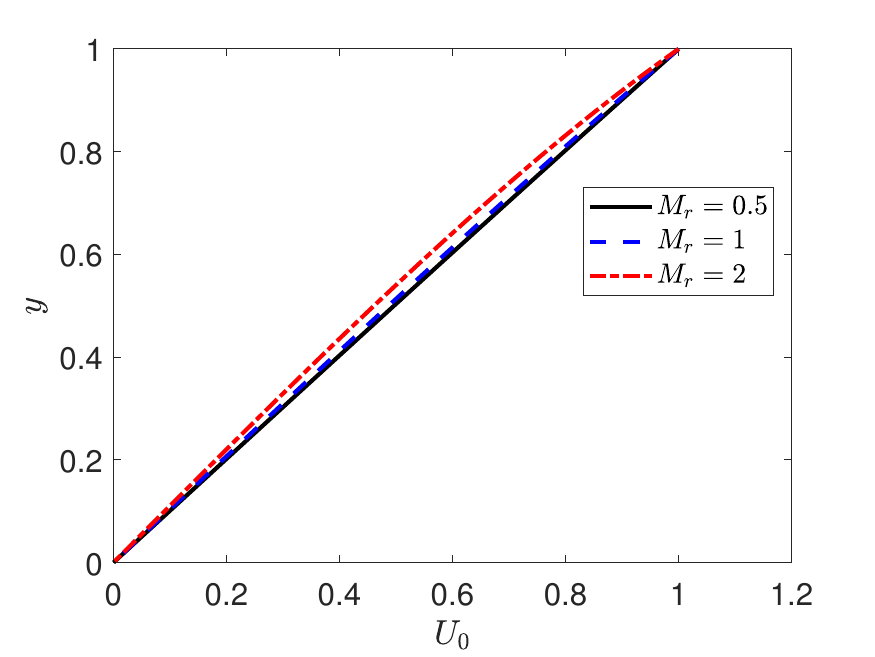}
\caption{Velocity}
\end{subfigure}
\begin{subfigure}[t]{0.495\linewidth}
\includegraphics[width=1\textwidth]{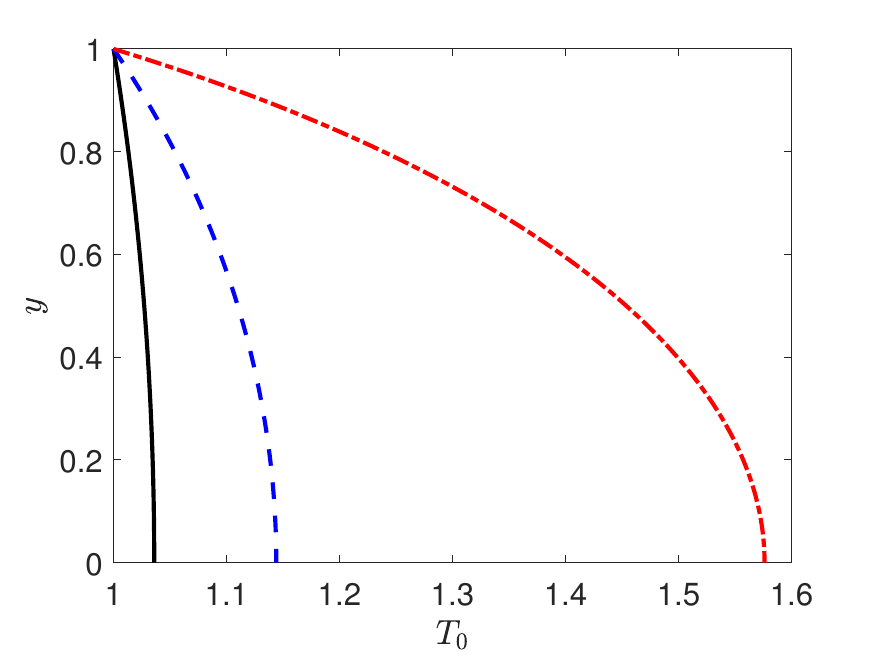}
\caption{Temperature}
\end{subfigure}
\begin{subfigure}[t]{0.495\linewidth}
\includegraphics[width=1\textwidth]{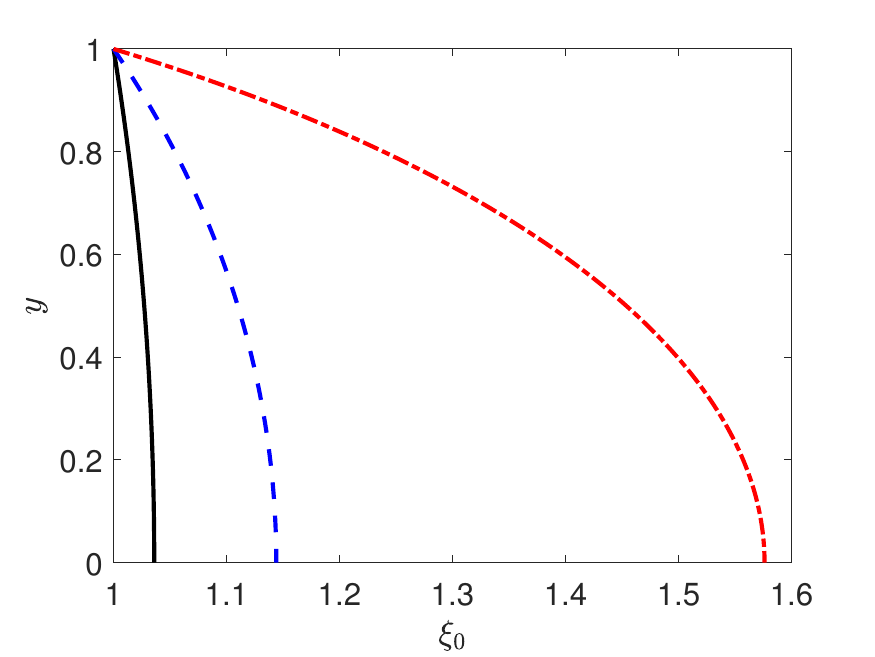}
\caption{Specific volume}
\end{subfigure}
\begin{subfigure}[t]{0.495\linewidth}
\includegraphics[width=1\textwidth]{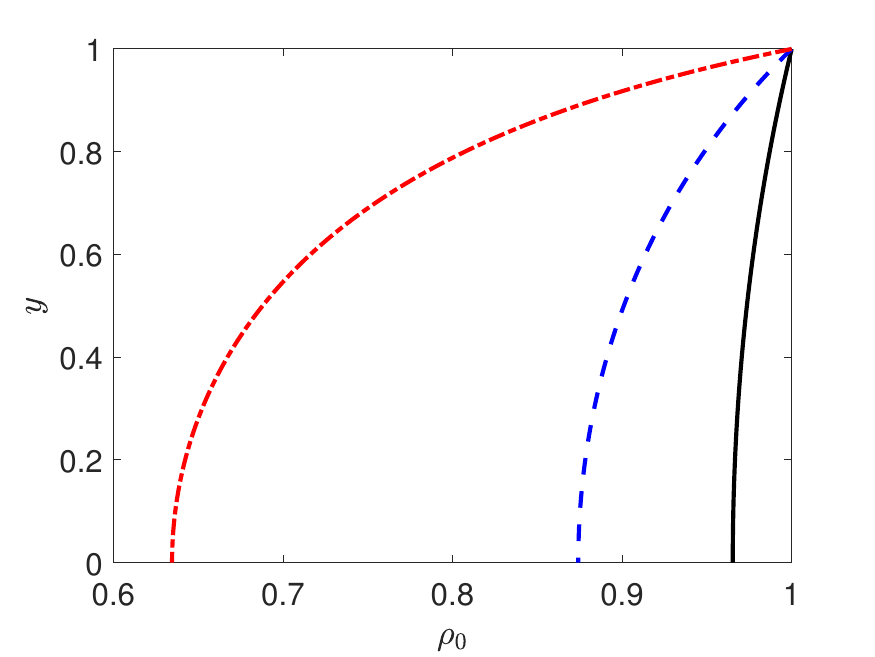}
\caption{Density}
\end{subfigure}
\begin{subfigure}[t]{0.495\linewidth}
\includegraphics[width=1\textwidth]{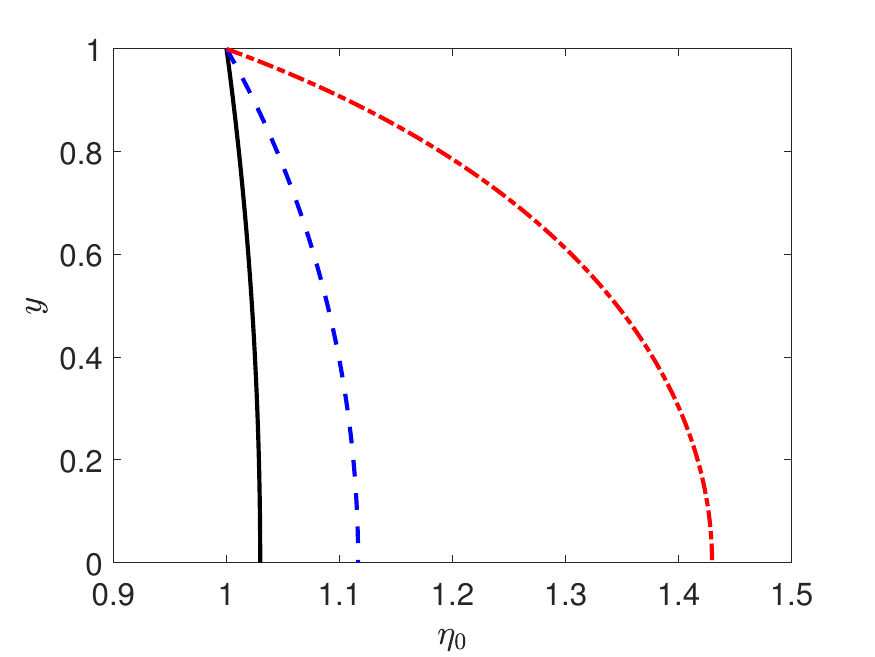}
\caption{Viscosity coefficient}
\end{subfigure}
\caption{Steady base flow profiles of compressible plane Couette flow at different Mach numbers.}
\label{fig:baseflow}
\end{figure}
\subsection{Steady Base Flow}
The base profile can be computed with relative ease: it can be shown that the base temperature profile~$T_0(y)$ will be a second-order polynomial function of the streamwise base velocity profile~$U_0(y)$~\citep{duckJFM1994}.
Further, this base flow profile will be independent of the Reynolds number $Re$.
Here we will show that the same conclusion can be drawn from the non-dimensional quadratic representation of the compressible NSE.
{Assume a base profile with $\left( \xi_0(y),\vec{u}_0,p_0(y) \right)=\left( \xi_0(y),U_0(y),0,0,p_0(y) \right)$,  
%
where the wall-normal coordinate is normalized using the channel height and the base velocity is normalized by the velocity difference between top and bottom wall.
Then, the continuity and $z$-momentum equations will be satisfied automatically.}
It follows from the $y$-momentum equation that the base pressure profile will actually be a constant, which we will take to be unity: $p_0(y)=p_0=1$.
The $x$-momentum equation shows that the base shear-stress profile $\tau_0=\eta_0\left(\partial U_0/\partial y\right)$ will be constant:
\begin{equation}
    \frac{\mathrm{d}}{\mathrm{d}y}\left(\eta_0\frac{\mathrm{d}U_0}{\mathrm{d}y}\right) = 0. \label{eq:basevel}
\end{equation}
The energy equation reduces to the condition: 
\begin{equation} \label{eq:baseflow energy equation}
    \frac{\mathrm{d}}{\mathrm{d}y} \left( (\gamma - 1) M_r^2 \tau_0 U_0 + \frac{\eta_0}{Pr} \frac{\mathrm{d} T_0}{\mathrm{d}y}  \right) = 0,
\end{equation}
where we have utilized the equation of state $\xi_0(y) = T_0(y)$ since $p_0 = 1$. The boundary conditions are taken to be
\begin{align}
    U_0(0)=0,\quad U_0(1)=1,\quad T_0(0)=T_L,\quad T(1)=1,
\end{align}
where $T_L$ is the mean temperature of the lower wall. 
Now, we can integrate \eqref{eq:baseflow energy equation} to obtain the baseflow temperature profile as \citep{duckJFM1994, malikPOF2006}
\begin{equation}
    T_0 = T_{rec} \left[r+(1-r)U_0-(1-T_{rec}^{-1})U_0^2\right],\label{eq:basetemp}
\end{equation}
where we have defined the recovery temperature $T_{rec}=1+(\gamma-1)PrM_r^2/2$ and recovery factor $r=T_L/T_{rec}$.
For consistency with this prior work, we assume temperature dependence of the base viscosity according to Sutherland's law:
\begin{equation}
    \eta_0 = \frac{T_0^{3/2}(1+C)}{T_0+C},\quad \text{with } C=0.5. \label{eq:sutherland}
\end{equation}
The base velocity profile can be computed from~\eqref{eq:basevel}.
Since the shear stress $\tau_0$ is an unknown constant, this can be done iteratively together with \eqref{eq:basetemp} and \eqref{eq:sutherland}.
Here, we use a shooting method composed of a fourth order Runge-Kutta integration scheme in conjunction with a secant method for determining the initial condition for the next iterate.
The process is then repeated until convergence.
Base profiles for $Pr=0.72$, $r=1$ (adiabatic lower wall), and $M_r=(0.5,1,2)$ are shown in Fig.~\ref{fig:baseflow}.
%
%
As noted in~\cite{duckJFM1994}, beginning with a monotone initial guess for $U_0(y)$ facilitates convergence.
For the cases considered in our study, convergence of the shooting method to an absolute error of $\epsilon\le10^{-8}$ between iterates required $\sim\mathcal{O}(10)$ total iterations.
%
\subsection{Perturbation Dynamics and Boundary Conditions} \label{Sec:Couette Perturbation Dynamics}
The perturbation dynamics about the base flow described above is of the form shown in \eqref{eq:perturbation_dynamics_1} with the linear operators given by
\begin{align} 
L_\sv (\vec{q}) &= -U_0\partial_x\sv - v\sv'_0  + \xi_0 \nabla\cdot\vec{u} \notag \\
L_u (\vec{q}) &=-U_0\partial_xu - U'_0v - \frac{1}{\gamma M_r^2} \sv_0\partial_xp \notag \\
& \quad + \frac{\sv_0}{Re}\left[\eta_0 \left(\nabla^2 u +\frac13(\partial_{xx}u+\partial_{xy}v+\partial_{xz}w) \right) + (\partial_y u + \partial_x v) \eta'_0 \right] \notag \\
L_v (\vec{q}) &=-U_0\partial_xv - \frac{1}{\gamma M_r^2} \sv_0 \partial_y p \label{Linear operators - perturbation dynamics} \\ 
&\quad + \frac{\sv_0}{Re}\left[\eta_0 \left( \nabla^2 v +\frac13(\partial_{xy}u+\partial_{yy}v+\partial_{yz}w) \right) + \left( \frac{4}{3} \partial_y v - \frac{2}{3} (\partial_z w + \partial_x u)  \right) \eta'_0 \right] \notag \\
L_w (\vec{q}) &=-U_0\partial_x w - \frac{1}{\gamma M_r^2} \sv_0\partial_z p \notag \\
&\quad + \frac{\sv_0}{Re}\left[\eta_0 \left( \nabla^2 w +\frac13(\partial_{xz}u+\partial_{yz}v+\partial_{zz}w) \right) + (\partial_z v + \partial_y w) \eta'_0 \right] \notag \\
L_p (\vec{q}) &= -U_0\partial_x p - \gamma p_0 \nabla\cdot\vec{u}  + \frac{\gamma(\gamma-1)M_r^2}{Re} \left( 2\eta_0 U'_0 \left(\partial_y u + \partial_x v \right) \right) \notag \\
    &\quad + \frac{\gamma}{Re Pr}\left(\eta_0 p_0 \nabla^2 \sv + \eta_0 \sv_0 \nabla^2p + 2 \eta_0 \sv'_0 \partial_y p + \eta_0 \sv_0'' p + \eta'_0 \left( p_0 \partial_y \xi + p \xi'_0 + \xi_0 \partial_y p  \right) \right) \notag 
\end{align}
where $(\cdot)'=\mathrm{d}(\cdot)/\mathrm{d}y$ and $(\cdot)''=\mathrm{d}^2(\cdot)/\mathrm{d}y^2$ for the associated base flow quantities.
While the structured I/O modeling for this flow remains mostly as described in Section \ref{Sec: Compressible Structured I/O}, the $C_{\Pi_{ij}}$ entries of the $\vec{C}_{1_{\chi}}$ matrix (compare with \eqref{eq:C_Pi_general}) simplify to the following expressions:
\begin{align*}
C_{\Pi_{11}} &= \eta_0 \nabla^2 + \frac{1}{3} \eta_0 \partial_{xx} + \eta'_0 \partial_y, \quad C_{\Pi_{12}} = \frac{1}{3} \eta_0 \partial_{xy} + \eta'_0 \partial_x, \quad C_{\Pi_{13}} = \frac{1}{3} \eta_0 \partial_{xz}, \\
C_{\Pi_{21}} &= \frac{1}{3} \eta_0 \partial_{xy} - \frac{2}{3} \eta'_0 \partial_x, \quad C_{\Pi_{22}} = \eta_0 \nabla^2 + \frac{1}{3} \eta_0 \partial_{yy} + \frac{4}{3} \eta'_0 \partial_y, \quad C_{\Pi_{23}} = \frac{1}{3} \eta_0 \partial_{yz} - \frac{2}{3} \eta'_0 \partial_z, \\
C_{\Pi_{31}} &= \frac{1}{3} \eta_0 \partial_{xz}, \quad C_{\Pi_{32}} = \frac{1}{3} \eta_0 \partial_{yz} + \eta'_0 \partial_z, \quad C_{\Pi_{33}} = \eta_0 \nabla^2 + \frac{1}{3} \eta_0 \partial_{zz} + \eta'_0 \partial_y .
\end{align*}
See Appendix~\ref{app:operators-2} for details on the discretized operators.
No-slip and impermeability boundary conditions are applied to velocity perturbations at both walls, i.e.,~$u(0)=u(1)=v(0)=v(1)=w(0)=w(1)=0$.
%
%
In this work, we assume an adiabatic lower wall (i.e.,~$\partial_y T(0)=0$) and an isothermal upper wall (i.e.,~$T(1)=0$). 
These assumptions on the temperature are imposed via appropriate boundary conditions on the pressure and specific volume at the walls.
{To this end, we impose homogeneous Dirichlet and Neumann boundary conditions, i.e., $\sv(1)=p(1)=0$ and $\partial_y\sv(0)=\partial_y p(0)=0$.}

\begin{figure}
\captionsetup[subfigure]{justification=centering}
    \centering
\begin{subfigure}[t]{0.495\linewidth}
\includegraphics[width=1\textwidth]{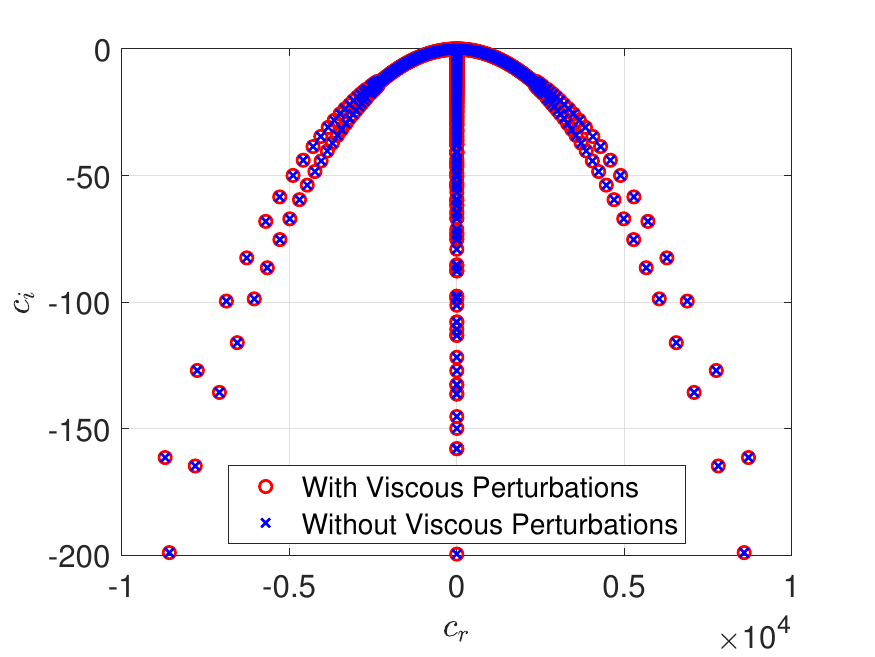}
\caption{}
\label{fig:eigenvalue-1}
\end{subfigure}
\begin{subfigure}[t]{0.495\linewidth}
\includegraphics[width=1\textwidth]{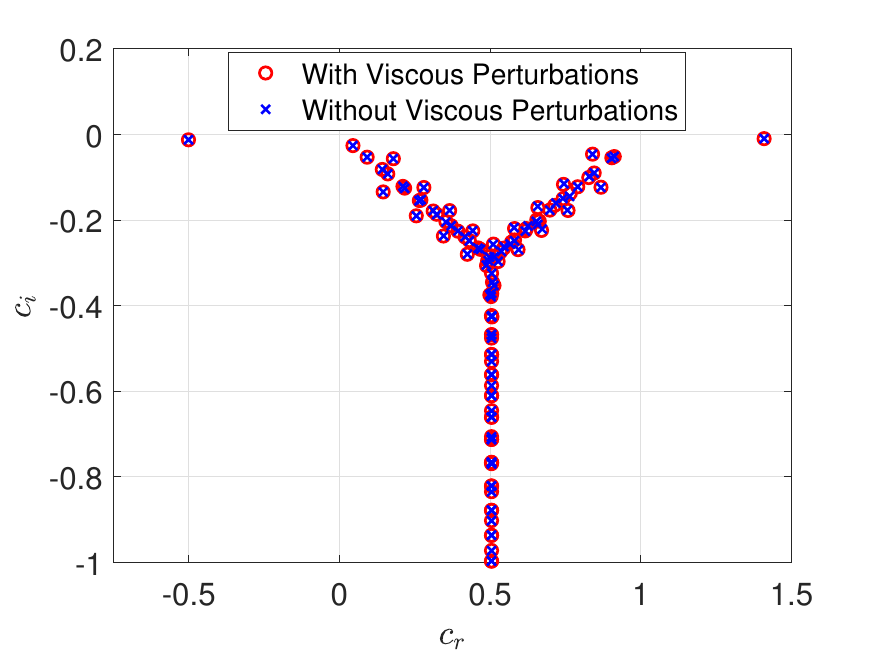}
\caption{}
\label{fig:eigenvalue-2}
\end{subfigure}
\caption{\label{fig:eigenvalues} Eigenvalue spectra of the linear operator $\hat{\vec{L}}(k_x,k_z)$---with and without accounting for the perturbations in viscosity about base flow---for $k_x = k_z = 0.1$, $Re = 2 \times 10^5$, $M_r = 2$, $Pr = 0.72$, $N_y = 200$.
The eigenvalues $\omega^{(e)}$ are plotted in terms of complex wavespeeds $c_w = c_r + \vec{i} c_i = \omega^{(e)}/k_x$ with $\omega^{(e)}$ satisfying $\vec{\hat{L}} (k_x,k_z)\vec{q}^{(e)} = -\vec{i} \omega^{(e)} \vec{q}^{(e)}$, where the negative sign is utilized to be consistent with the temporal frequency sign convention in the existing literature.}
\end{figure} 

\subsection{Code Validation: Eigenvalues of $\hat{\vec{L}}(k_x,k_z)$}
The eigenvalue spectrum obtained through our numerical implementation is shown in Fig. \ref{fig:eigenvalues}.
The `Y' shape of the eigenvalues in Fig. \ref{fig:eigenvalue-2} are consistent with the existing literature~\citep{dawsonAIAA2019, duckJFM1994, malikPOF2006, huPOF1998}. Note that these eigenvalues are associated with the viscous eigenmodes, i.e., modes that arise due to the viscous terms in the momentum and energy equations \citep{duckJFM1994}. 
In addition, the distribution of the eigenvalues  depicted in Fig. \ref{fig:eigenvalue-1} away from the imaginary axis is also expected (see, for example, similar trends reported in \cite{malikPOF2006}). 
These eigenvalues belong to the inviscid or acoustic eigenmodes~\citep{dawsonAIAA2019, duckJFM1994, malikPOF2006}.  
{We note that the viscous perturbations about the base flow do not have significant influence on 
the eigenvalues of the linear operator, as depicted in the comparison plots in Fig. \ref{fig:eigenvalues}. 
Therefore, conclusions drawn from classical stability analysis based on eigenvalues of the linear operator would be independent of the viscosity perturbations.}


%
\subsection{Numerical Results} \label{Sec: Results}
In this section, we will present numerical results for the compressible plane Couette flow by considering a  $n_{k_{x}} \times n_{k_{y}} \times n_\omega$ grid, where $n_{k_{x}}$, $n_{k_{z}}$, $n_\omega$ are the total number of grid points for $k_x$, $k_z$, and $\omega$, respectively.
We choose $(n_{k_{x}}, n_{k_{z}}) = (60,80)$ logarithmically spaced points for the spatial wavenumbers in the range $k_x \in [10^{-3}, 10^{2}]$ and $k_z \in [10^{-4}, 10^{3}]$, and {take $n_\omega = 50$ logarithmically spaced points for the temporal frequency in the range $\omega \in [-1, 1]$}\footnote[4]{{We note here that the temporal frequency grid contains $n_\omega/2$ logarithmically spaced points in the range $\omega \in [\omega_\epsilon, 1]$ with some $\omega_\epsilon > 0$ (for the results in this section, we have taken $\omega_\epsilon = 0.01$). 
The negative frequencies 
are then chosen symmetrically about the origin with the sign altered, i.e., the negative half of the grid is given by $\omega \in [-1, -\omega_\epsilon]$. 
Finally, the total grid is defined by sorting the frequencies in ascending order such that $\omega \in [-1, -\omega_\epsilon] \cup [\omega_\epsilon, 1]$.}}.
%
%
{It should be noted that we consider negative temporal frequencies as the system matrices are complex-valued and the corresponding frequency response 
is not symmetric about the $\omega =0$ line.}
Also, we choose other parameter values as $Re = 2 \times 10^5, \ Pr = 0.72, \ N_y = 100$. 
We will illustrate the results for a subsonic ($M_r=0.5$), a sonic ($M_r=1$), and a supersonic ($M_r=2$) Mach number. 
Note that the procedures for computing $\mu$ bounds at each wavenumber pair $(k_x, k_z)$ are as outlined in Section \ref{Sec:Structured I/O Analysis}.
For comparison with those $\mu$ bounds, we also compute the resolvent gain at each wavenumber pair $(k_x, k_z)$, denoted by $\sigma_R (k_x, k_z)$, using a similar procedure as 
\begin{equation*}
    \sigma_R (k_x, k_z) = \max_{\omega \in \Omega} \| \mathcal{R} (k_x,k_z,\omega) \|_2
\end{equation*}
where $\mathcal{R} (k_x,k_z,\omega) = \left(\vec{i} \omega \vec{I}_{n_q} - \hat{\vec{L}}(k_x, k_z) \right)^{-1}$ is the resolvent operator \citep{mckeon2017engine, dawsonAIAA2019, baeJFM2019}.
It should be noted that the resolvent gain calculations are carried out using the quadratic representation of the perturbation equations presented in Section \ref{Sec:Couette Perturbation Dynamics} and not the standard formulation typically utilized in the literature (see, cf. \cite{dawsonAIAA2019}).
This is due to the fact that I/O results are contingent upon the choice of variables used to describe the flow \citep{karban2020ambiguity}. 
See \cite{bhattacharjee2024formulation} for a detailed resolvent analysis of the compressible plane Couette flow using two sets of flow variables: the one utilized here in this paper (i.e., $\vec{q}=(\xi,u,v,w,p)$) and the one typically used in the literature (i.e., $\vec{q}^c=(\rho,u,v,w,T)$). 
%
%
%
For a fixed value of the tuple $(k_x, k_z, \omega)$, the right and left singular vectors associated with the largest singular value of the resolvent operator $\mathcal{R} (k_x, k_z, \omega)$ correspond to the associated forcing and response modes, respectively,
%
and the structured I/O modes are obtained by executing the lower bound power iteration algorithm on $\mathcal{H}(k_x, k_z, \omega)$, as described in Section \ref{Sec:Lower bound and mode shapes}. 
Also note that all the structured I/O modes are scaled to be unit norm to facilitate comparison with the resolvent modes.
All the results included in this section are generated using MATLAB R2022a, and the $\mu$ upper bounds are computed using the `osbal' command available in MATLAB's Robust Control Toolbox.
{Structured I/O analysis increases computational complexity over resolvent analysis, but only modestly: computation times associated with both methods scale super-quadratically and sub-cubically with the number of wall-normal collocation/grid points $N_y$ (see Appendix \ref{app:computation_times_structuredIO_reslvent} for details).}

\begin{figure}
\captionsetup[subfigure]{justification=centering}
  \centering
  \begin{subfigure}[b]{0.495\linewidth}
    \includegraphics[width=1\textwidth]{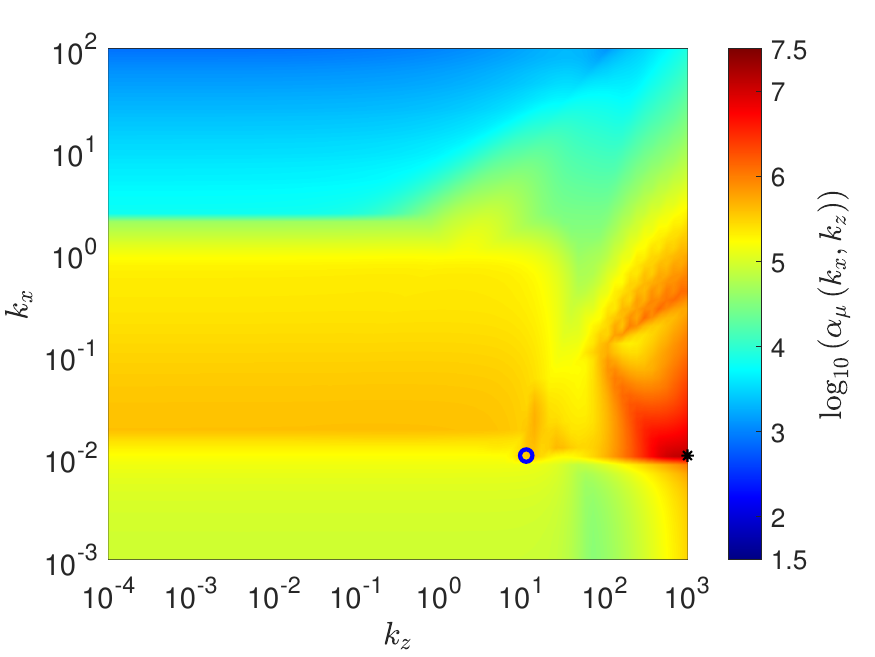}
    \caption{$\mu$ upper bound}
    \label{fig:Muub_Mach_half}
  \end{subfigure}%
  \begin{subfigure}[b]{0.495\linewidth}
    \includegraphics[width=1\textwidth]{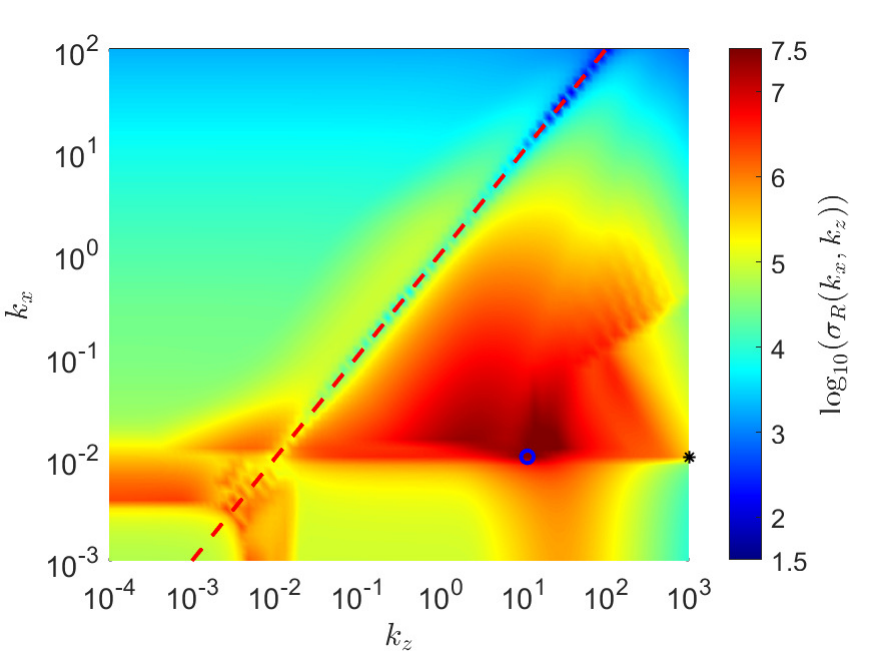}
    \caption{Resolvent gain}
    \label{fig:Resolvent_Mach_half}
  \end{subfigure}
  \begin{subfigure}[b]{0.495\linewidth}
    \includegraphics[width=1\textwidth]{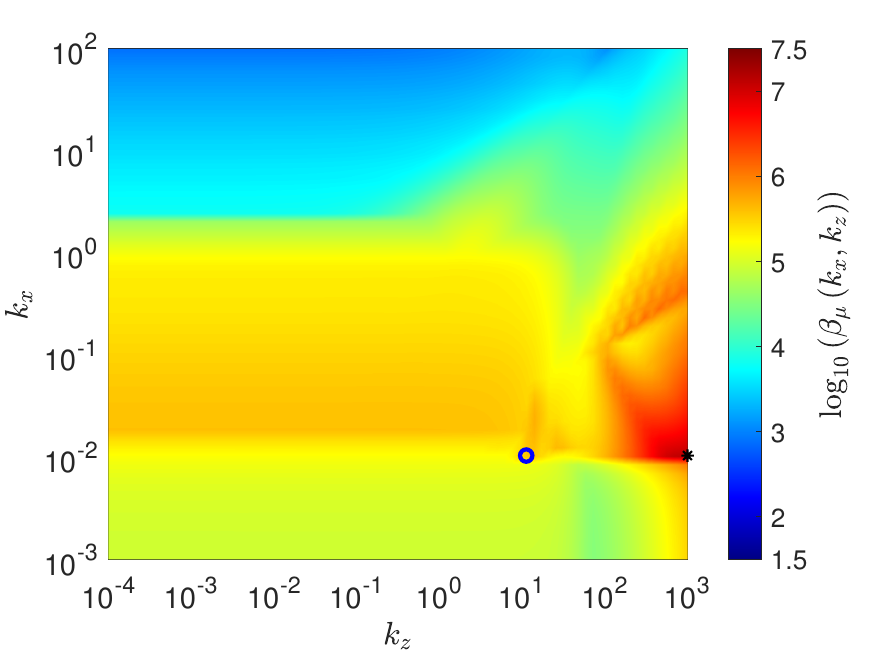}
    \caption{$\mu$ lower bound}
    \label{fig:Mulb_Mach_half}
  \end{subfigure}
  \begin{subfigure}[b]{0.495\linewidth}
    \includegraphics[width=1\textwidth]{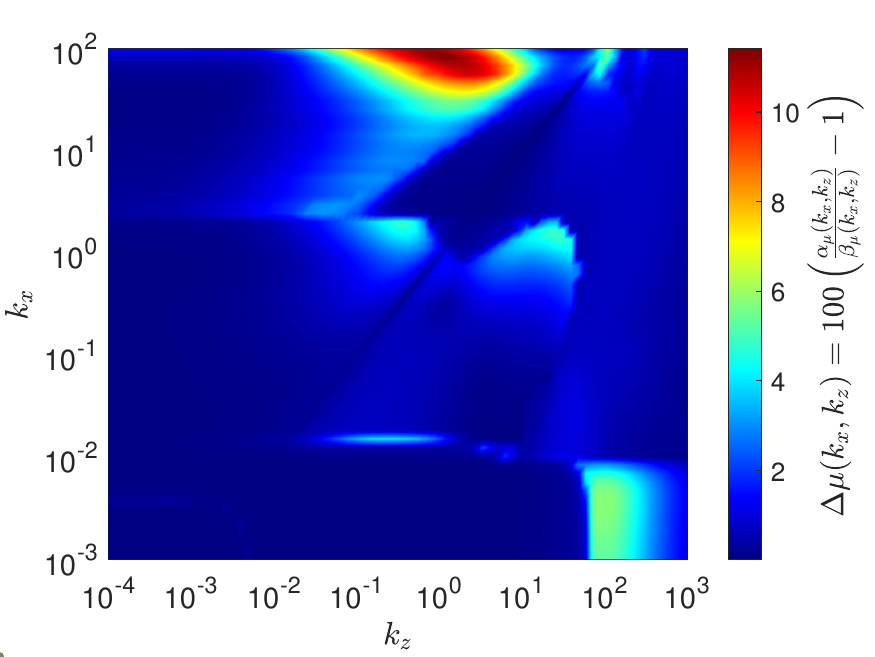}
    \caption{Gap between $\mu$ upper and lower bounds}
    \label{fig:Mugap_Mach_half}
  \end{subfigure}
 \caption{Distributions of the $\mu$ upper and lower bounds (log-scaled), percentage gap between the $\mu$ bounds, and the resolvent gain (log-scaled) over the wavenumber pair $(k_x, k_z)$ grid for $M_r=0.5$. 
 The circle and the asterisk denote the $(k_x, k_z)$ values associated with the largest computed resolvent gain and $\mu$ bounds, respectively.
 {Moreover, the dashed line in resolvent gain plot denotes $k_x=k_z$.}}
  \label{fig:Heat_plots_Mach_half}
\end{figure} 

The I/O gains (both structured I/O and resolvent) for the subsonic Mach number ($M_r=0.5$) are shown in Fig. \ref{fig:Heat_plots_Mach_half}. 
The quality of the upper and lower bounds can be assessed through the gap between these bounds, and the results in Fig. \ref{fig:Mugap_Mach_half} show that the gap is fairly small for a majority of the wavenumber pairs on the grid. 
More precisely, the gap is less than 5\% at approximately 96.02\% of the grid points, and the average gap is 1.002\%. 
Thus, the upper and lower bound computations in this case have led to tight estimates of the exact value of $\mu$ at these wavenumber pairs. 
%
%
The largest gap, which lies within the localized region of relatively higher gaps for $k_x \approx (10^1,10^2)$ and $k_z \approx (10^{-2},10^1)$, is 11.41\% at $(k_x,k_z)=(82.27,1.19)$.
Also, the maxima in both the bounds occur at $(k_x,k_z)=(0.01,10^3)$ which is denoted using an asterisk in Figs. \ref{fig:Muub_Mach_half}, \ref{fig:Mulb_Mach_half}, and the gap between the bounds at this $(k_x,k_z)$ is approximately 1\% (compare with Fig. \ref{fig:Mugap_Mach_half}).
Due to the tightness of the $\mu$ bounds at the maxima, the corresponding structured I/O forcing and response modes can be computed from the upper bounds, as discussed in \cite{mushtaq2023structured}.

The resolvent gains for this case, as shown in Fig. \ref{fig:Resolvent_Mach_half}, are higher overall compared to the $\mu$ bounds, which indicates that utilization of the structure of the nonlinearity leads to a reduction of the I/O gain and a larger estimated margin of stability via the small-gain theorem \citep{packard1993, zhou1996robust}.
The regions of maximum amplifications in both sets (i.e., structured I/O and resolvent) of I/O gains are associated with oblique structures, which correspond to smaller values of $k_z$ (i.e., larger length scales) in case of the resolvent gains and vice versa in the $\mu$ bounds. 
Despite the differences in the distributions, the global peak in the resolvent gain corresponds to a local peak in the $\mu$ bounds and vice versa. 
The `horizontal band' of constant $k_x$ of higher gains predicted by the structured I/O results indicate a spanwise elongated structure and the resolvent gains appear in qualitative agreement; however the streamwise and spanwise wavenumbers associated with these localized regions are different between the structured I/O and resolvent gains, and the associated forcing and response modes---to be discussed next---highlight differences in the instability mechanisms between the two analyses.

It is interesting to note that the streamwise elongated structure in the resolvent gain for $k_x\approx (10^{-3},10^{-2.5})$ and $k_z \approx (10^{-4},10^{-2})$ is not predicted by the structured I/O results.
This particular feature is not present in the structured I/O result, which means that accounting for the structure of the nonlinearity eliminates these instability mechanisms otherwise predicted through the unstructured analysis.
It is likely that these flow structures are not consistent with the admissible forcing and response pairs through the actual nonlinearity in the compressible NSE.
Structured I/O analysis can, therefore, prove helpful in eliminating potentially redundant/non-physical instability mechanisms.
%
%
{Furthermore, the region of smallest resolvent gain can be approximately captured via a straight line in the wavenumber space that represents equality of the streamwise and spanwise wavenumbers (i.e., the line $k_x=k_z$, see Fig. \ref{fig:Resolvent_Mach_half}).}
It is noteworthy that the particular choice of flow variables in this paper reveals this feature and the flow variables typically used for resolvent analysis does not capture this flow feature (see the comparison results in \cite{bhattacharjee2024formulation}).
Thus, the choice of flow variables is directly linked with this observation and we will investigate this further in our future work.
%

\begin{figure}
\captionsetup[subfigure]{justification=centering}
  \centering
  \begin{subfigure}[b]{0.5\linewidth}
    \includegraphics[width=1\textwidth]{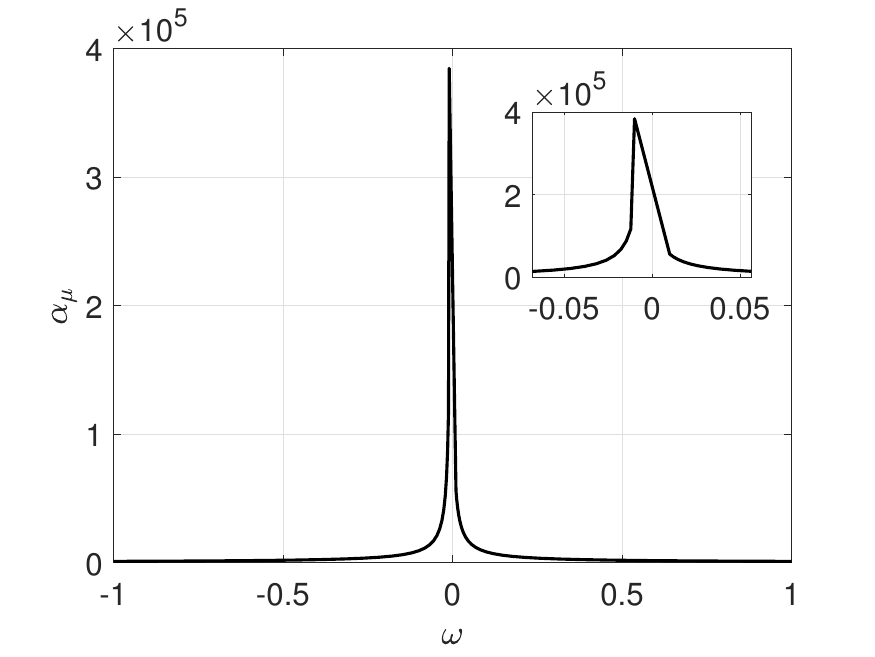}
    \caption{$\mu$ upper bound} \label{fig:muub_min_freq_1e-2_nf_25}
  \end{subfigure}%
  \begin{subfigure}[b]{0.5\linewidth}
    \includegraphics[width=1\textwidth]{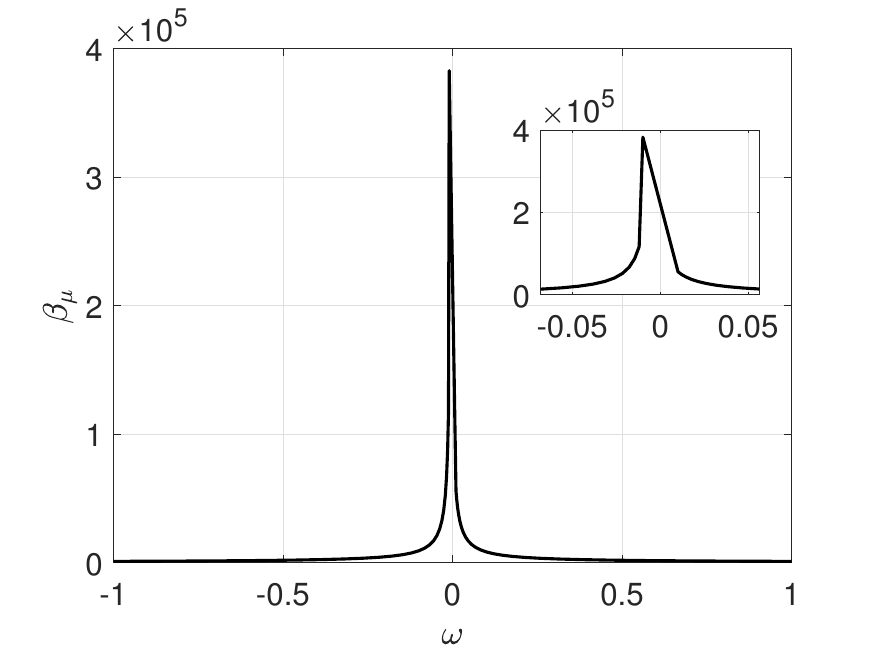}
    \caption{$\mu$ lower bound} \label{fig:mulb_min_freq_1e-2_nf_25}
  \end{subfigure}
    \begin{subfigure}[b]{0.5\linewidth}
    \includegraphics[width=1\textwidth]{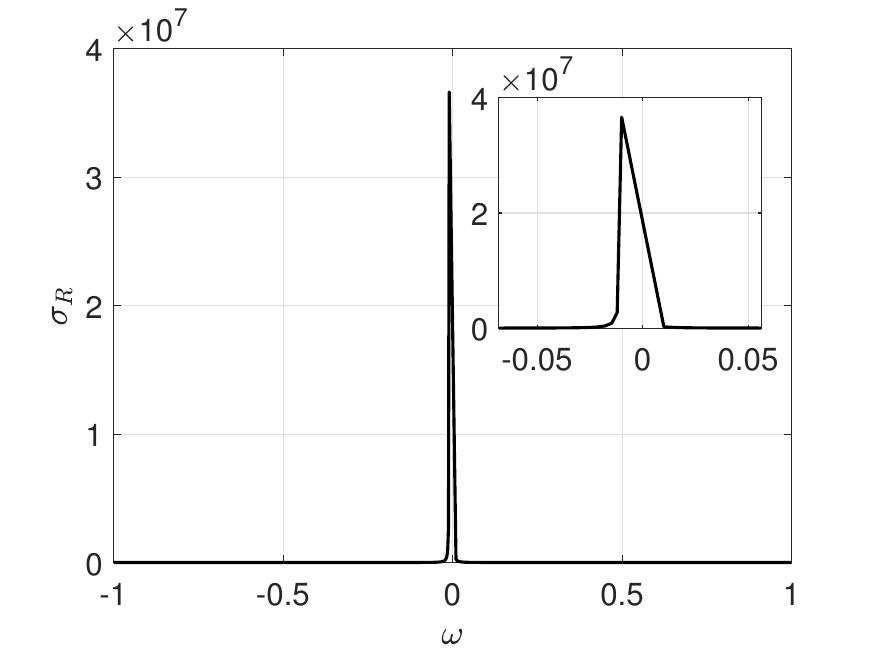}
    \caption{Resolvent gain} \label{fig:Res_gain_min_freq_1e-2_nf_25}
  \end{subfigure}%
\caption{{The $\mu$ bounds and resolvent gain as functions of the temporal frequency for $M_r=0.5$ and $(k_x,k_z)=(0.01,11.24)$, which corresponds to the maximum resolvent gain on the wavenumber pair grid considered for the results in Fig. \ref{fig:Resolvent_Mach_half}. }}
  \label{fig:I/O_gains_vs_temporal_freq}
\end{figure}

{Note that the temporal frequency associated with maximum gain (i.e., the $\mu$ bounds and the resolvent gain) at a fixed wave-number pair is typically a small negative value and the gains are close zero at higher magnitudes (positive or negative) of temporal frequencies.
For example, the $\mu$ upper and lower bounds for $M_r =0.5$ and $(k_x,k_z)=(0.01,11.24)$ (which corresponds to the largest resolvent gain on the wavenumber pair grid considered, as shown in Fig. \ref{fig:Resolvent_Mach_half}) are respectively shown in Figs. \ref{fig:muub_min_freq_1e-2_nf_25} and \ref{fig:mulb_min_freq_1e-2_nf_25}. 
The zoomed-in plots clearly display 
single peaks in both the upper and lower bounds as well as the above-mentioned `trail off' behavior of the bounds as the frequency increases in magnitude. 
%
Resolvent gain variation over temporal frequency, as shown in Fig. \ref{fig:Res_gain_min_freq_1e-2_nf_25}, is qualitatively similar.
Furthermore, these trends are observed at other wavenumber pairs and flow parameter settings as well.}

\begin{figure}
\captionsetup[subfigure]{justification=centering}
  \centering
  \begin{subfigure}[b]{0.495\textwidth}
    \includegraphics[width=1\textwidth]{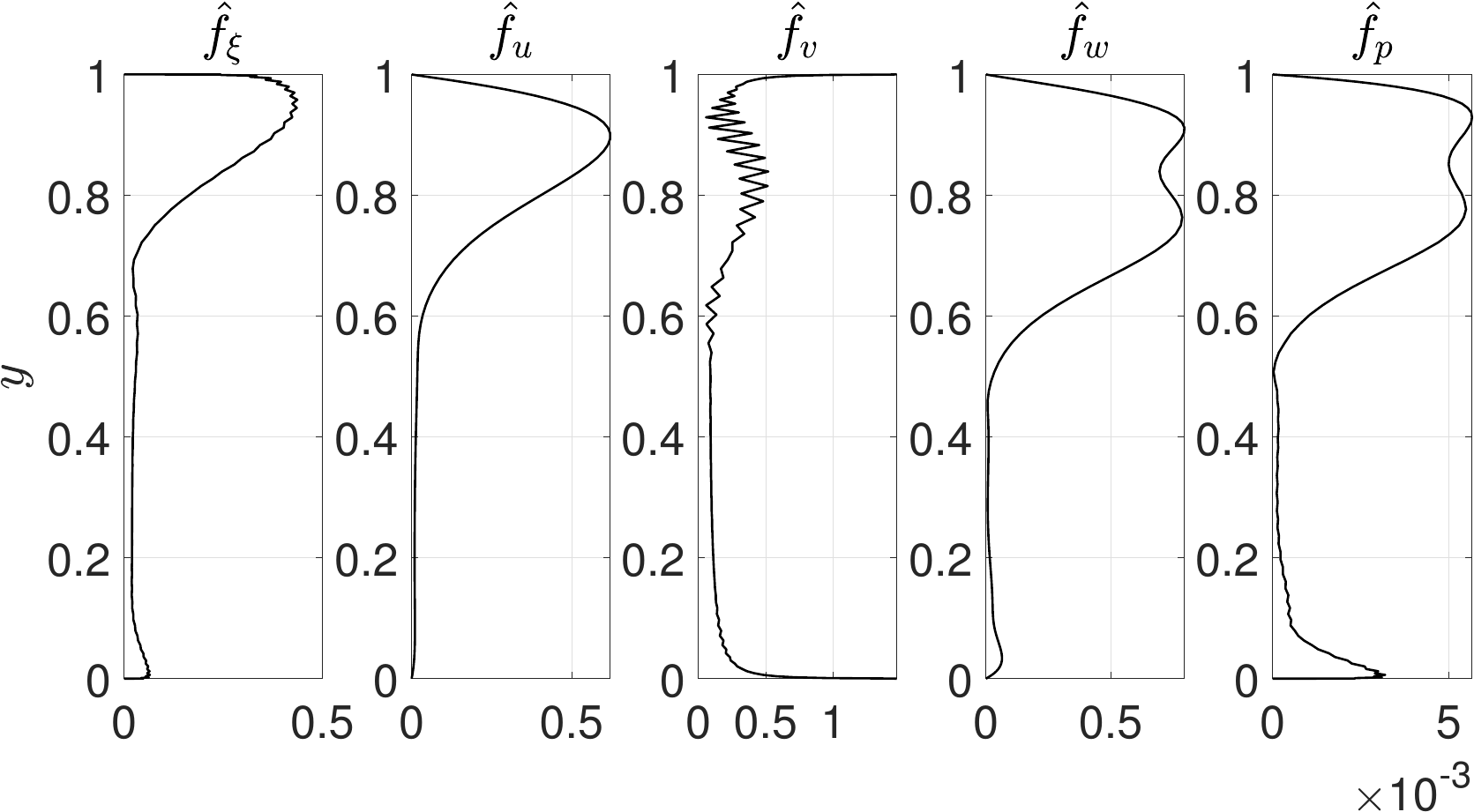}
    \caption{Structured I/O forcing modes} \label{fig:structured_forcingmodes_Mach_half_1}
  \end{subfigure}
\begin{subfigure}[b]{0.495\textwidth}
    \includegraphics[width=1\textwidth]{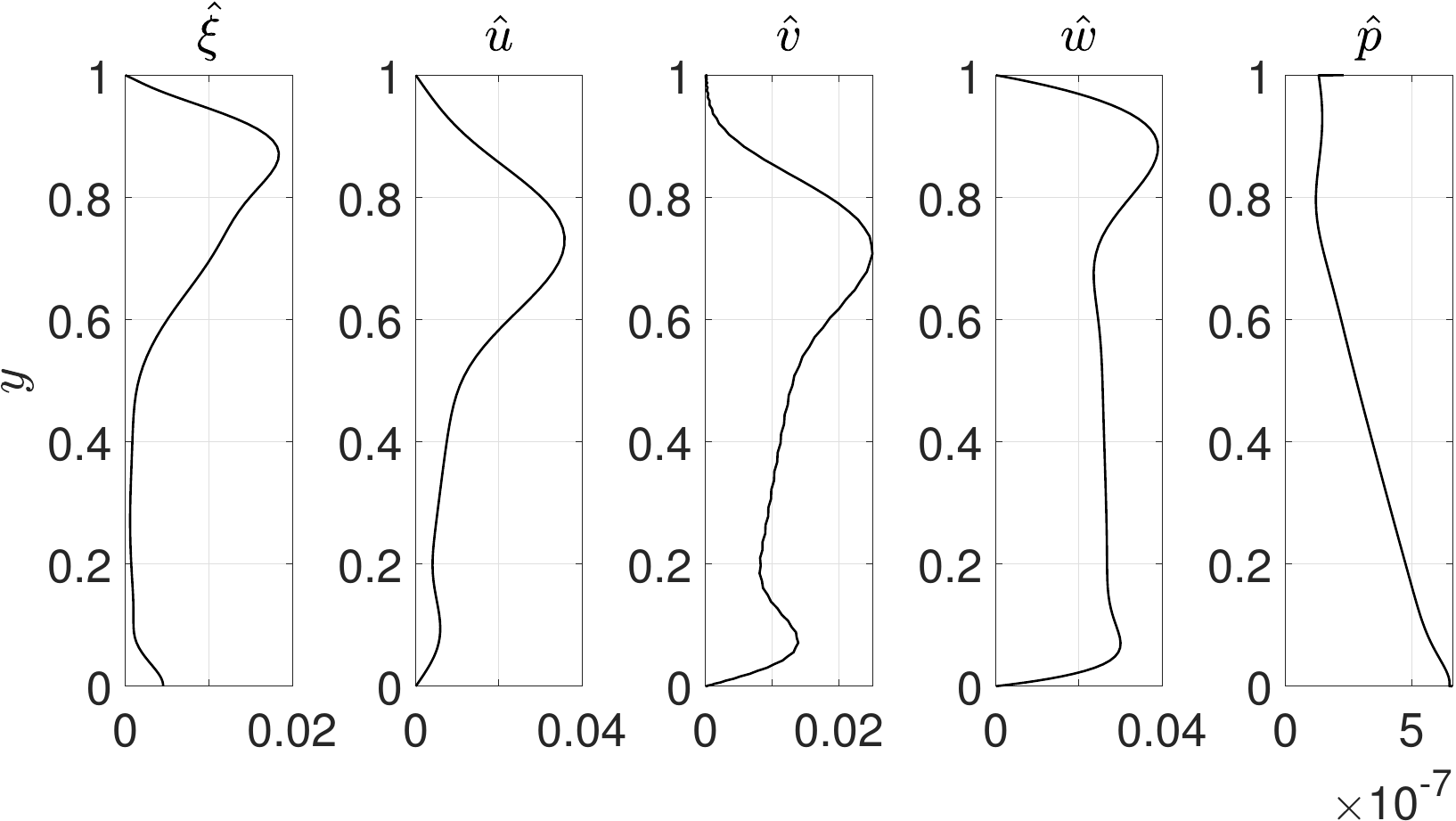}
    \caption{Structured I/O response modes} \label{fig:structured_responsemodes_Mach_half_1}
  \end{subfigure}
  \begin{subfigure}[b]{0.495\textwidth}
    \includegraphics[width=1\textwidth]{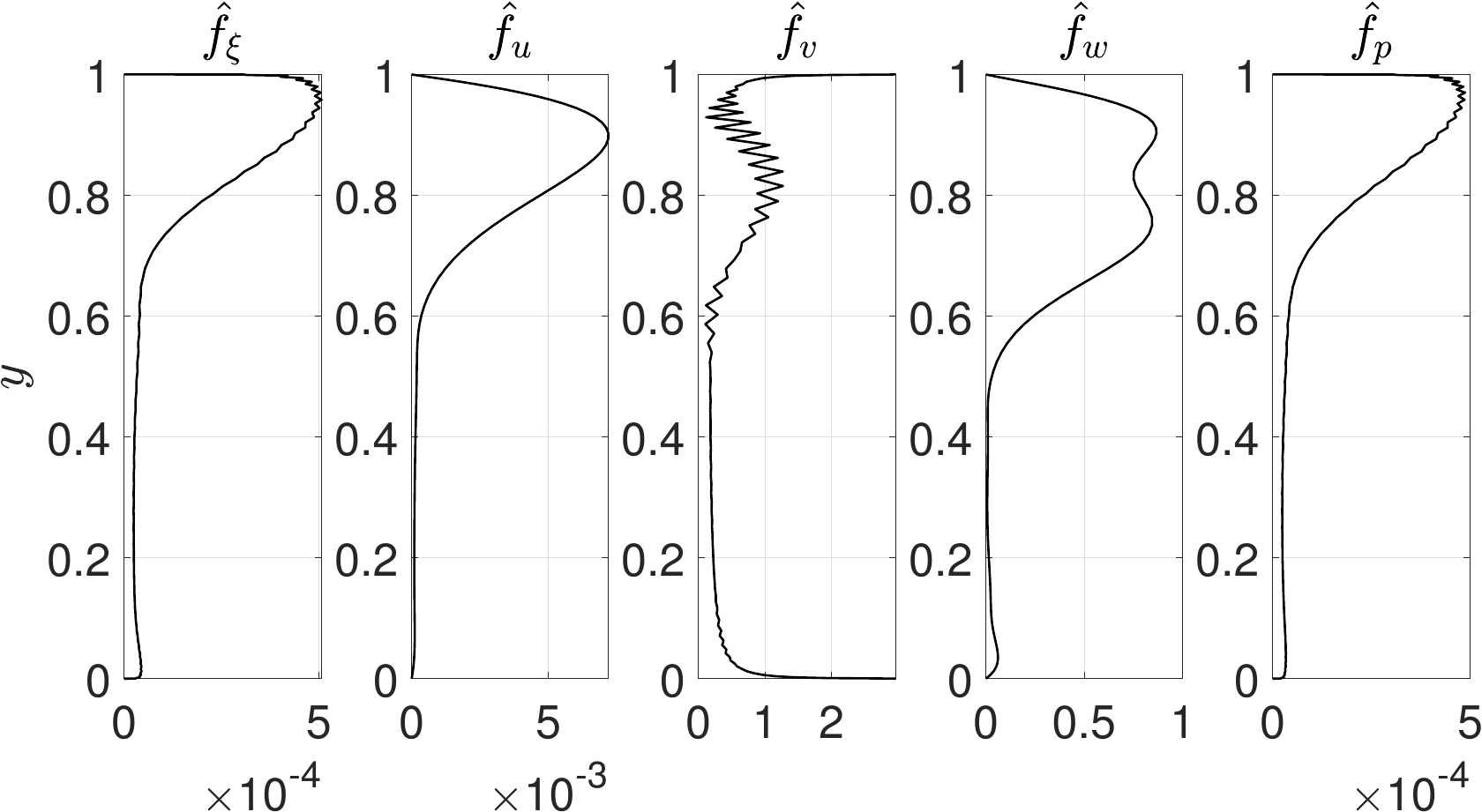}
    \caption{Resolvent forcing modes} \label{fig:resolvent_forcingmodes_Mach_half_1}
  \end{subfigure}
  \begin{subfigure}[b]{0.495\textwidth}
    \includegraphics[width=1\textwidth]{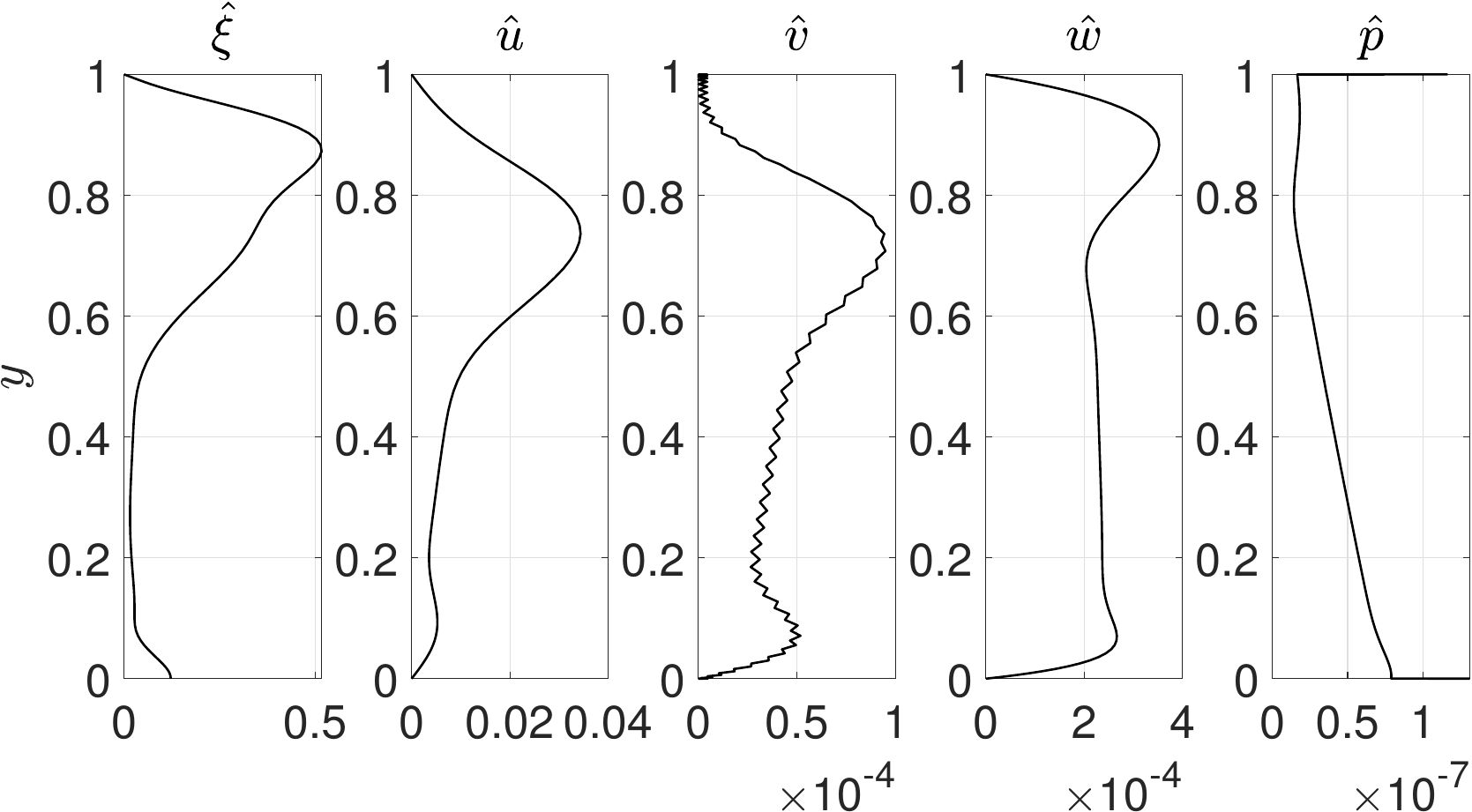}
    \caption{Resolvent response modes} \label{fig:resolvent_responsemodes_Mach_half_1}
  \end{subfigure}
 \caption{Absolute values of the structured I/O and resolvent modes for $M_r=0.5$, $(k_x, k_z, \omega)=(0.01,11.24,-0.01)$, which corresponds to the largest resolvent gain in Fig. \ref{fig:Resolvent_Mach_half}.}
  \label{fig:forcing_and_response_modes_Mach_half_1}
\end{figure}

\begin{figure}
\captionsetup[subfigure]{justification=centering}
  \centering
  \begin{subfigure}[b]{0.495\textwidth}
    \includegraphics[width=1\textwidth]{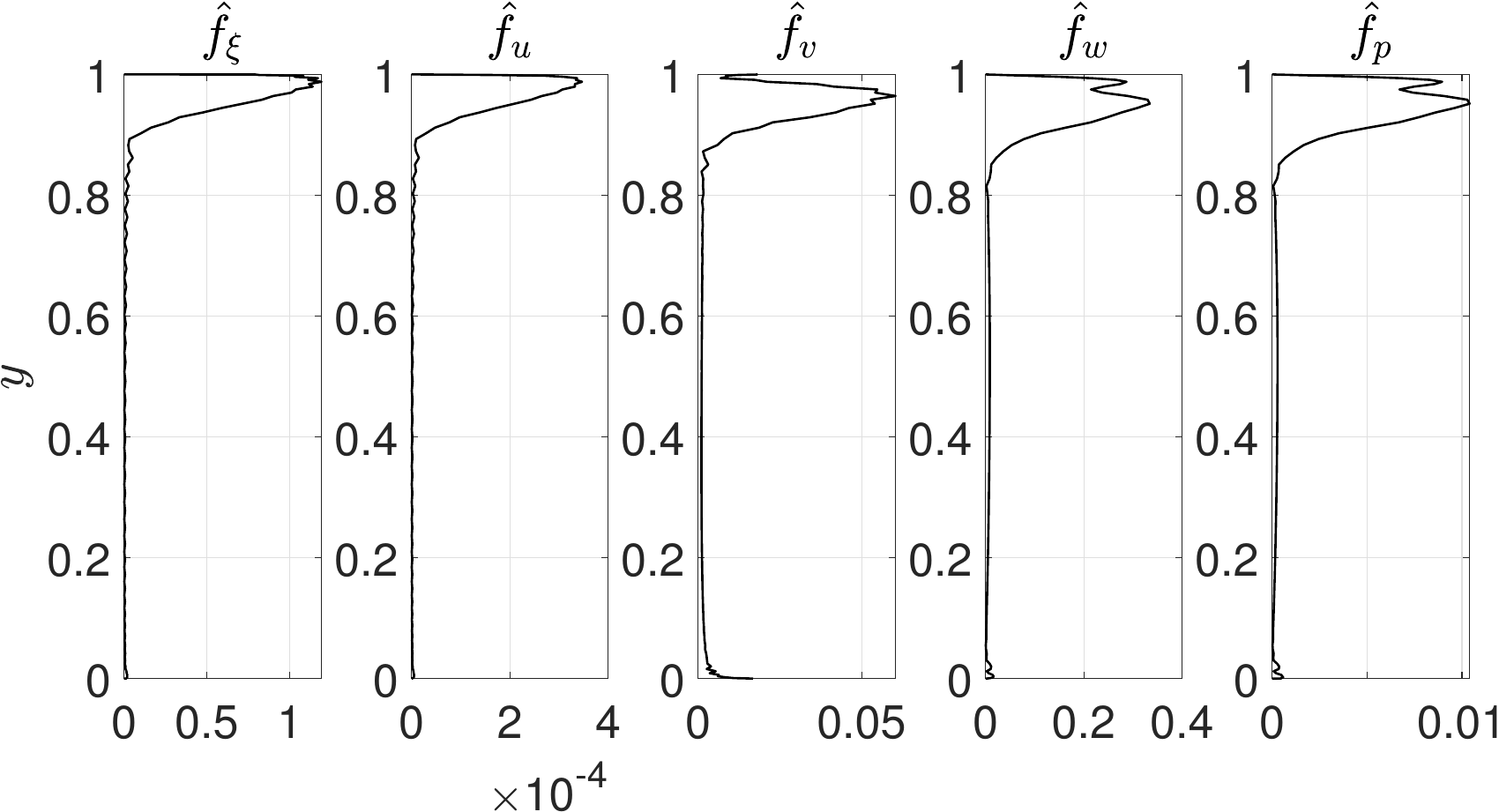}
    \caption{Structured I/O forcing modes} \label{fig:structured_forcingmodes_Mach_half_2}
  \end{subfigure}
\begin{subfigure}[b]{0.495\textwidth}
    \includegraphics[width=1\textwidth]{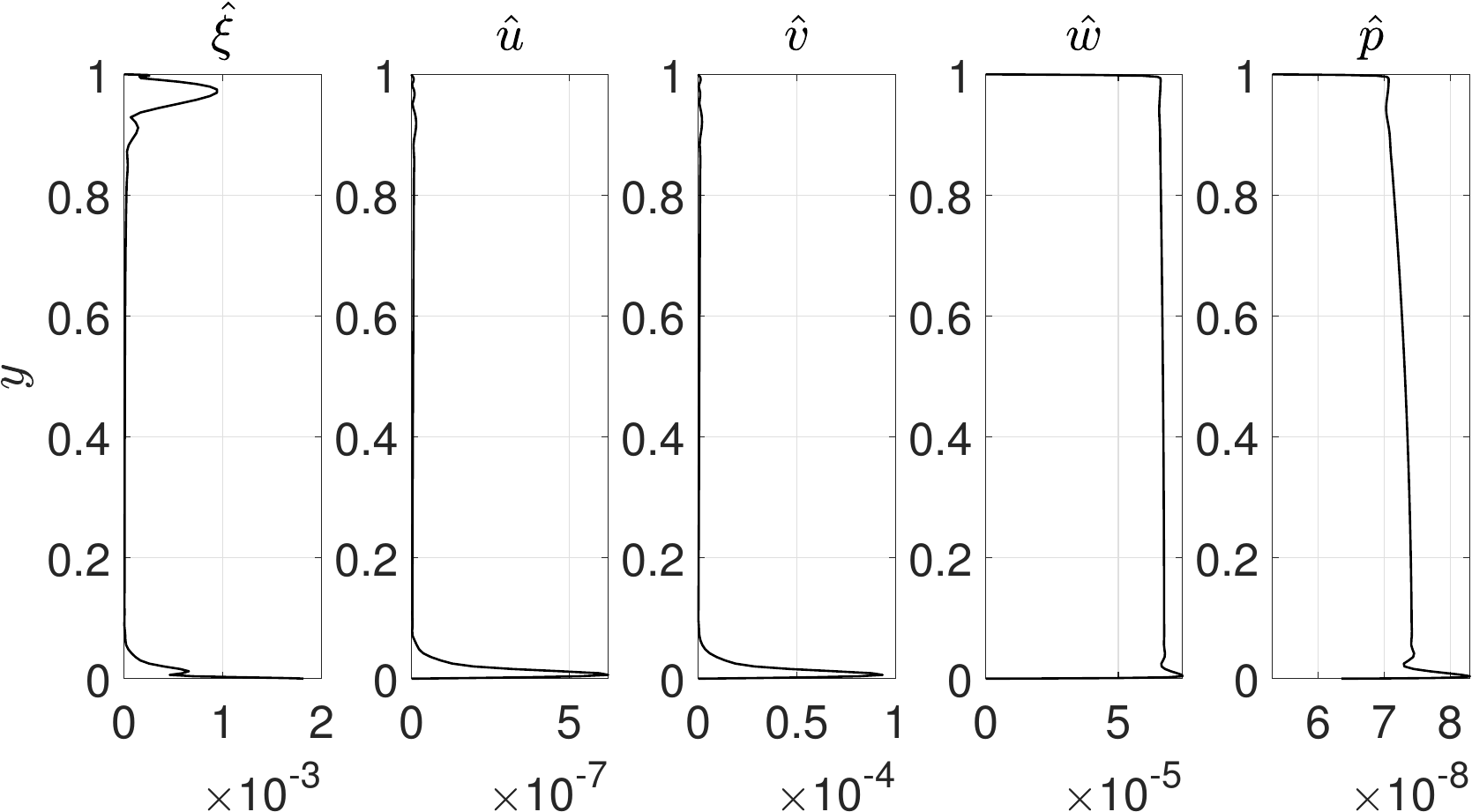}
    \caption{Structured I/O response modes} \label{fig:structured_responsemodes_Mach_half_2}
  \end{subfigure}
  \begin{subfigure}[b]{0.495\textwidth}
    \includegraphics[width=1\textwidth]{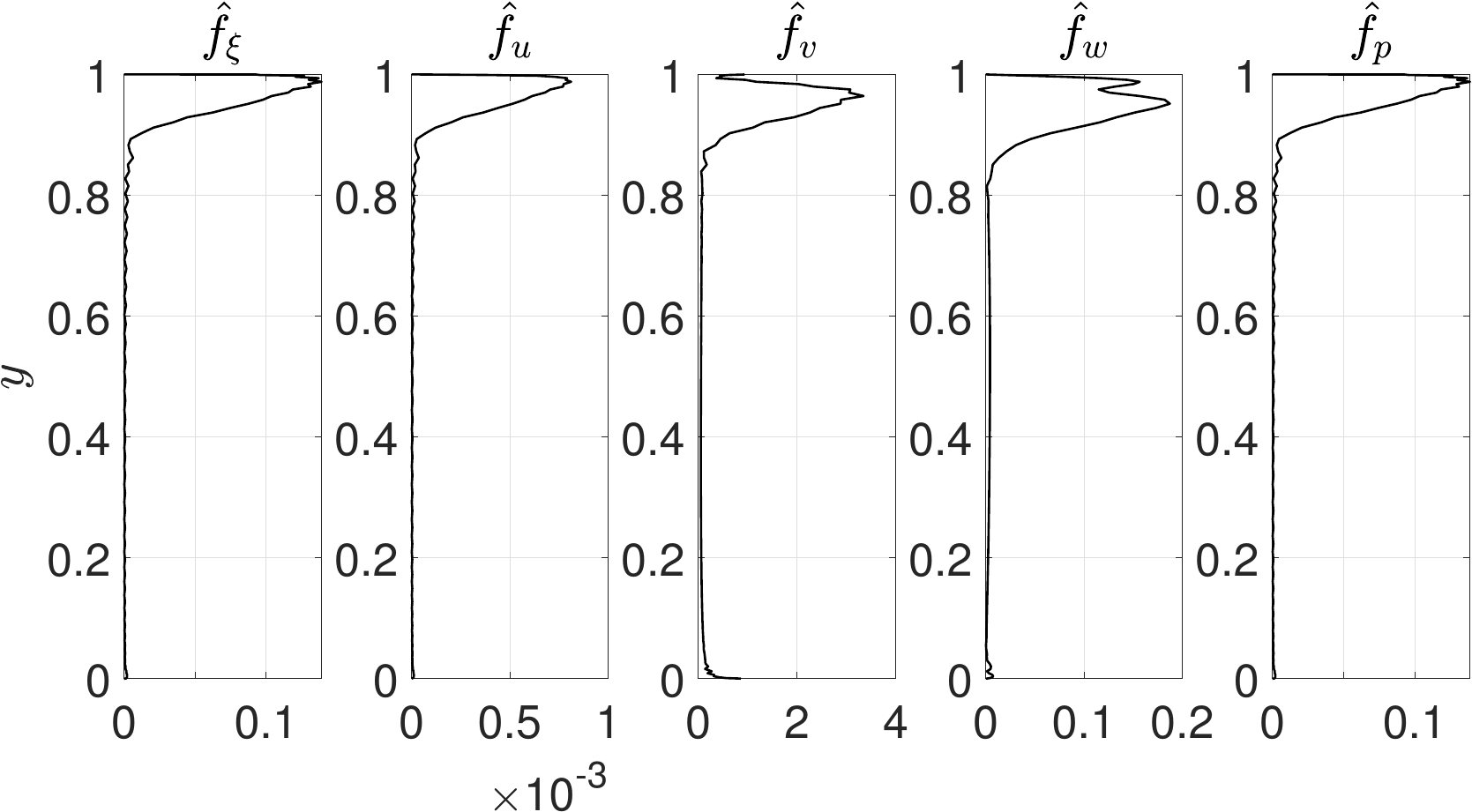}
    \caption{Resolvent forcing modes} \label{fig:resolvent_forcingmodes_Mach_half_2}
  \end{subfigure}
  \begin{subfigure}[b]{0.495\textwidth}
    \includegraphics[width=1\textwidth]{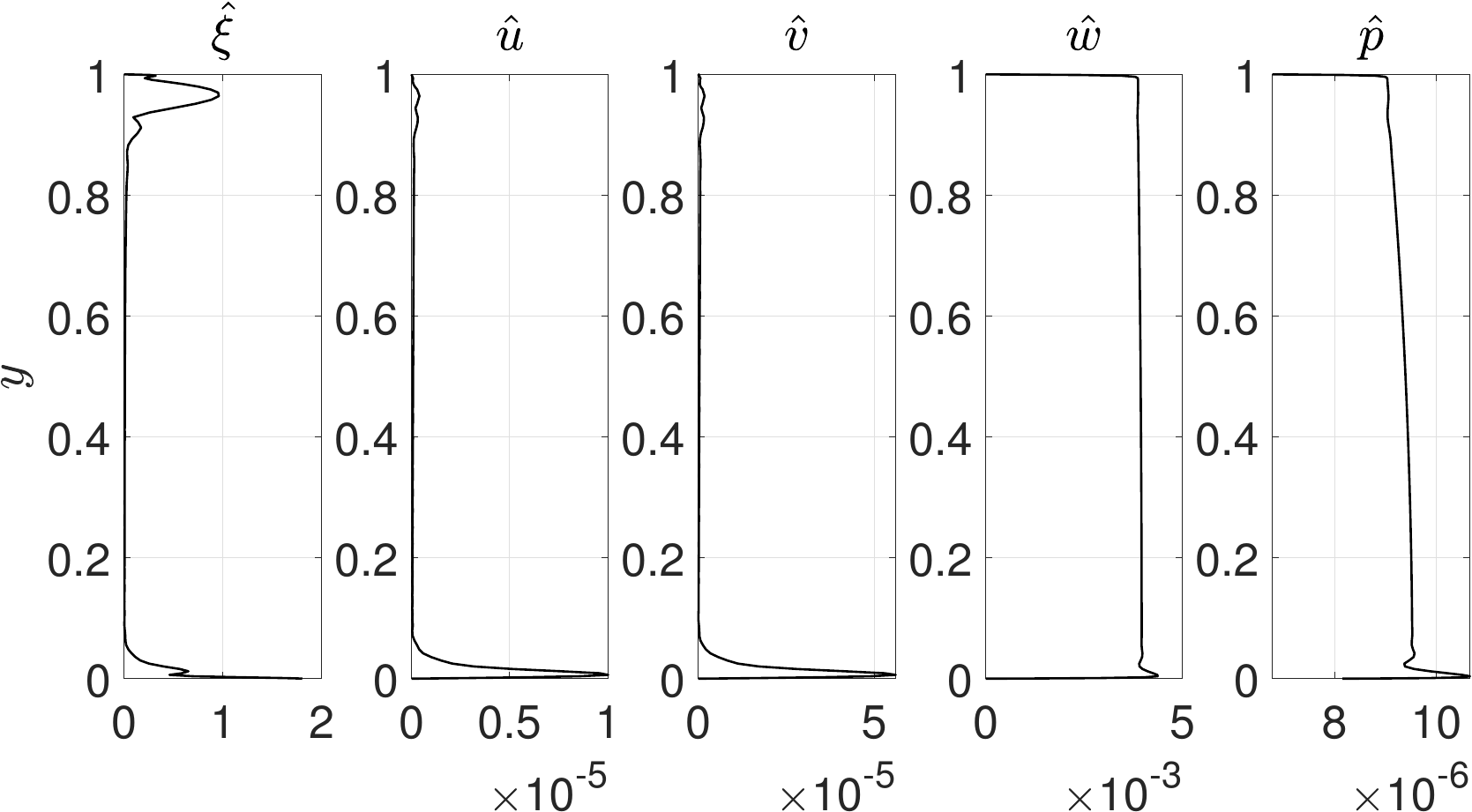}
    \caption{Resolvent response modes} \label{fig:resolvent_responsemodes_Mach_half_2}
  \end{subfigure}
 \caption{Absolute values of the structured I/O and resolvent modes for $M_r=0.5$, $(k_x, k_z, \omega)=(0.01,10^3,-0.01)$, which corresponds to the largest values in both the $\mu$ bounds in Fig. \ref{fig:Heat_plots_Mach_half}.}
\label{fig:forcing_and_response_modes_Mach_half_2}
\end{figure}

The structured I/O and resolvent modes are shown in Fig. \ref{fig:forcing_and_response_modes_Mach_half_1}, which are computed for $(k_x,k_z,\omega)= (0.01,11.24,-0.01)$ and correspond to the largest resolvent gain in Fig. \ref{fig:Resolvent_Mach_half}.
The instability associated with the largest resolvent gain is also captured by the $\mu$ bounds, albeit as a local maxima at the same $(k_x,k_z,\omega)$ tuple.
Thus, both analyses are able to consistently identify a potential mechanism of flow instability at $(k_x,k_z,\omega)= (0.01,11.24,-0.01)$. 
%
%
A comparison of the plots in Fig. \ref{fig:forcing_and_response_modes_Mach_half_1} reveals that the `shape' of the forcing and response modes across the two sets of results are similar for the velocity components and the specific volume.
Despite the qualitative similarities, the resolvent forcing are dominated by the wall-normal and spanwise velocity modes, while the structured I/O forcing is dictated by the specific volume and all three velocity modes. 
Especially, the structured I/O modes indicate a significantly larger forcing in the streamwise velocity component and leads to an amplified response in the spanwise velocity component compared to the resolvent modes, which indicate momentum exchange through vorticity-like effects.
%
%
While the forcing modes for the wall-normal velocity component are similar, the corresponding resolvent modes show a significantly diminished responses in the wall-normal velocity compared to the corresponding structured I/O response mode (compare Figs. \ref{fig:structured_responsemodes_Mach_half_1}, \ref{fig:resolvent_responsemodes_Mach_half_1}). 
%
%
Also, the pressure modes have different shapes close to the upper wall, and the magnitude of forcing and response in the specific volume are substantially different across the two sets of modes.
%
Overall, the resolvent modes predict that the instability identified by the maximum resolvent gain in  Fig. \ref{fig:Resolvent_Mach_half} is 
primarily related to momentum forcing in the wall-normal and spanwise directions whose dominant responses show up in the specific volume 
perturbations. 
%
%
The structured I/O modes, by contrast, illustrate the dominant modal forcing and response behavior associated with the same instability as a combination of momentum and thermodynamic properties of the flow. 

The modes corresponding to the largest structured I/O gain (i.e., the location of maximum $\mu$ bounds at  $(k_x,k_z,\omega)= (0.01,10^3,-0.01)$) are plotted in Fig. \ref{fig:forcing_and_response_modes_Mach_half_2}. 
Note that the resolvent gain is also able to identify this instability in terms of a local maximum in the resolvent gain at $(k_x,k_z,\omega)= (0.01,10^3,-0.01)$ (see Fig. \ref{fig:Resolvent_Mach_half}).
The structured I/O modes characterize the associated instability to be primarily driven by the specific volume forcing near the upper wall (see Fig. \ref{fig:structured_forcingmodes_Mach_half_2}), and these forcing result in specific volume responses localized near both the walls, as shown in Fig. \ref{fig:structured_responsemodes_Mach_half_2}.
Also, the forcing associated with the momentum equations are largely in the wall-normal and spanwise coordinates and are concentrated near the upper wall (see Fig. \ref{fig:structured_forcingmodes_Mach_half_2}). 
These translate into fluctuations concentrated near the lower wall in the streamwise and wall-normal velocity components and a fluctuation in the spanwise velocity component that remains roughly invariant across the width of the channel (see Fig. \ref{fig:structured_responsemodes_Mach_half_2}).
The forcing in the pressure is mainly localized near the upper wall and the corresponding response mode remains approximately invariant across the width of the channel. 
The resolvent modes in Figs. \ref{fig:resolvent_forcingmodes_Mach_half_2}, \ref{fig:resolvent_responsemodes_Mach_half_2} are qualitatively consistent with the structured I/O ones in Fig. \ref{fig:structured_forcingmodes_Mach_half_2}, \ref{fig:structured_responsemodes_Mach_half_2}; however, the resolvent forcing and response modes are dominated by the wall-normal velocity component and the specific volume, respectively.
%
%
Thus, structured I/O and resolvent modes lead to conflicting conclusions on the underlying physics---the former suggests that the instability is primarily thermodynamic in nature while the latter indicates that the instability is mainly caused by fluctuations in the wall-normal momentum and amplifies the associated response in the specific volume, thereby indicating interactions between mechanisms related to momentum and thermodynamic properties of the flow.
%
%
%
Therefore, we again conclude that although the two sets of I/O results might point towards the same instability in terms of localized amplification in the corresponding I/O gains at a fixed wavenumber pair and temporal frequency, the modal inferences obtained regarding the underlying physics causing that instability as well as the corresponding response mechanisms can differ significantly between the structured and unstructured analysis.
%

\begin{figure}
\captionsetup[subfigure]{justification=centering}
  \centering
  \begin{subfigure}[b]{0.495\linewidth}
    \includegraphics[width=1\textwidth]{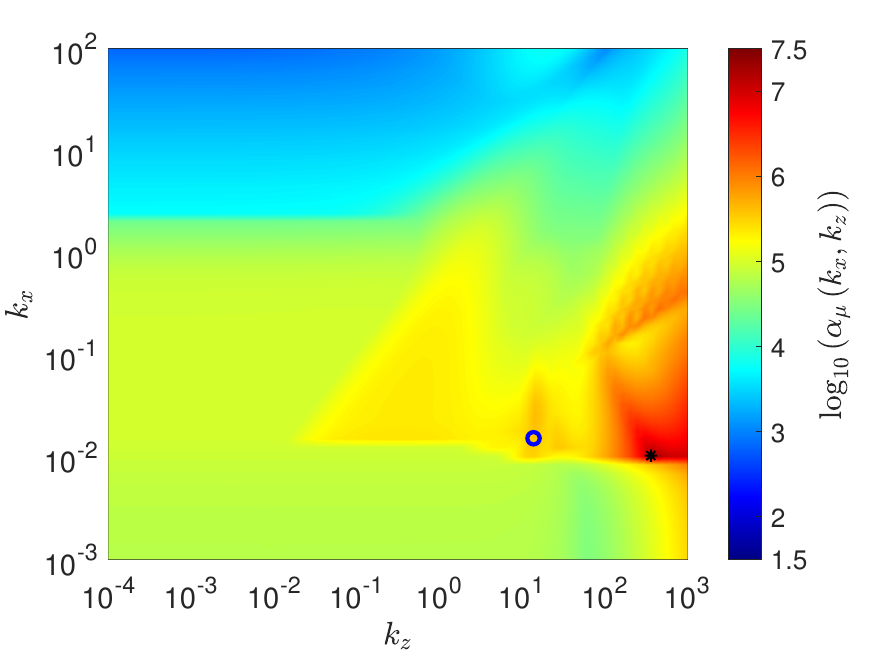}
    \caption{$\mu$ upper bound} \label{fig:Muub_Mach_one}
  \end{subfigure}
  \begin{subfigure}[b]{0.495\linewidth}
    \includegraphics[width=1\textwidth]{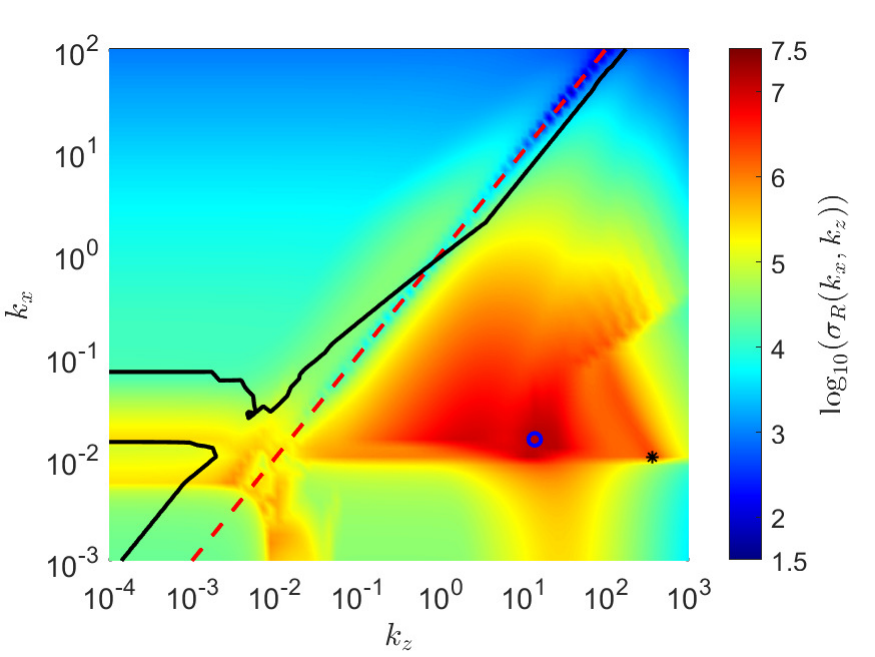}
    \caption{Resolvent gain} \label{fig:Resolvent_Mach_one}
  \end{subfigure}
  \begin{subfigure}[b]{0.495\linewidth}
    \includegraphics[width=1\textwidth]{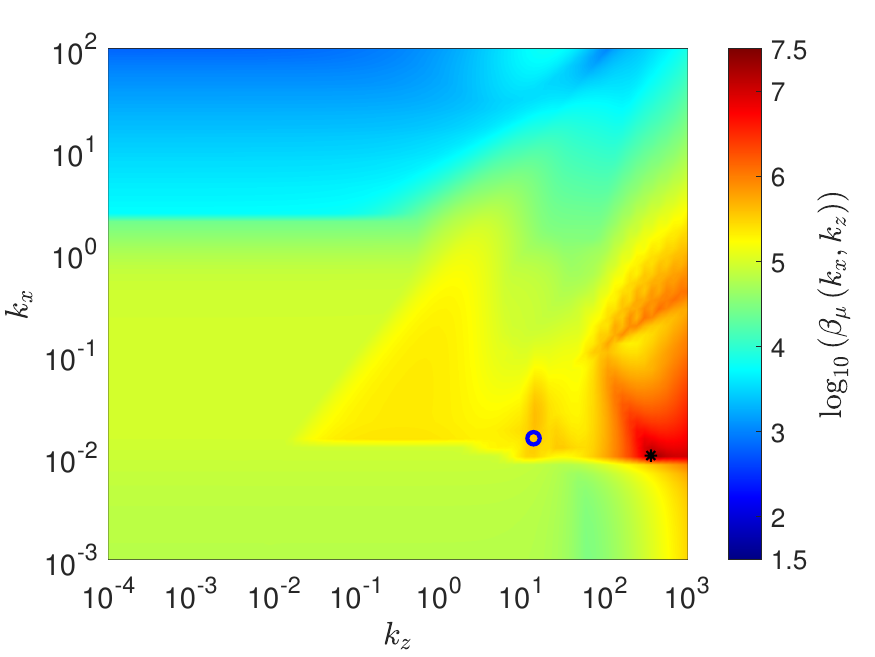}
    \caption{$\mu$ lower bound} \label{fig:Mulb_Mach_one}
  \end{subfigure}
  \begin{subfigure}[b]{0.495\linewidth}
    \includegraphics[width=1\textwidth]{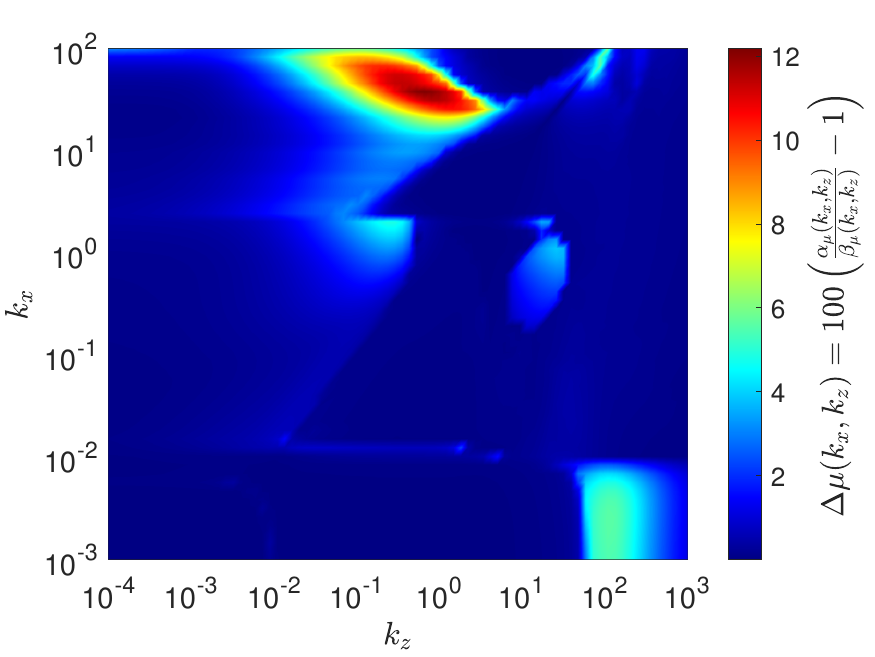}
    \caption{Gap between $\mu$ upper and lower bounds} \label{fig:Mugap_Mach_one}
  \end{subfigure} 
 \caption{Distributions of the $\mu$ upper and lower bounds (log-scaled), percentage gap between the $\mu$ bounds, and the resolvent gain (log-scaled) over the wavenumber pair $(k_x, k_z)$ grid for $M_r=1$. 
 The circle and the asterisk denote the $(k_x, k_z)$ values associated with the largest computed resolvent gain and $\mu$ bounds, respectively.
 {Moreover, the solid and dashed lines in resolvent gain plot respectively denote the unity relative Mach number contour and the $k_x=k_z$ line.}}
  \label{fig:Heat_plots_Mach_one}
\end{figure} 
\begin{figure}
\captionsetup[subfigure]{justification=centering}
  \centering
  \begin{subfigure}[b]{0.495\linewidth}
    \includegraphics[width=1\textwidth]{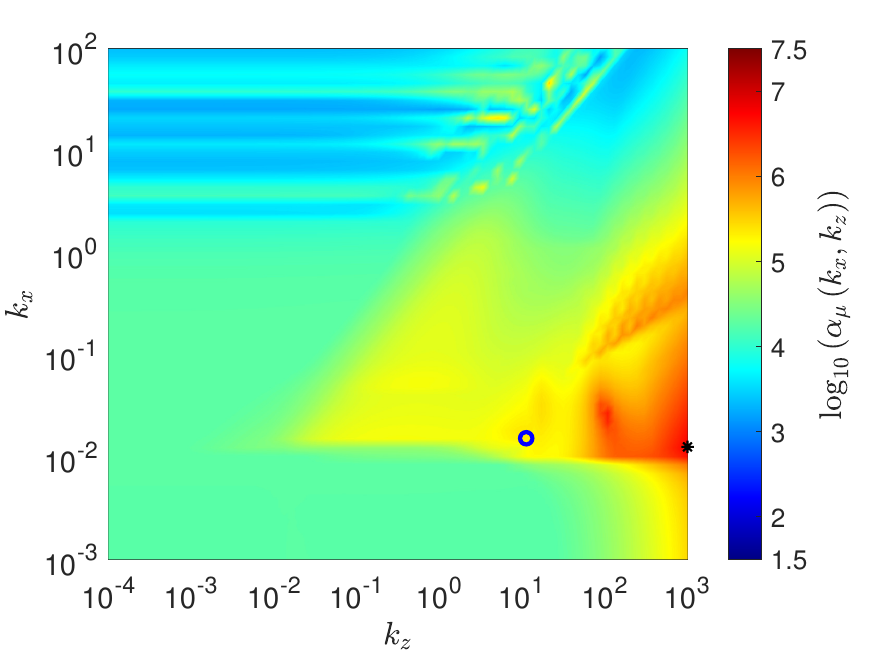}
    \caption{$\mu$ upper bound} \label{fig:Muub_Mach_two}
  \end{subfigure}
   \begin{subfigure}[b]{0.495\linewidth}
    \includegraphics[width=1\textwidth]{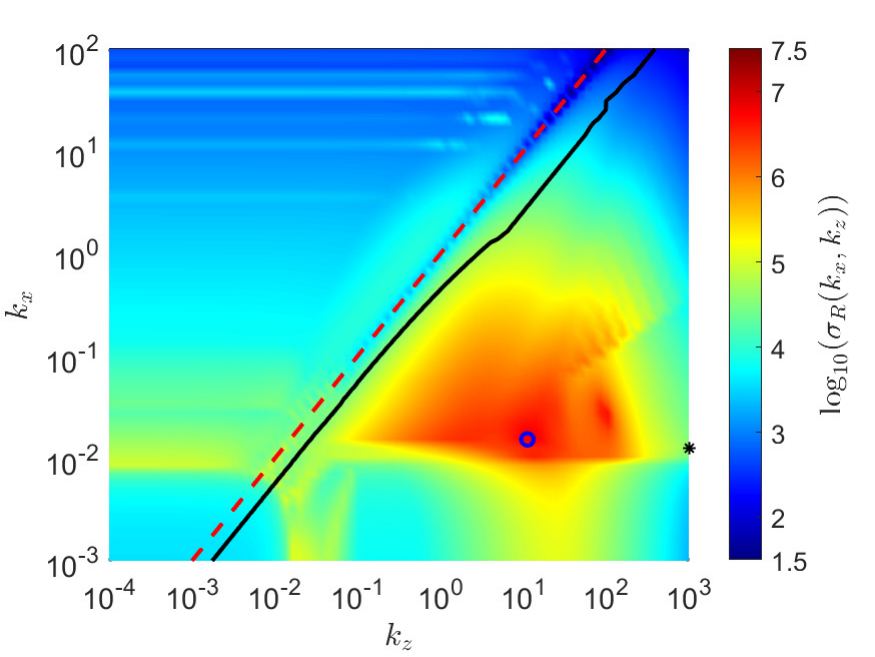}
    \caption{Resolvent gain} \label{fig:Resolvent_Mach_two}
  \end{subfigure}
  \begin{subfigure}[b]{0.495\linewidth}
    \includegraphics[width=1\textwidth]{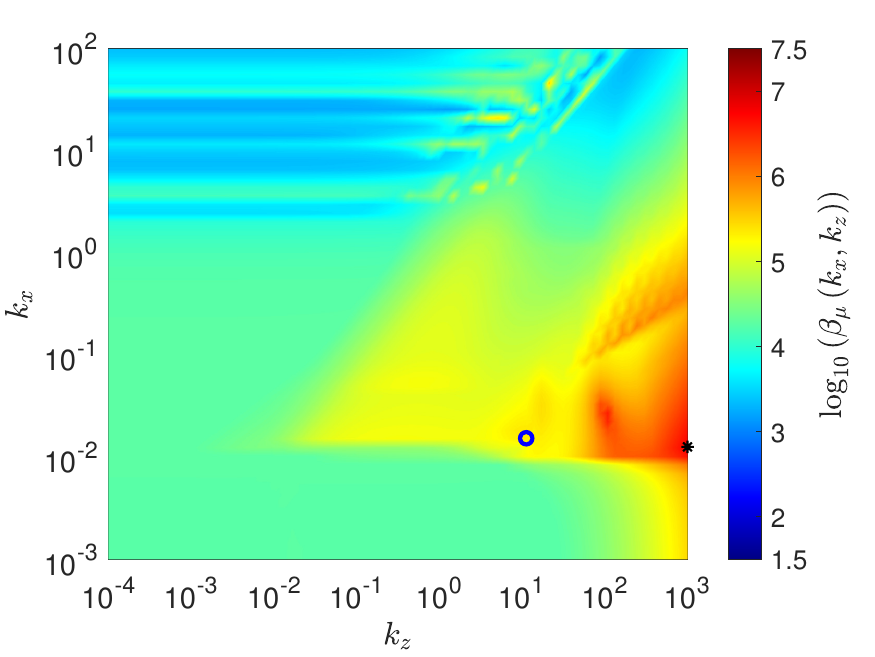}
    \caption{$\mu$ lower bound} \label{fig:Mulb_Mach_two}
  \end{subfigure}
  \begin{subfigure}[b]{0.495\linewidth}
    \includegraphics[width=1\textwidth]{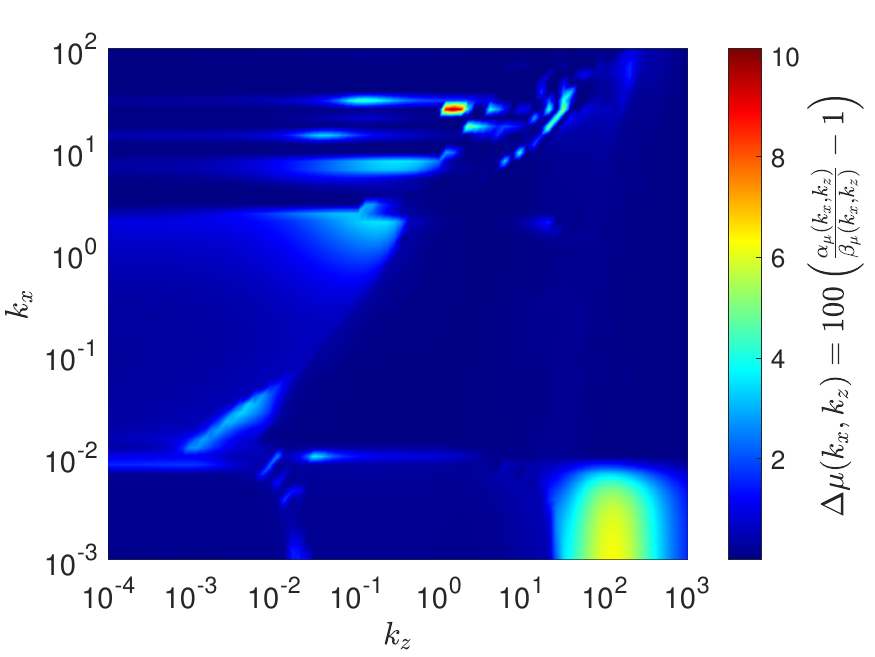}
    \caption{Gap between $\mu$ upper and lower bounds} \label{fig:Mugap_Mach_two}
  \end{subfigure}
 \caption{Distributions of the $\mu$ upper and lower bounds (log-scaled), percentage gap between the $\mu$ bounds, and the resolvent gain (log-scaled) over the wavenumber pair $(k_x, k_z)$ grid for $M_r=2$. 
 The circle and the asterisk denote the $(k_x, k_z)$ values associated with the largest computed resolvent gain and $\mu$ bounds, respectively.
 {Moreover, the solid and dashed lines in resolvent gain plot respectively denote the unity relative Mach number contour and the $k_x=k_z$ line.}}
  \label{fig:Heat_plots_Mach_two}
\end{figure} 

The I/O gains for a sonic ($M_r=1$) and a supersonic ($M_r=2$) condition are shown in Figs. \ref{fig:Heat_plots_Mach_one} and \ref{fig:Heat_plots_Mach_two}, respectively.
Similar to the subsonic results earlier, the upper and lower bounds on $\mu$ here are tight for a large fraction of the $(k_x,k_z)$ grid points, as illustrated by the gap plots in Fig. \ref{fig:Mugap_Mach_one}, \ref{fig:Mugap_Mach_two}, with the gap being less than 5\% for approximately 95.79\% (in Fig. \ref{fig:Mugap_Mach_one} for $M_r=1$) and 98.67\% (in Fig. \ref{fig:Mugap_Mach_two} for $M_r=2$) of the total $(k_x,k_z)$ grid points.
Also, the quality of the bounds improves with an increase in the Mach number, as reflected by the average gaps which are approximately 0.896\% and 0.523\% for $M_r=1$ and $M_r=2$, respectively.
Going by the upper and lower bound values, the structured I/O gains (i.e., the exact $\mu$) get smaller overall as the Mach number increases, which is true for the resolvent gains as well. 
These trends are consistent with the existing resolvent analysis results for the compressible Couette flow \citep{dawsonAIAA2019}. 
Also, as remarked by the authors in \cite{dawsonAIAA2019}, this variation in the I/O gains with Mach number is consistent with the linear transient energy growth results reported for this flow \citep{malikPOF2006, malik2008linear}.
{Note that the region of smallest resolvent gain, which approximately corresponds to $k_x=k_z$, also aligns close to the streamwise and spanwise wavenumber pairs associated with the relative Mach number \citep{mack1984boundary,schmid2002stability} equal to one for the results in Figs. \ref{fig:Resolvent_Mach_one}, \ref{fig:Resolvent_Mach_two}.
This concept is frequently utilized for studying and specifying subsonic, sonic and supersonic disturbances in compressible boundary layer flows \citep{mack1984boundary, baeJFM2019, baeAIAA2020}.}

\begin{figure}
\captionsetup[subfigure]{justification=centering}
  \centering
  \begin{subfigure}[b]{0.495\textwidth}
    \includegraphics[width=1\textwidth]{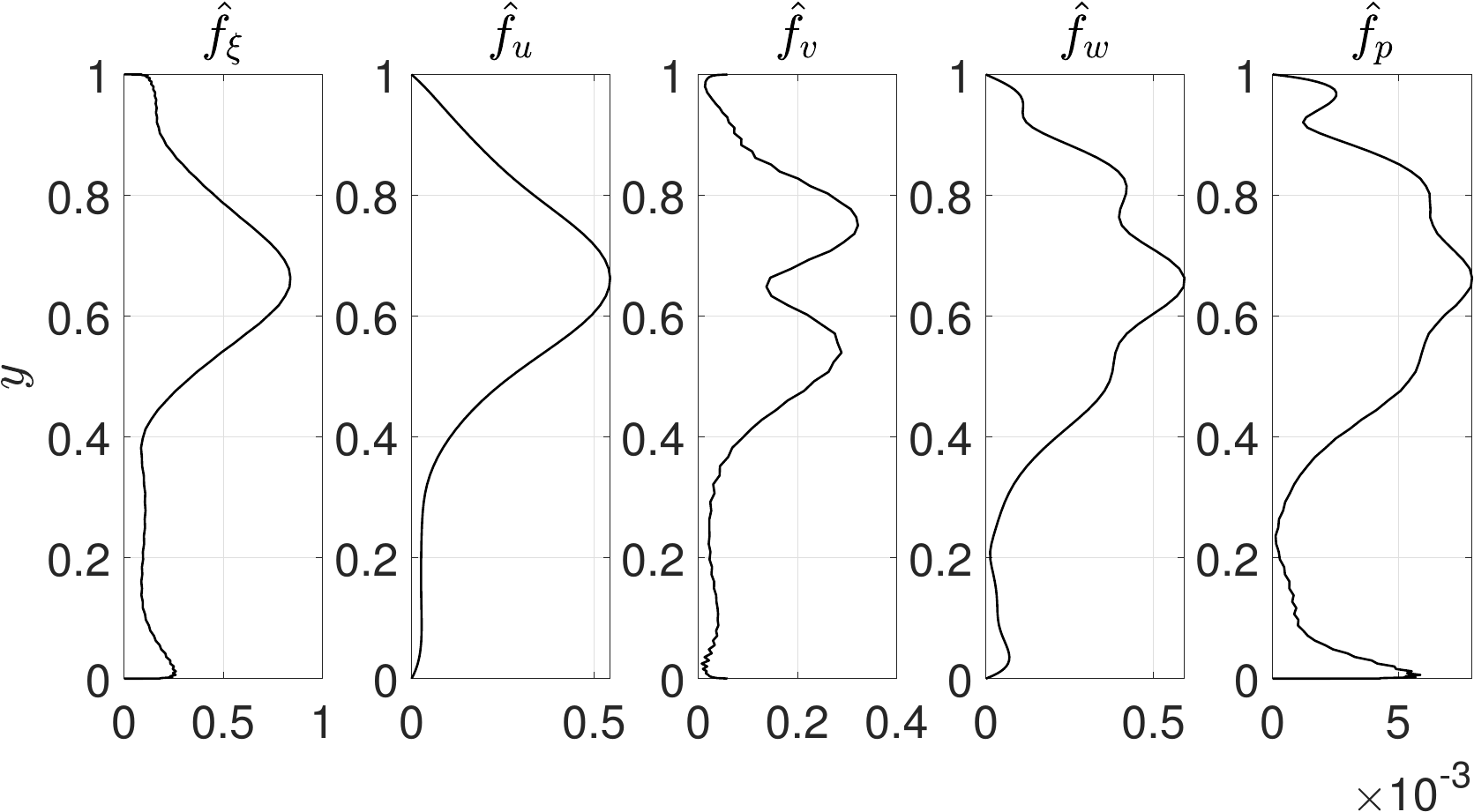}
    \caption{Structured I/O forcing modes} \label{fig:structured_forcingmodes_Mach_one_maxresolvent}
  \end{subfigure}
\begin{subfigure}[b]{0.495\textwidth}
    \includegraphics[width=1\textwidth]{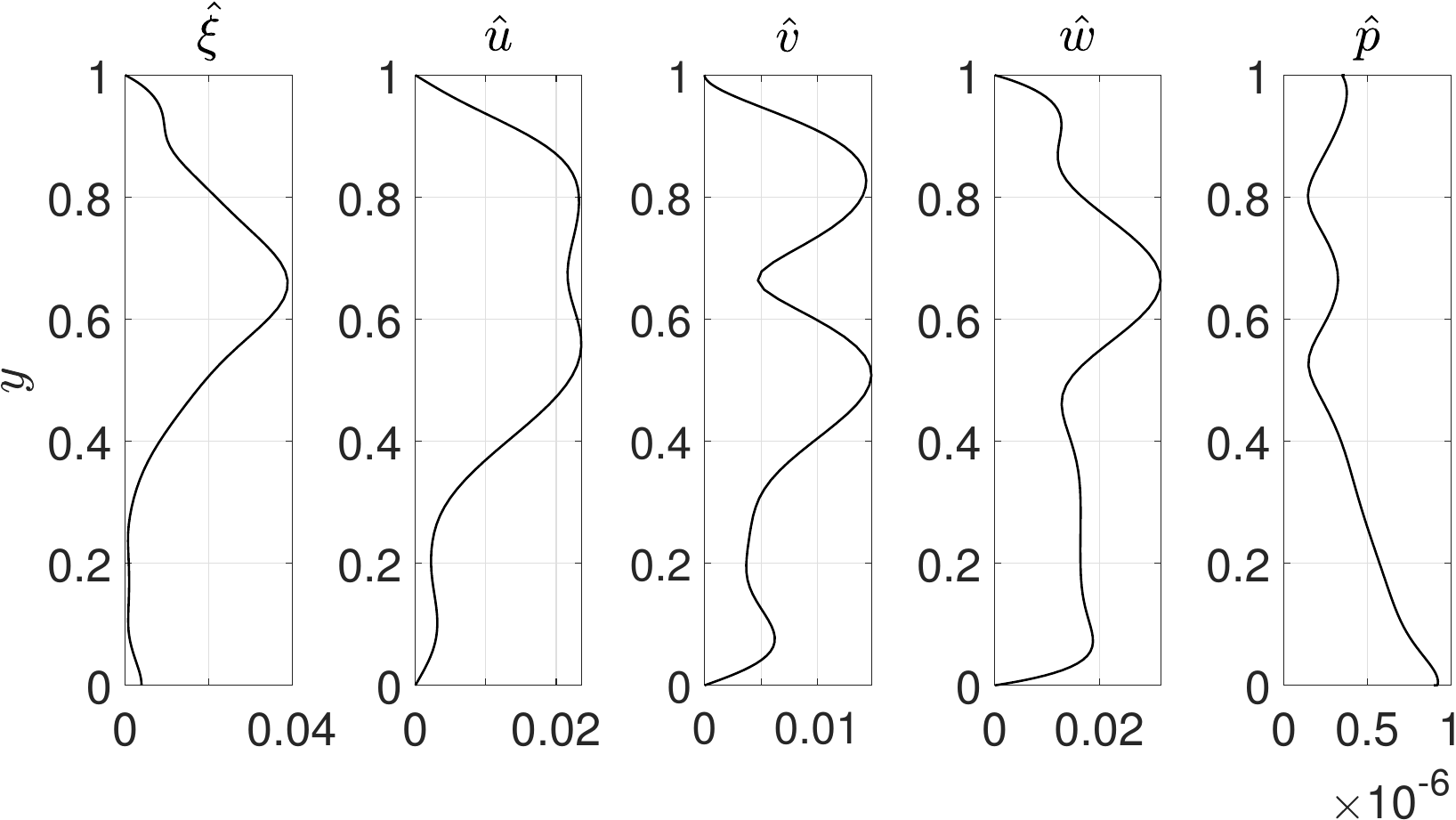}
    \caption{Structured I/O response modes} \label{fig:structured_responsemodes_Mach_one_maxresolvent}
  \end{subfigure}
  \begin{subfigure}[b]{0.495\textwidth}
    \includegraphics[width=1\textwidth]{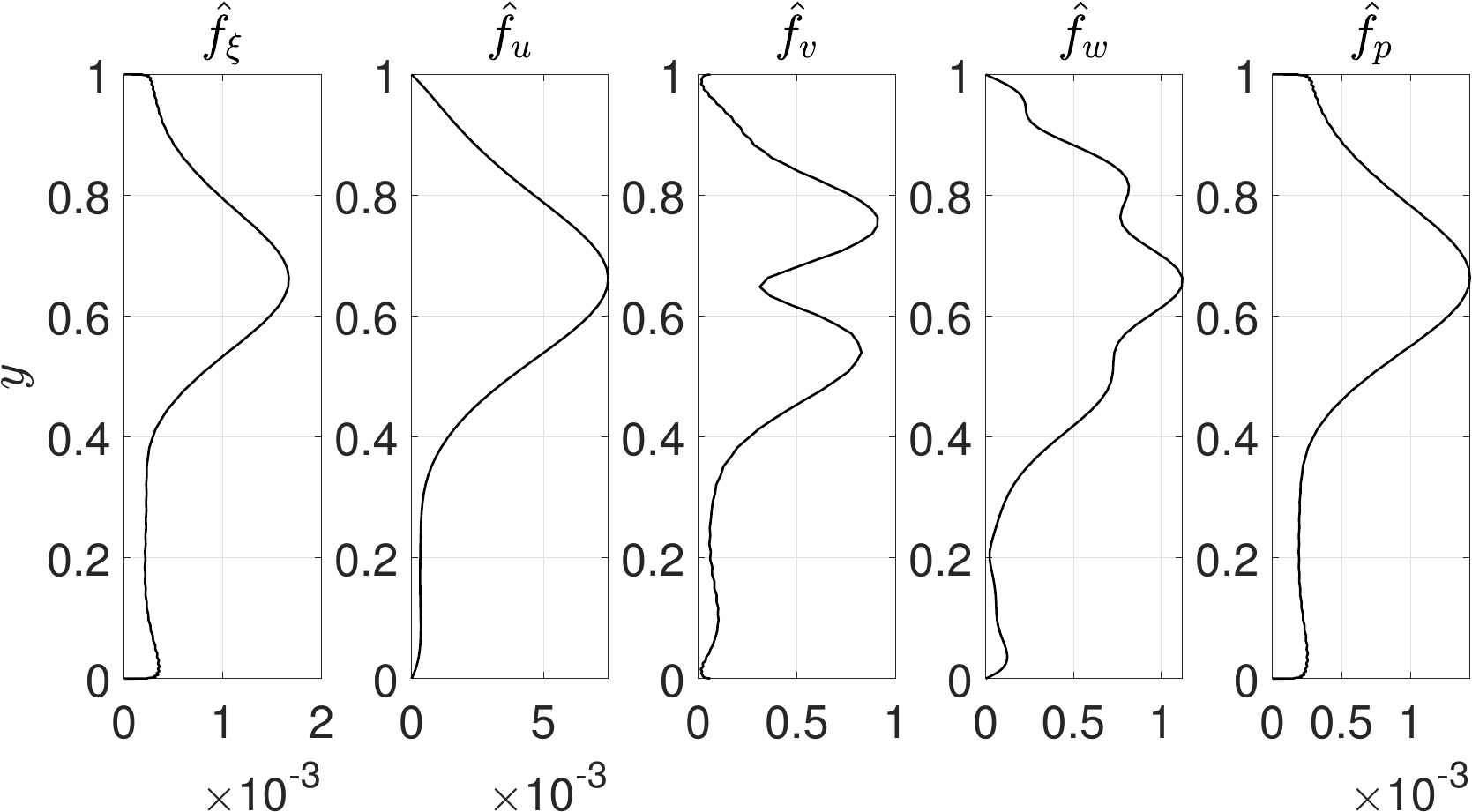}
    \caption{Resolvent forcing modes} \label{fig:resolvent_forcingmodes_Mach_one_maxresolvent}
  \end{subfigure}
  \begin{subfigure}[b]{0.495\textwidth}
    \includegraphics[width=1\textwidth]{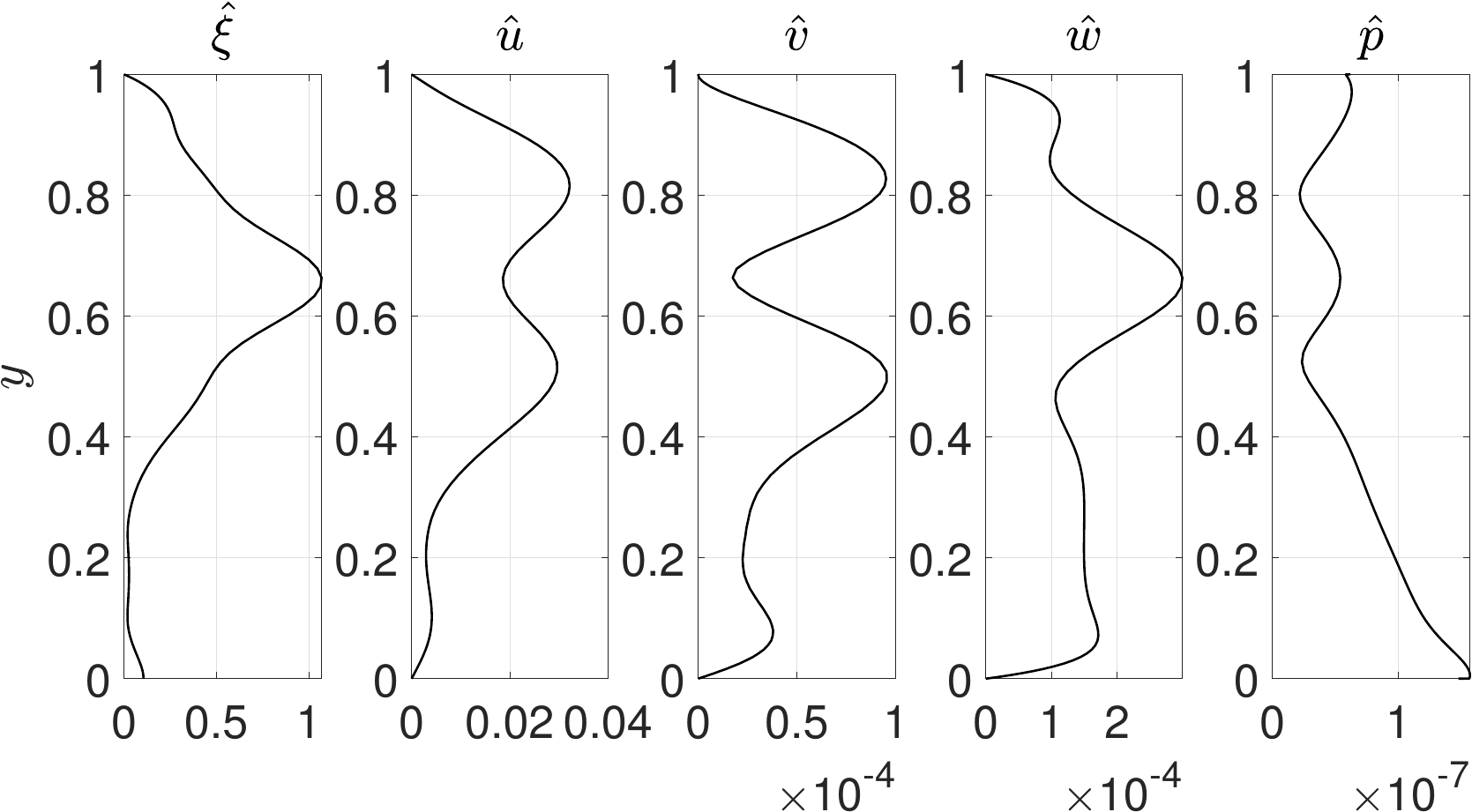}
    \caption{Resolvent response modes} \label{fig:resolvent_responsemodes_Mach_one_maxresolvent}
  \end{subfigure}
 \caption{Absolute values of the structured I/O and resolvent modes for $M_r=1$, $(k_x, k_z, \omega)=(0.015,13.78,-0.01)$, which corresponds to the largest resolvent gain in Fig. \ref{fig:Heat_plots_Mach_one}.}
\label{fig:forcing_and_response_modes_Mach_one_maxresolvent}
\end{figure}
\begin{figure}
\captionsetup[subfigure]{justification=centering}
  \centering
  \begin{subfigure}[b]{0.495\textwidth}
    \includegraphics[width=1\textwidth]{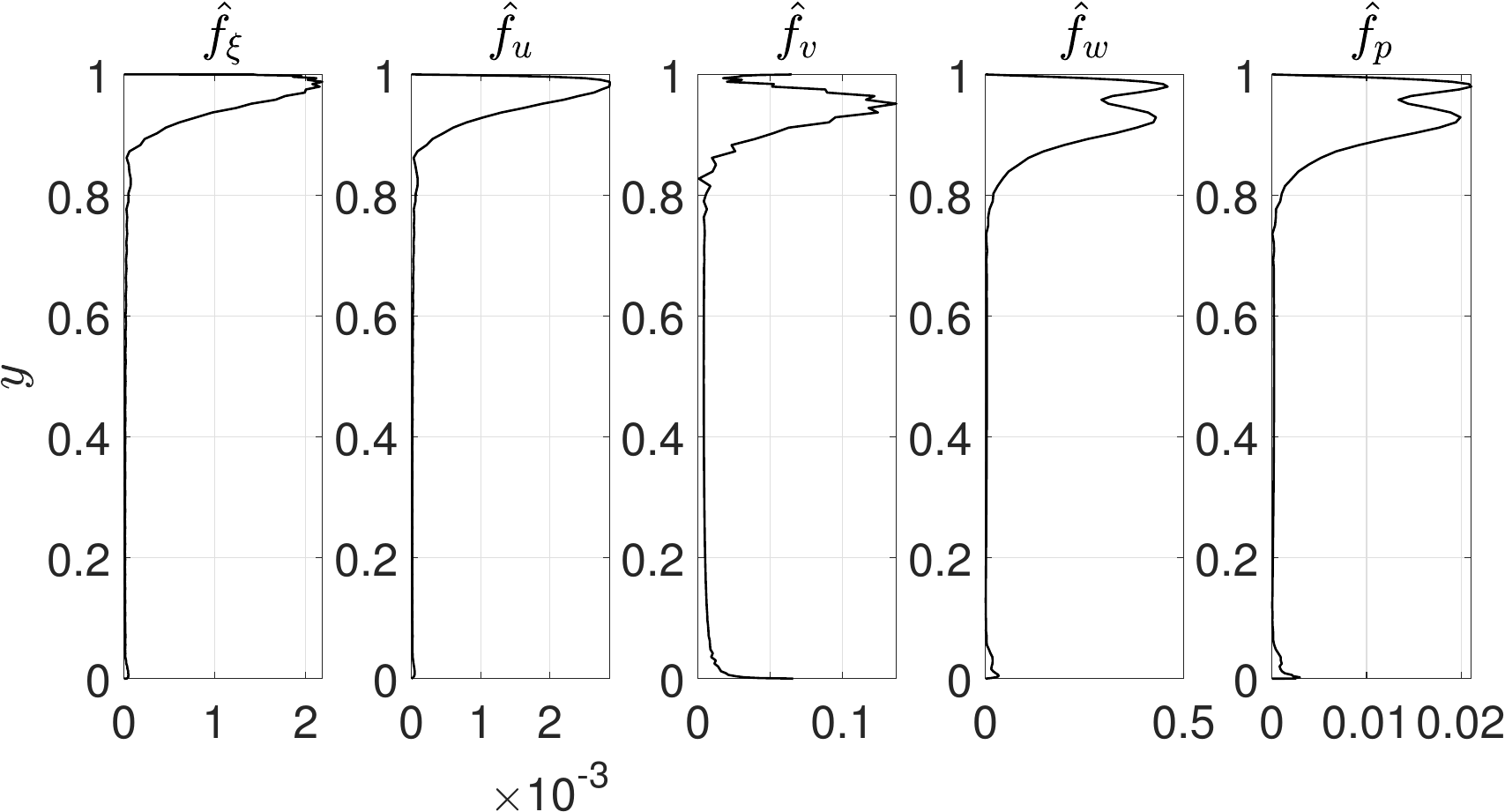}
    \caption{Structured I/O forcing modes} \label{fig:structured_forcingmodes_Mach_one_maxSSV}
  \end{subfigure}
\begin{subfigure}[b]{0.495\textwidth}
    \includegraphics[width=1\textwidth]{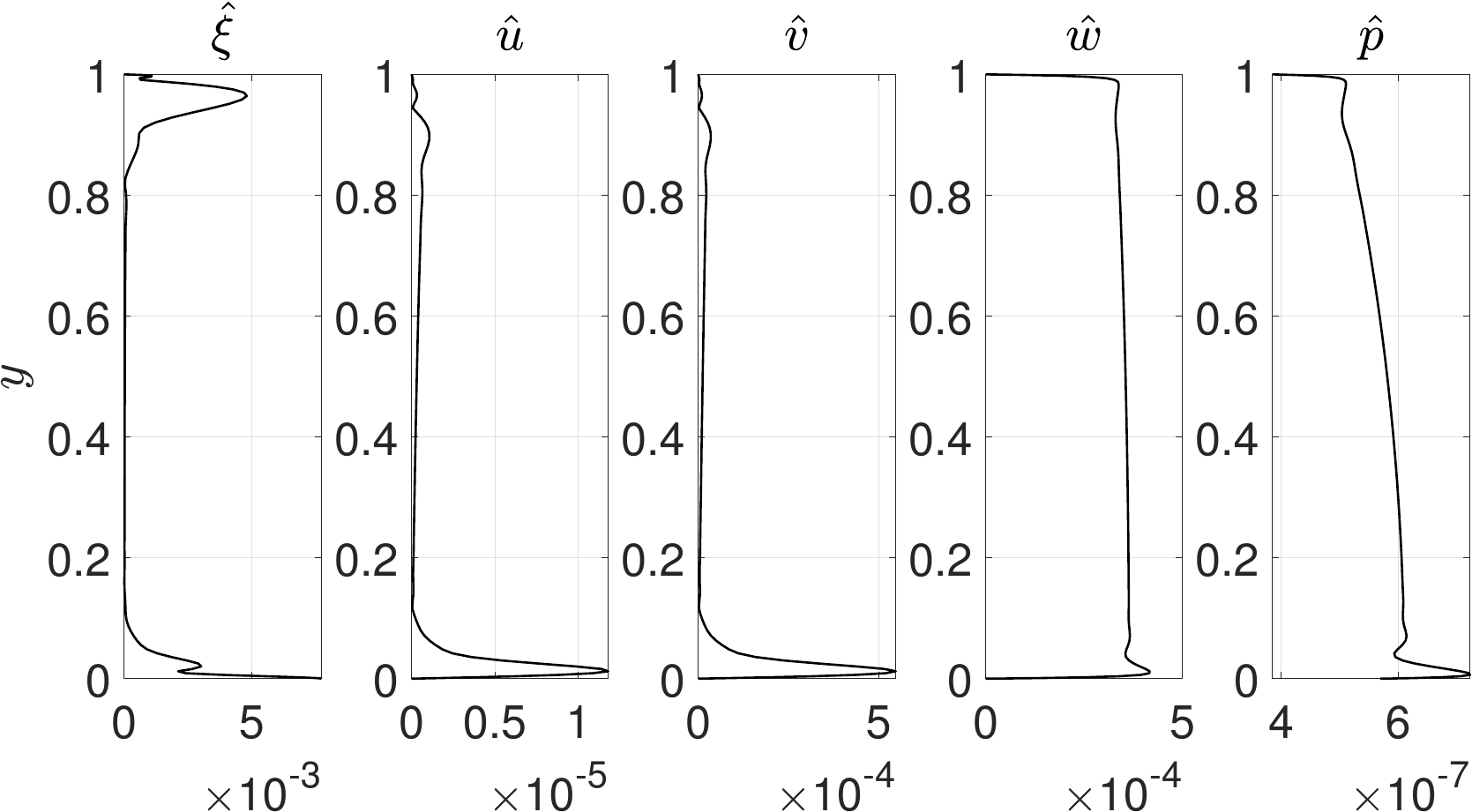}
    \caption{Structured I/O response modes} \label{fig:structured_responsemodes_Mach_one_maxSSV}
  \end{subfigure}
  \begin{subfigure}[b]{0.495\textwidth}
    \includegraphics[width=1\textwidth]{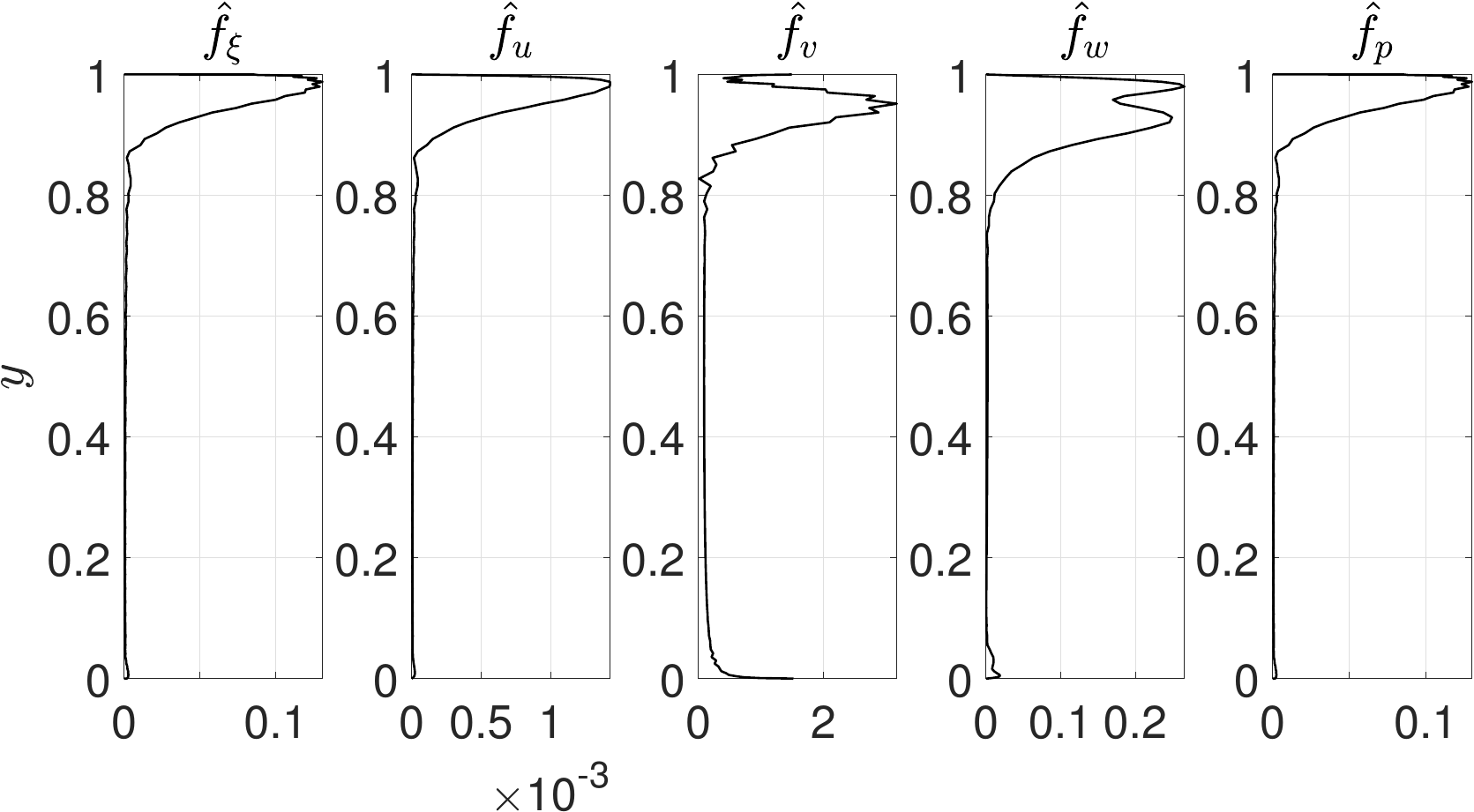}
    \caption{Resolvent forcing modes} \label{fig:resolvent_forcingmodes_Mach_one_maxSSV}
  \end{subfigure}
  \begin{subfigure}[b]{0.495\textwidth}
    \includegraphics[width=1\textwidth]{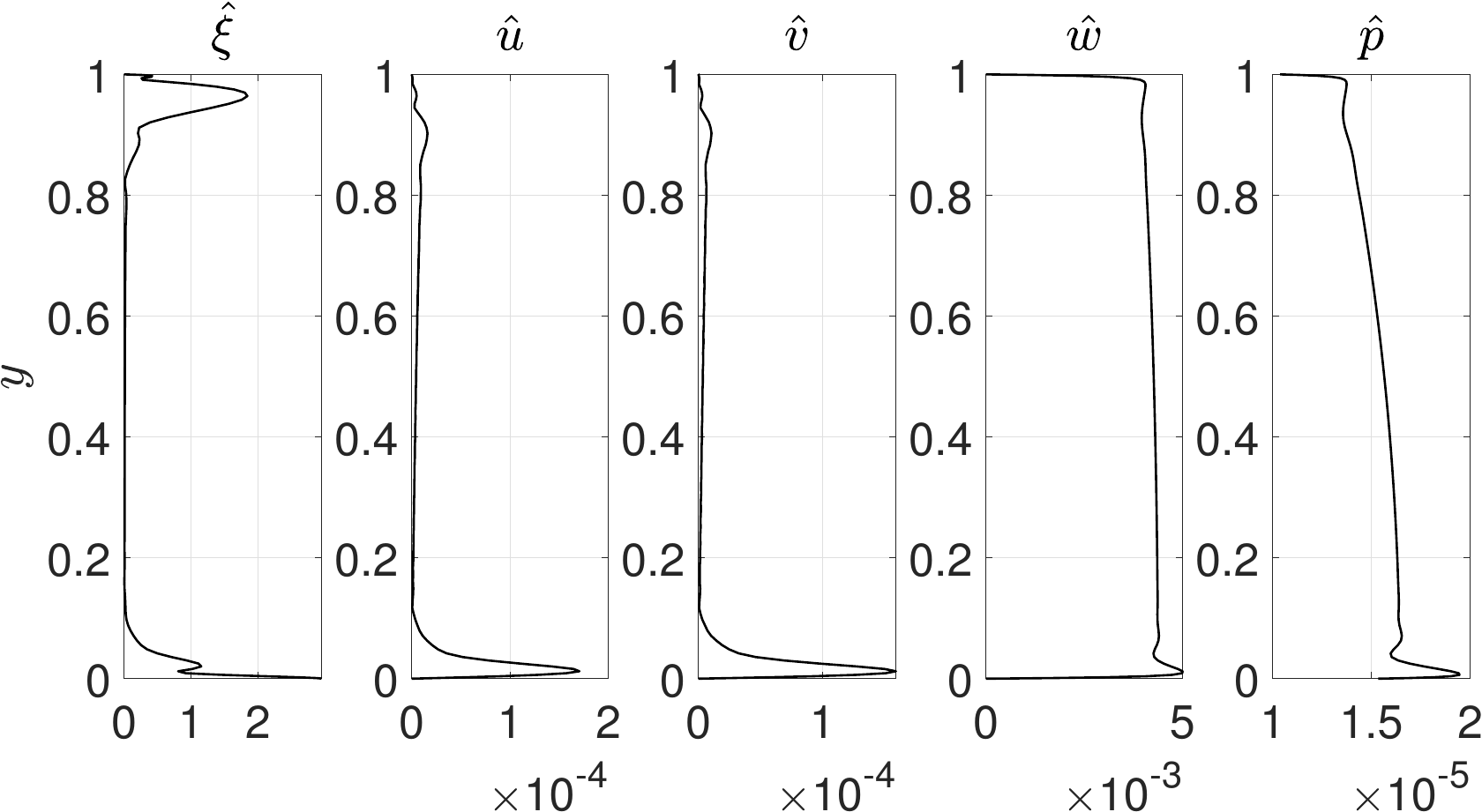}
    \caption{Resolvent response modes} \label{fig:resolvent_responsemodes_Mach_one_maxrSSV}
  \end{subfigure}
 \caption{{Absolute values of the structured I/O and resolvent modes for $M_r=1$, $(k_x, k_z, \omega)=(0.01,360.55,-0.01)$, which corresponds to the largest values in both the $\mu$ bounds in Fig. \ref{fig:Heat_plots_Mach_one}.}}
\label{fig:forcing_and_response_modes_Mach_one_maxSSV}
\end{figure}

While the regions of high structured I/O gain for $k_z$ approximately greater than $10^2$ in the subsonic results carry over to the supersonic case, it is interesting to note that the spanwise elongated structure (i.e., the `horizontal band') in the subsonic results takes the form of an oblique structure at these higher Mach numbers.
Thus, structured I/O is able to identify an instability that depends on the Mach number, as well as predicting the evolution of its characteristics over different Mach number regimes considered here.
On the other hand, resolvent analysis is unable to capture this evolution and consistently predicts spanwise elongated structures even for $M_r=2$, which, similar to our earlier remarks for the subsonic results, might not represent actual flow behavior.
It is also noteworthy that the structured I/O analysis is able to better distinguish between the amplified flow features locally (i.e., localized regions of high I/O gain) compared to the resolvent analysis. 
For example, in Figs. \ref{fig:Muub_Mach_one}, \ref{fig:Mulb_Mach_one}, four distinct regions/features of high I/O gain are highlighted: (a) the oblique structure (triangular region) for $k_x \approx (10^{-2},10^0)$ and $k_z \approx (10^{-2},5\times 10^0)$; (b) the streamwise structure (rectangular region) for $k_x \approx (10^{-2},10^{-1})$ and $k_z \approx (5\times 10^0, 5 \times 10^1)$; (c) the oblique structure associated with the largest gains (highlighted in red); (d) the streamwise structure for $k_x \approx (10^{-3},10^{-2})$ and $k_z \approx (7\times 10^2, 10^3)$.
Although resolvent analysis is able to capture all the above-mentioned features (see Fig. \ref{fig:Resolvent_Mach_one} for comparison), the features mentioned in (a)-(c) above are all lumped into a single, relatively large region of high gain in the wavenumber space, and it might be challenging to identify these separate features from the resolvent gain plot alone without referring to the structured I/O results.
Therefore, structured I/O analysis can be utilized to determine intricate details associated with amplified flow structures, thereby providing additional insight into the flow physics.

\begin{figure}
\captionsetup[subfigure]{justification=centering}
  \centering
  \begin{subfigure}[b]{0.495\textwidth}
    \includegraphics[width=1\textwidth]{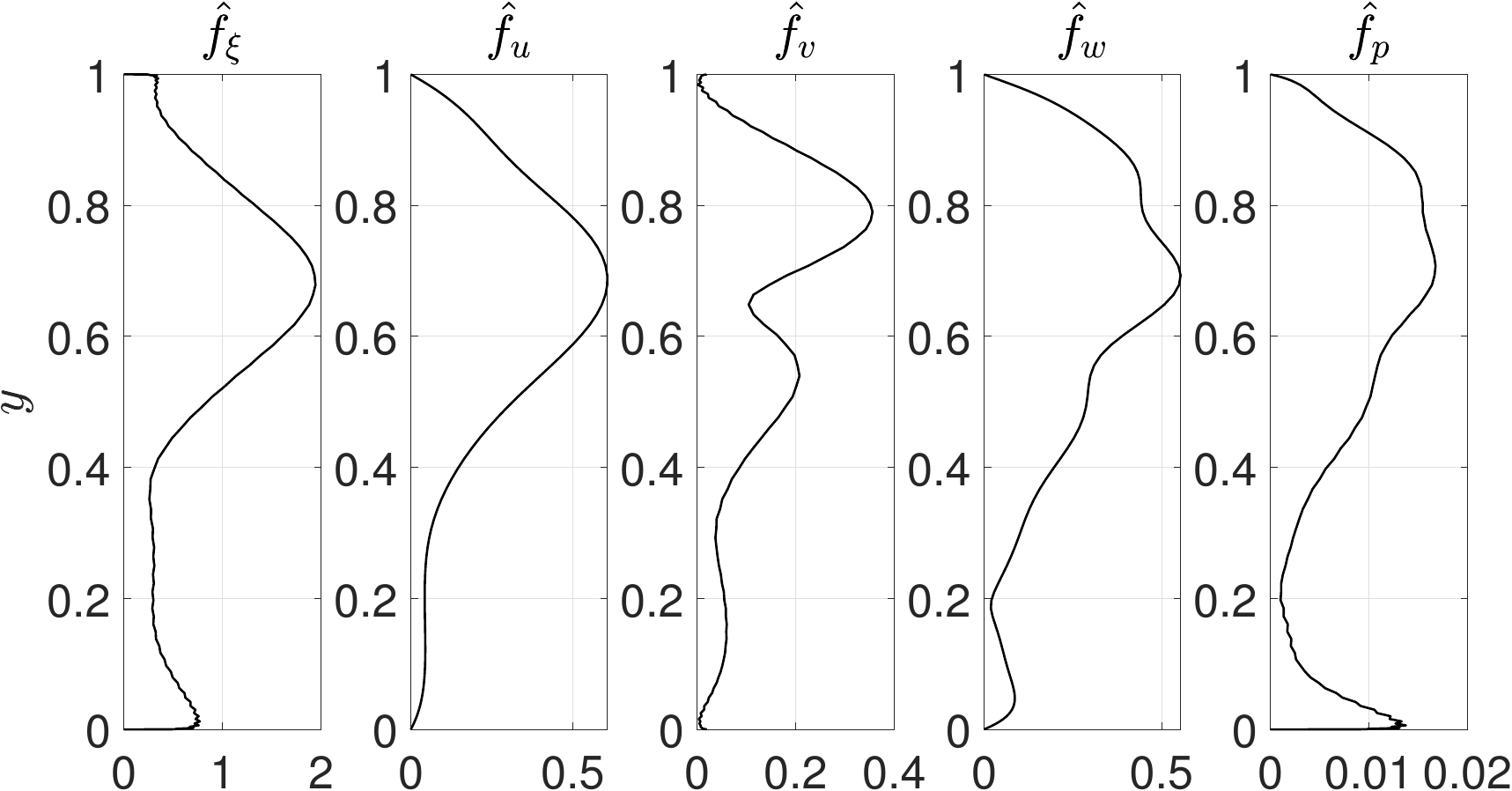}
    \caption{Structured I/O forcing modes} \label{fig:structured_forcingmodes_Mach_two_maxresolvent}
  \end{subfigure}
\begin{subfigure}[b]{0.495\textwidth}
    \includegraphics[width=1\textwidth]{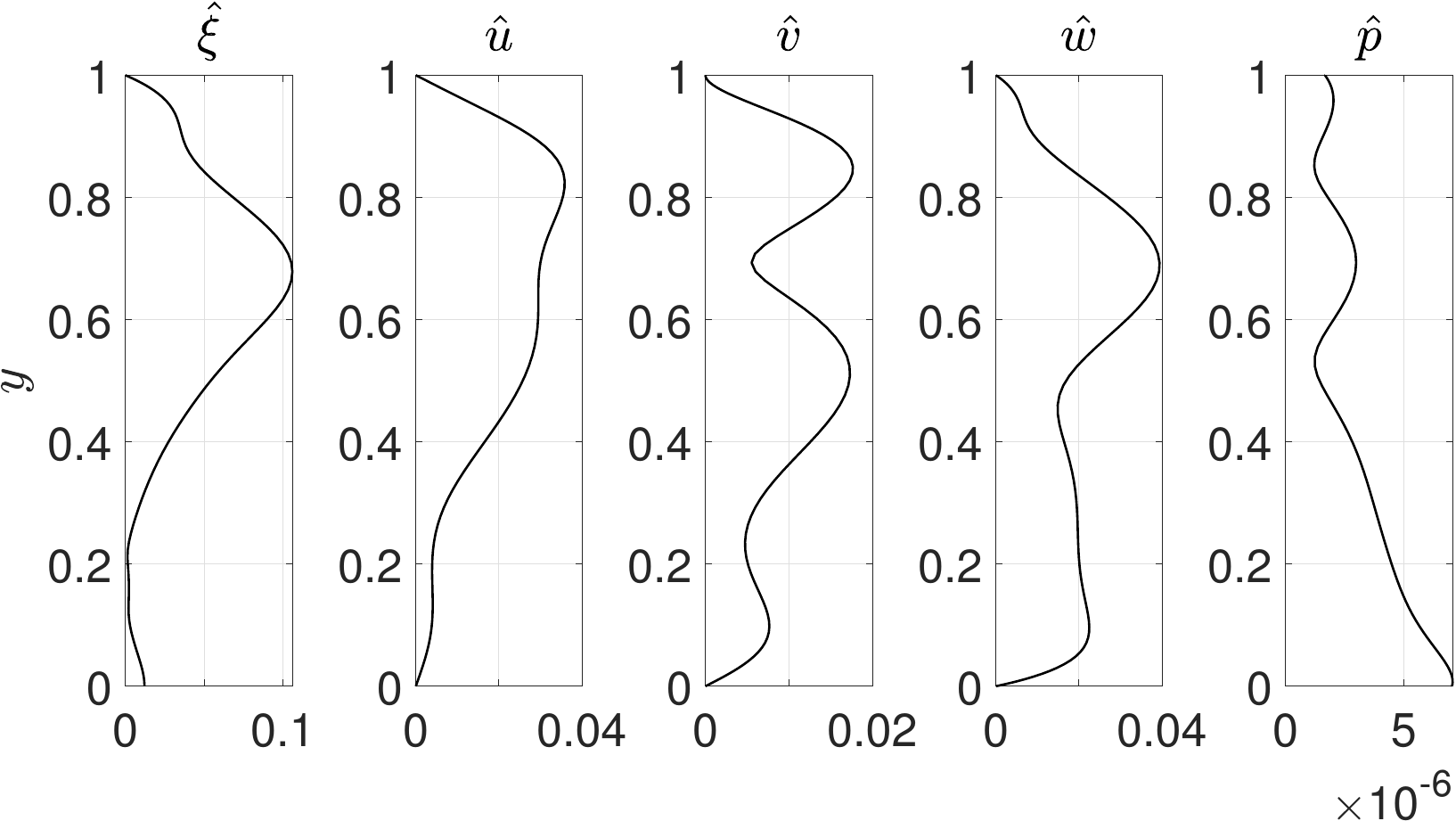}
    \caption{Structured I/O response modes} \label{fig:structured_responsemodes_Mach_two_maxresolvent}
  \end{subfigure}
  \begin{subfigure}[b]{0.495\textwidth}
    \includegraphics[width=1\textwidth]{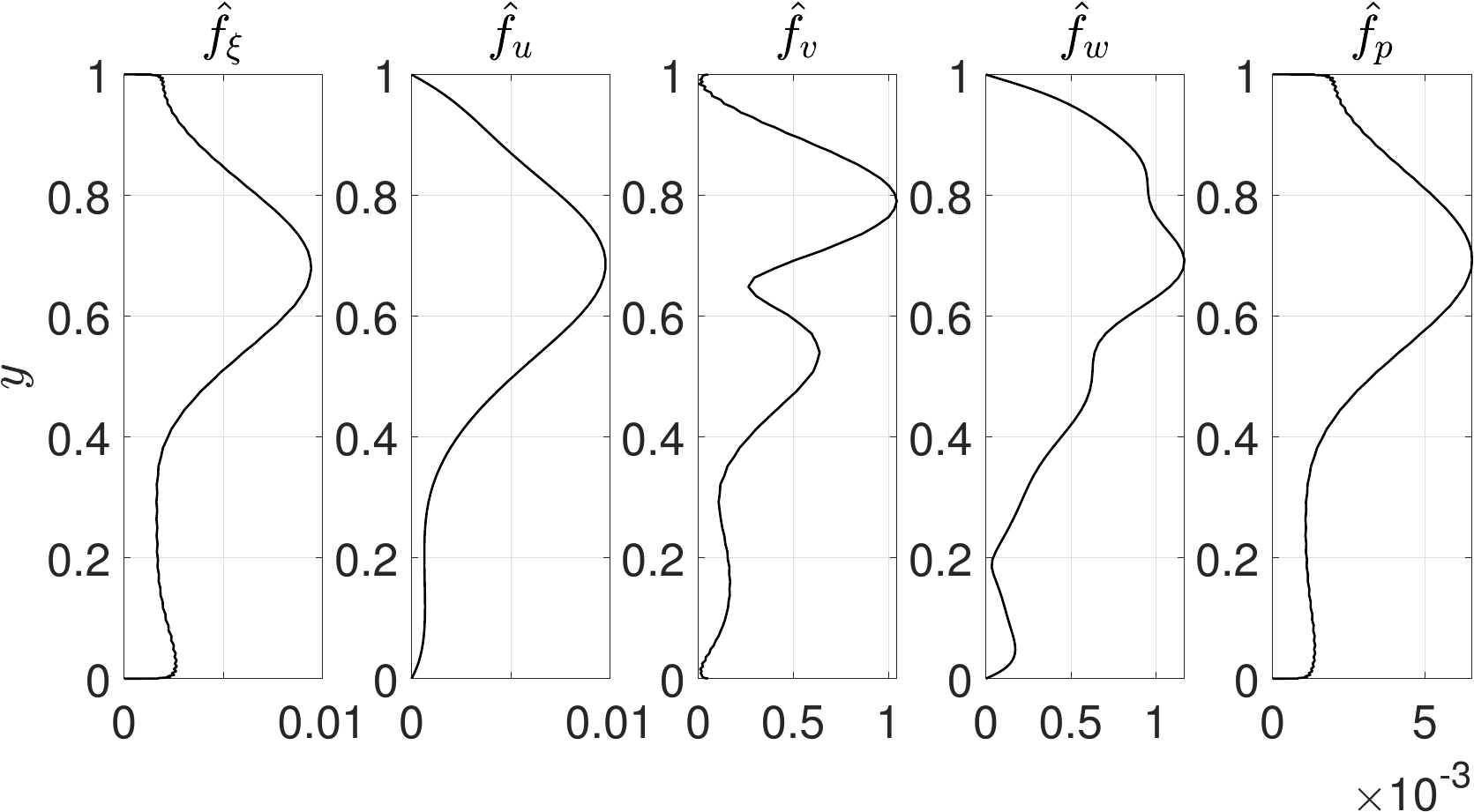}
    \caption{Resolvent forcing modes} \label{fig:resolvent_forcingmodes_Mach_two_maxresolvent}
  \end{subfigure}
  \begin{subfigure}[b]{0.495\textwidth}
    \includegraphics[width=1\textwidth]{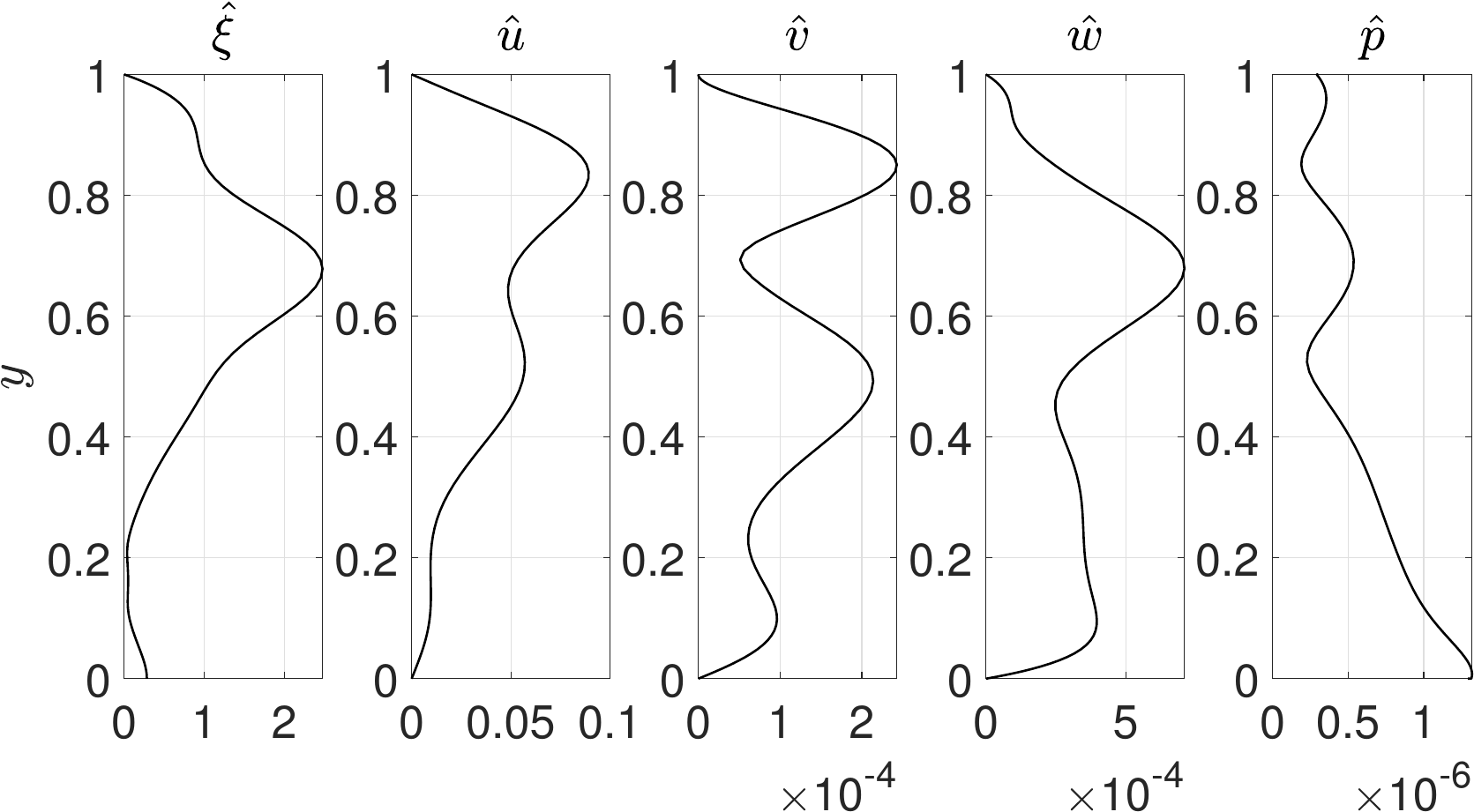}
    \caption{Resolvent response modes} \label{fig:resolvent_responsemodes_Mach_two_maxresolvent}
  \end{subfigure}
 \caption{Absolute values of the structured I/O and resolvent modes for $M_r=2$, $(k_x, k_z, \omega)=(0.015,11.24,-0.01)$, which corresponds to the largest resolvent gain in Fig. \ref{fig:Heat_plots_Mach_two}.}
\label{fig:forcing_and_response_modes_Mach_two_maxresolvent}
\end{figure}

\begin{figure}
\captionsetup[subfigure]{justification=centering}
  \centering
  \begin{subfigure}[b]{0.495\textwidth}
    \includegraphics[width=1\textwidth]{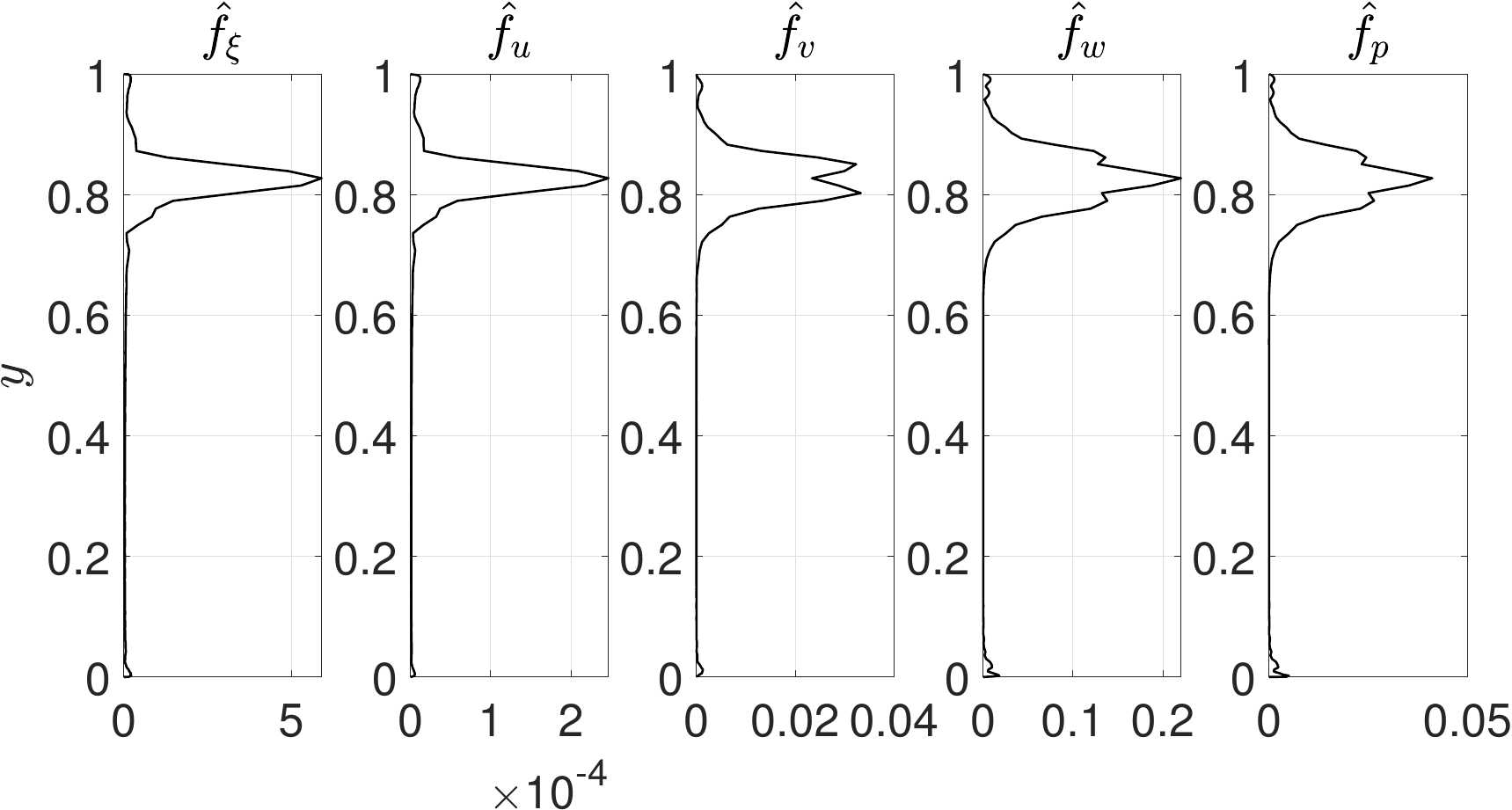}
    \caption{Structured I/O forcing modes} \label{fig:structured_forcingmodes_Mach_two_maxSSV}
  \end{subfigure}
\begin{subfigure}[b]{0.495\textwidth}
    \includegraphics[width=1\textwidth]{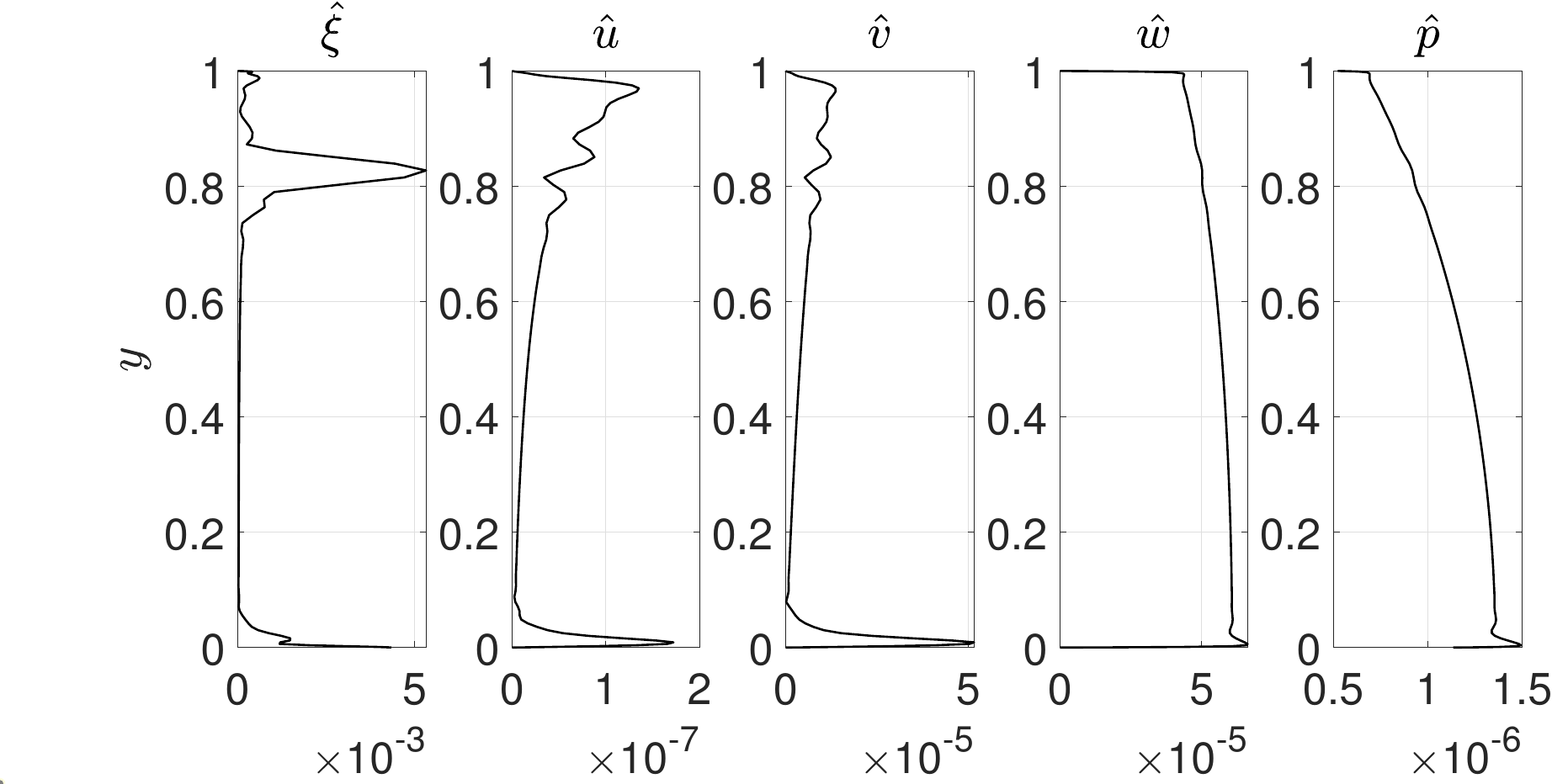}
    \caption{Structured I/O response modes} \label{fig:structured_responsemodes_Mach_two_maxSSV}
  \end{subfigure}
  \begin{subfigure}[b]{0.495\textwidth}
    \includegraphics[width=1\textwidth]{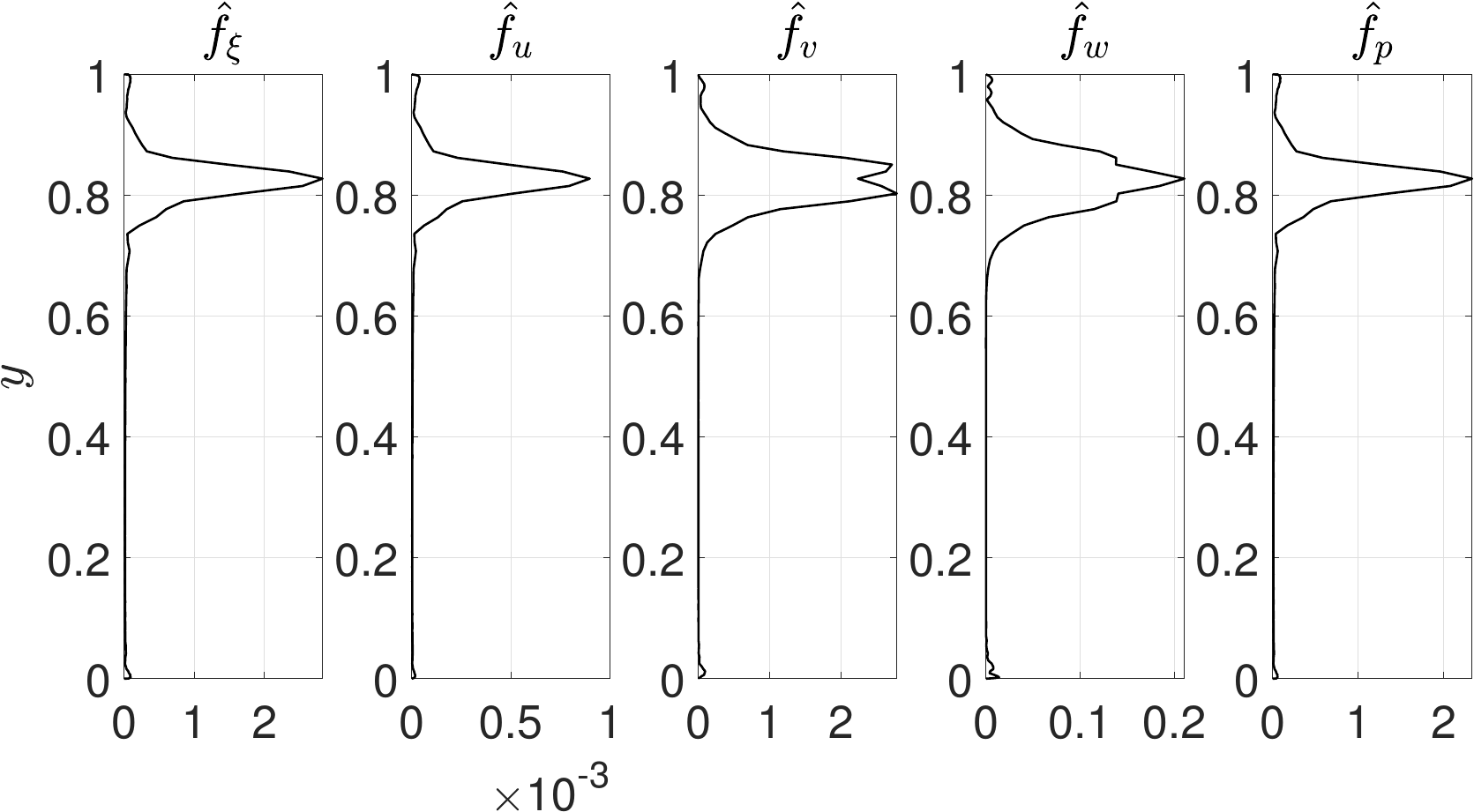}
    \caption{Resolvent forcing modes} \label{fig:resolvent_forcingmodes_Mach_two_maxSSV}
  \end{subfigure}
  \begin{subfigure}[b]{0.495\textwidth}
    \includegraphics[width=1\textwidth]{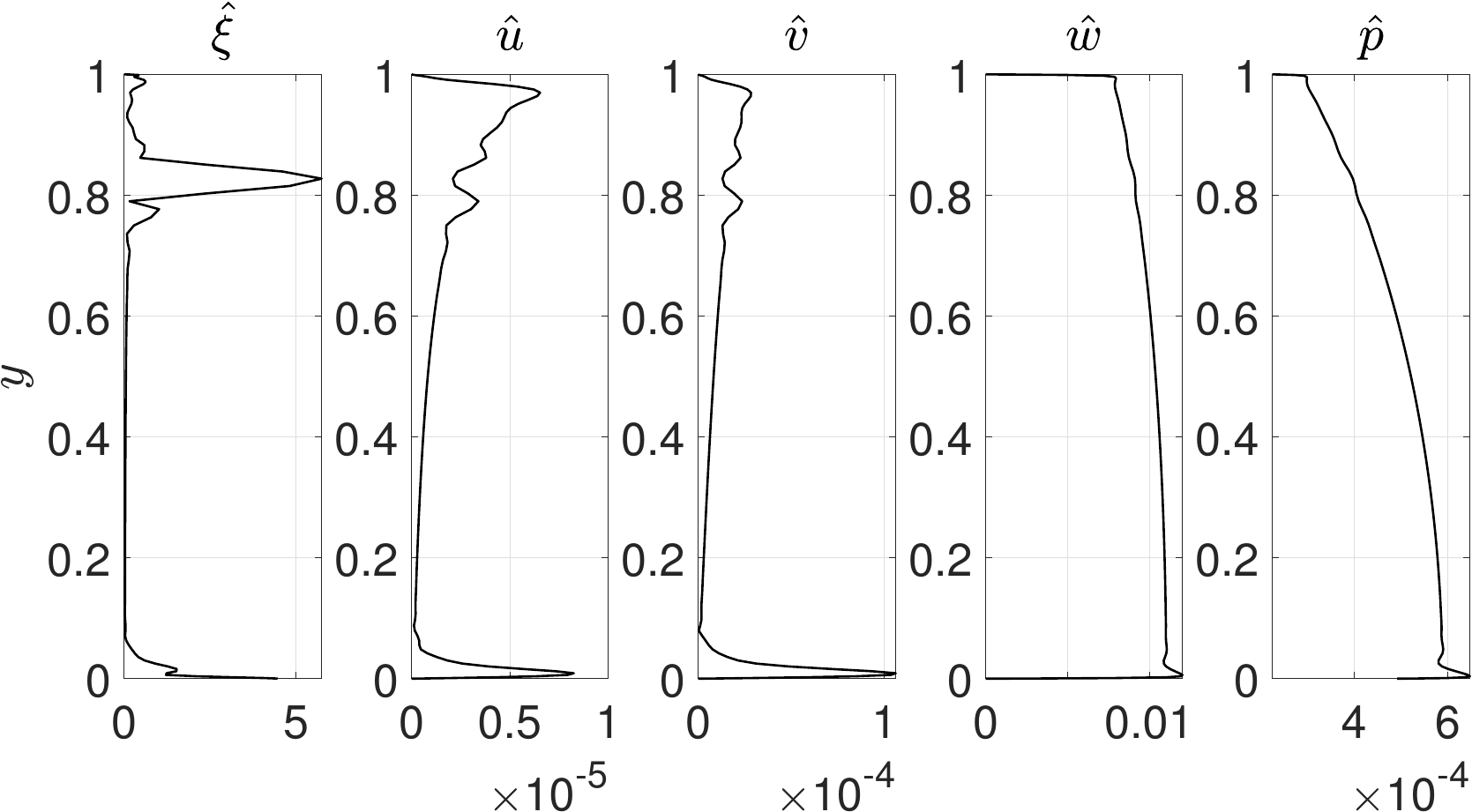}
    \caption{Resolvent response modes} \label{fig:resolvent_responsemodes_Mach_two_maxrSSV}
  \end{subfigure}
 \caption{Absolute values of the structured I/O and resolvent modes for $M_r=2$, $(k_x, k_z, \omega)=(0.013,10^3,-0.01)$, which corresponds to the largest values in both the $\mu$ bounds in Fig. \ref{fig:Heat_plots_Mach_two}.}
\label{fig:forcing_and_response_modes_Mach_two_maxSSV}
\end{figure}

Next, we analyze the modal behavior of the instability 
associated with the largest resolvent and structured I/O gain for the sonic and supersonic conditions, which are shown in Figs. \ref{fig:forcing_and_response_modes_Mach_one_maxresolvent}, \ref{fig:forcing_and_response_modes_Mach_one_maxSSV}, \ref{fig:forcing_and_response_modes_Mach_two_maxresolvent}, \ref{fig:forcing_and_response_modes_Mach_two_maxSSV}.
These results essentially depict the evolution of the forcing and response modes in Figs. \ref{fig:forcing_and_response_modes_Mach_half_1}, \ref{fig:forcing_and_response_modes_Mach_half_2} as the Mach number increases, and provides insight into the modifications of modal characteristics of the associated instability. 
%
%
Specifically, Figs. \ref{fig:forcing_and_response_modes_Mach_one_maxresolvent}, \ref{fig:forcing_and_response_modes_Mach_two_maxresolvent} illustrate the evolution of the modes associated with the 
global maximum in resolvent gain and the accompanying local maximum in the structured I/O gain.
In other words, these are higher Mach number variants of the modes in Fig. \ref{fig:forcing_and_response_modes_Mach_half_1}. 
The flow variables characterizing the dominant forcing and response modes in Fig. \ref{fig:forcing_and_response_modes_Mach_half_1} remain consistent in case of the sonic (Fig. \ref{fig:forcing_and_response_modes_Mach_one_maxresolvent}) and supersonic (Fig. \ref{fig:forcing_and_response_modes_Mach_two_maxresolvent}) flow conditions too. 
However, the fluctuations in the forcing modes, especially for the wall-normal velocity component, move closer to the center of the channel as the Mach number increases. 
The response modes in Figs. \ref{fig:forcing_and_response_modes_Mach_one_maxresolvent}, \ref{fig:forcing_and_response_modes_Mach_two_maxresolvent} also exhibit increased fluctuations near the center of the channel compared to the response modes in Fig. \ref{fig:forcing_and_response_modes_Mach_half_1}. 
More crucially, structured I/O modes in Figs. \ref{fig:structured_forcingmodes_Mach_one_maxresolvent}, \ref{fig:structured_forcingmodes_Mach_two_maxresolvent} show that the dominant forcing is in the specific volume 
at sonic and supersonic Mach numbers, which is different from the subsonic modes in Fig. \ref{fig:structured_forcingmodes_Mach_half_1} where the dominant forcing mode corresponds to the wall-normal velocity component.
This indicates a shift in the characteristics of the underlying physical process responsible for the instability at higher Mach numbers. 
While the relative magnitude of forcing in the resolvent specific volume modes increase with an increase in the Mach number, the wall-normal mode dominates the forcing for all the Mach numbers considered here.
The structured I/O forcing in the pressure fluctuations also change over the Mach number regime considered, especially for the supersonic results in Fig. \ref{fig:structured_forcingmodes_Mach_two_maxresolvent} where the pressure modes are significantly more dominant than the pressure forcing modes at the subsonic and sonic conditions (compare with Figs. \ref{fig:structured_forcingmodes_Mach_half_1}, \ref{fig:structured_forcingmodes_Mach_one_maxresolvent}).
Again, this is an aspect not captured by the resolvent forcing modes.   
Thus, the resolvent forcing modes do not capture the Mach number dependence/evolution of the fundamental properties of the forcing that causes the instability associated with the global maximum of the resolvent gain.
This is likely due to the implicit nature of the forcing in the resolvent analysis.
%


Similar to the above discussion, the modes in Figs. \ref{fig:forcing_and_response_modes_Mach_one_maxSSV}, \ref{fig:forcing_and_response_modes_Mach_two_maxSSV} represent the sonic and supersonic versions of the modes shown in Fig. \ref{fig:forcing_and_response_modes_Mach_half_2}, all of which correspond to the global maxima in the structured I/O gain and the accompanying local maxima in the resolvent gain.
Fluctuations in all the forcing and specific volume response modes move more towards the center of the channel at higher Mach numbers. 
Apart from this feature, the structured I/O modes remain qualitatively consistent across the different Mach numbers, indicating that the flow properties of associated instability might largely be invariant of the Mach numbers considered. 
However, resolvent forcing in the specific volume and wall-normal velocity component becomes comparable (in terms of magnitude) at supersonic speeds, which is in contrast with the subsonic and sonic cases where the wall-normal velocity dominates the forcing.
The resolvent spanwise velocity component response mode also increases in magnitude significantly at supersonic condition compared to the subsonic and sonic results. 
Thus, on the one hand, the structured I/O modes point towards an instability whose modal behavior likely remains invariant within the Mach number range considered. 
On the other hand, resolvent modes paint a different picture about the modal behavior of same instability at different Mach numbers.  
%
%
These inconsistencies observed here suggest that any insights drawn from structured input-output analysis should be accompanied by computational and/or experimental studies to corroborate findings and interpretations of the underlying physics.


\section{Conclusion}  \label{Sec: Conclusion}
We proposed a structured input-output analysis tool for compressible flows by reformulating the compressible Navier-Stokes equations. 
This reformulation facilitated the subsequent pseudo-linear modeling of the quadratic nonlinearity for 
analysis using the structured singular value formalism. 
Several discrepancies were highlighted in the numerical results of the proposed method when compared with those of an unstructured I/O (i.e., resolvent) analysis.
The key takeaways from the comparison were that the proposed method could effectively: (a) remove potentially non-physical/redundant flow behavior and instabilities; (b) distinctly characterize amplified flow features in the streamwise-spanwise wavenumber space locally; (c) extend the estimated stability margin of the flow perturbations by reducing the conservatism in the unstructured input-output gain. 
In addition, the two sets of forcing and response modes, computed for a potential source of instability identified through local or global maxima in both structured and unstructured I/O gains, revealed different 
momentum and thermodynamic characteristics driving the instability as well as disparate amplifications in the associated flow perturbations.
Thus, accounting for the structure of the nonlinearity can have profound influence on the eventual interpretation of underlying flow physics.
Still, the disparate nature of the results obtained through structured and unstructured analysis here point towards the need for computational and/or experimental data---which, unfortunately, is not available in the open literature for the compressible plane Couette flow we studied---to extract physical interpretations and insight.
Our future work would involve extending and applying the proposed tool to study compressible turbulence in boundary layer flows.
Furthermore, the quadratic formulation presented here is 
suitable for a wide array of recently proposed system-theoretic tools for quadratic nonlinear systems (see, e.g., \cite{kalur2021estimating, liao2022quadratic, liao2024convex}), which can open up potential avenues of future research on nonlinear analysis of compressible fluid flows. 
%


\backsection[Acknowledgements]{The authors acknowledge the Minnesota Supercomputing Institute (MSI) at the University of Minnesota for providing the computational resources utilized to generate the numerical results. Diganta Bhattacharjee and Maziar S. Hemati are grateful to Scott T.M. Dawson for insightful discussions on compressible resolvent analysis.}

\backsection[Funding]{This material is based upon work supported by the Air Force Office of Scientific Research under award number FA9550-21-1-0106, the Army Research Office under award number W911NF-20-1-0156, the National Science Foundation under award number CBET-1943988, and the Office of Naval Research under award number N00014-22-1-2029.}

\backsection[Declaration of interests]{The authors report no conflict of interest.}




\appendix

\bibliographystyle{jfm}
\bibliography{Ref}

\newpage

\section{Linear Operators of Structured Input-Output Modeling} \label{app:operators-1}
The operators $\vec{B}_{i_{\chi}}, i=1,2,3,$ in \eqref{eq:truth_to_modeled_nonlinearity} are as follows:
\begin{align*}
\vec{B}_{1_{\chi}} &= \begin{bmatrix}
0  &  0  &  1  & 0 & 0 & 0 & 0 \\
0 &  -\frac{1}{\gamma M_r^2} \vec{I}_3  &  0  & \frac{1}{Re} \vec{I}_3 & 0 & 0 & 0 \\
c_3 \eta_0 &  c_3 (\nabla \eta_0)^\text{T}  &  0  & 0 & c_3 \eta_0 & c_3 (\nabla \eta_0)^\text{T} & -\gamma \\
\end{bmatrix}, \\
\vec{B}_{2_{\chi}} &= - \vec{I}_5, \\
\vec{B}_{3_{\chi}} &= \begin{bmatrix}
0  &  0  &  0  & 0 & 0 & 0 & 0 & 0 \\
0  &  0  &  0  & 0 & 0 & 0 & 0 & 0 \\
c_1  &  c_1  &  c_1  & c_1 & c_1 & c_1 & 2 \eta_0 c_3 & c_2 \\
\end{bmatrix}
\end{align*}
with constants $c_1 = \frac{\gamma (\gamma - 1) M_r^2}{Re} \frac{{\eta_0}}{2}$, $c_2 = - \frac23 \eta_0 \frac{\gamma (\gamma - 1) M_r^2}{Re}$, and $c_3 = \frac{\gamma}{Re Pr}$.
Also, the operators mapping the perturbed flow states $\vec{q}$ to the output quantities $\vec{y}_{i_{\chi}}, i=1,2$ are given by
\begin{equation} \label{eq:output_operators_continuous_form_1}
\vec{C}_{1_{\chi}} = \begin{bmatrix}
   0     & 0 & 0 & 0 & \nabla^2 \\
   0     & 0 & 0 & 0 & \nabla  \\
0   & \partial_x  & \partial_y & \partial_z & 0 \\   
0   & C_{\Pi_{11}} & C_{\Pi_{12}} & C_{\Pi_{13}} & 0\\
0   & C_{\Pi_{21}} & C_{\Pi_{22}} & C_{\Pi_{23}} & 0\\
0   & C_{\Pi_{31}} & C_{\Pi_{32}} & C_{\Pi_{33}} & 0\\
\nabla^2     & 0 & 0 & 0 & 0 \\
\nabla    & 0 & 0 & 0 & 0  \\
0   & \partial_x  & \partial_y & \partial_z & 0  \\
\end{bmatrix}, \quad
\vec{C}_{2_{\chi}} = \begin{bmatrix}
\nabla     & 0 & 0 & 0 & 0  \\
0 & \nabla & 0 & 0 & 0 \\
0 & 0 & \nabla     & 0 & 0  \\
0  & 0 & 0   & \nabla & 0   \\
0 & 0 & 0    & 0 & \nabla   \\
\end{bmatrix},
\end{equation}
with 
\begin{equation} \label{eq:C_Pi_general}
\begin{split}
C_{\Pi_{11}} &= \eta_0 \nabla^2 + \frac{1}{3} \eta_0 \partial_{xx} + \left(\nabla \eta_0\right) \cdot \nabla + \frac{1}{3} \left(\partial_x \eta_0 \right) \partial_x, \\
C_{\Pi_{12}} &= \frac{1}{3} \eta_0 \partial_{xy} + \left(\partial_y \eta_0 \right) \partial_x - \frac{2}{3} \left(\partial_x \eta_0 \right) \partial_y, \\
C_{\Pi_{13}} &= \frac{1}{3} \eta_0 \partial_{xz} + \left(\partial_z \eta_0 \right) \partial_x - \frac{2}{3} \left(\partial_x \eta_0 \right) \partial_z, \\
C_{\Pi_{21}} &= \frac{1}{3} \eta_0 \partial_{xy} + \left(\partial_x \eta_0 \right) \partial_y - \frac{2}{3} \left(\partial_y \eta_0 \right) \partial_x \\
C_{\Pi_{22}} &= \eta_0 \nabla^2 + \frac{1}{3} \eta_0 \partial_{yy} + \left(\nabla \eta_0\right) \cdot \nabla + \frac{1}{3} \left(\partial_y \eta_0 \right) \partial_y, \\
C_{\Pi_{23}} &= \frac{1}{3} \eta_0 \partial_{yz} + \left(\partial_z \eta_0 \right) \partial_y - \frac{2}{3} \left(\partial_y \eta_0 \right) \partial_z, \\
C_{\Pi_{31}} &= \frac{1}{3} \eta_0 \partial_{xz} + \left(\partial_x \eta_0 \right) \partial_z - \frac{2}{3} \left(\partial_z \eta_0 \right) \partial_x, \\
C_{\Pi_{32}} &= \frac{1}{3} \eta_0 \partial_{yz} + \left(\partial_y \eta_0 \right) \partial_z - \frac{2}{3} \left(\partial_z \eta_0 \right) \partial_y, \\
C_{\Pi_{33}} &= \eta_0 \nabla^2 + \frac{1}{3} \eta_0 \partial_{zz} + \left(\nabla \eta_0\right) \cdot \nabla + \frac{1}{3}  \left(\partial_z \eta_0 \right) \partial_z.
\end{split}
\end{equation}
%
Additionally, the operators $\vec{C}_{{3_{1}}_\chi}$ and $\vec{C}_{{3_{2}}_\chi}$ (mapping the perturbed flow states $\vec{q}$ to the output vector $\vec{y}_{3_{\chi}}$) are given by
\begin{align*}
\vec{C}_{{3_{1}}_\chi} &= \begin{bmatrix}
\vec{I}_3 & 0 & 0 & 0 & 0 & 0 & 0 \\
0 & \vec{I}_3 & 0 & 0 & 0 & 0 & 0 \\
0 & 0 & \vec{I}_3 & 0 & 0 & 0 & 0 \\
2 \vec{I}_3 & 0 & 0 & \vec{I}_3 & 0 & 0 & 0 \\
0 & 2 \vec{I}_3 & 0 & 0 & \vec{I}_3 & 0 & 0 \\
0 & 0 & 2 \vec{I}_3 & 0 & 0 & \vec{I}_3 & 0 \\
0 & 0 & 0 & 0 & 0 & 0 & \vec{I}_3 \\
\begin{bmatrix}
1 & 0 & 0
\end{bmatrix} & \begin{bmatrix}
0 & 1 & 0
\end{bmatrix} & \begin{bmatrix}
0 & 0 & 1
\end{bmatrix} & 0 & 0 & 0 &0
\end{bmatrix}, \\
\vec{C}_{{3_{2}}_\chi} &= \begin{bmatrix}
0 & \nabla & 0 & 0 & 0 \\
0 & 0 & \nabla & 0 & 0 \\
0 & 0 & 0 & \nabla & 0 \\
0 &\partial_x & 0 & 0 &  0 \\
0 & 0 & \partial_x & 0 &  0 \\
0 & 0 & 0 & \partial_x &  0 \\
0 & \partial_y & 0 & 0 &  0 \\
0 & 0 & \partial_y & 0 &  0 \\
0 & 0 & 0 & \partial_y &  0 \\
0 & \partial_z & 0 & 0 &  0 \\
0 & 0 & \partial_z & 0 &  0 \\
0 & 0 & 0 & \partial_z &  0\\
\nabla & 0 & 0 & 0 & 0 \\
\end{bmatrix} .
\end{align*}


\section{Discretized Linear Operators for Compressible Plane Couette Flow}
\label{app:operators-2}
The discretization is carried out using Chebyshev polynomials in the wall-normal direction and Fourier modes in the streamwise and spanwise directions. 
The linear operator $\hat{\vec{L}}(k_x, k_z)$ can be expressed as
\begin{equation*}
\hat{\vec{L}}(k_x, k_z) = \begin{bmatrix}
\hat{L}_{{\xi,\xi}} & \hat{L}_{{\xi,u}} & \hat{L}_{{\xi,v}} & \hat{L}_{{\xi,w}} & \hat{L}_{{\xi,p}}  \\
\hat{L}_{{u,\xi}} & \hat{L}_{{u,u}} & \hat{L}_{{u,v}} & \hat{L}_{{u,w}} & \hat{L}_{{u,p}}  \\
\hat{L}_{{v,\xi}} & \hat{L}_{{v,u}} & \hat{L}_{{v,v}} & \hat{L}_{{v,w}} & \hat{L}_{{v,p}}  \\
\hat{L}_{{w,\xi}} & \hat{L}_{{w,u}} & \hat{L}_{{w,v}} & \hat{L}_{{w,w}} & \hat{L}_{{w,p}}  \\
\hat{L}_{{p,\xi}} & \hat{L}_{{p,u}} & \hat{L}_{{p,v}} & \hat{L}_{{p,w}} & \hat{L}_{{p,p}}  \\
\end{bmatrix}
\end{equation*}
with the sub-operators given by
\begin{align*}
    \hat{L}_{\sv,\sv} &= -\vec{i} k_xU_0 \\
    \hat{L}_{\sv,u} &= \vec{i} k_x\sv_0\\
    \hat{L}_{\sv,v} &= -\sv'_0+\sv_0D_y\\
    \hat{L}_{\sv,w} &= \vec{i} k_z\sv_0\\
    \hat{L}_{\sv,p} &= 0\\
    \hat{L}_{u,\sv} &= 0\\
    \hat{L}_{u,u} &= -\vec{i} k_xU_0 - \frac{\sv_0}{Re}\left[\eta_0\left(\frac{4}{3}k_x^2 \vec{I}_{N_{y}} -D_{yy} + k_z^2 \vec{I}_{N_{y}} \right) - \eta_0'D_y\right]\\
    \hat{L}_{u,v} &= -U'_0+\vec{i} k_x\frac{\sv_0}{3Re}\left[\eta_0D_y+3\eta_0'\right]\\
    \hat{L}_{u,w} &= -k_x k_z\frac{\sv_0\eta_0}{3Re}\\
    \hat{L}_{u,p} &= -\vec{i} k_x\frac{\sv_0}{\gamma M_r^2}\\
    \hat{L}_{v,\sv} &= 0\\
    \hat{L}_{v,u} &= \vec{i} k_x \frac{\sv_0}{3Re}\left[\eta_0D_y-2\eta_0'\right]\\
    \hat{L}_{v,v} &= -\vec{i} k_xU_0-\frac{\sv_0}{Re}\left[\eta_0\left(k_x^2 \vec{I}_{N_{y}} -\frac{4}{3}D_{yy} + k_z^2 \vec{I}_{N_{y}} \right) -\frac{4}{3}\eta_0'D_y\right]\\
    \hat{L}_{v,w} &= \vec{i} k_z\frac{\sv_0}{3Re}\left[\eta_0D_y-2\eta_0'\right]\\
    \hat{L}_{v,p} &= -\frac{\sv_0}{\gamma M_r^2} D_y\\
    \hat{L}_{w,\sv} &= 0\\
    \hat{L}_{w,u} &= -k_x k_z\frac{\sv_0\eta_0}{3Re}\\
    \hat{L}_{w,v} &= \vec{i} k_z\frac{\sv_0}{3Re}\left[\eta_0D_y+3\eta_0'\right]\\
    \hat{L}_{w,w} &= -\vec{i} k_xU_0-\frac{\sv_0}{Re}\left[\eta_0\left(k_x^2 \vec{I}_{N_{y}} -D_{yy} + \frac{4}{3}k_z^2 \vec{I}_{N_{y}} \right)-\eta_0'D_y\right]\\
    \hat{L}_{w,p} &= -\vec{i} k_z\frac{\sv_0}{\gamma M_r^2}\\
    \hat{L}_{p,\sv} &= -\left(\frac{\gamma}{RePr}\right)\left[\eta_0\left(k_x^2 \vec{I}_{N_{y}} -D_{yy} + k_z^2 \vec{I}_{N_{y}} \right) - \eta_0'D_y\right]\\
    \hat{L}_{p,u} &= -\vec{i} k_x\gamma \vec{I}_{N_{y}} + \frac{\gamma (\gamma-1) M_r^2}{Re}\left(2U_0'\eta_0 \right)D_y\\ 
    \hat{L}_{p,v} &= -\gamma D_y + \vec{i} k_x\frac{\gamma (\gamma-1) M_r^2}{Re}\left(2U_0'\eta_0 \right)\\
    \hat{L}_{p,w} &= -\vec{i} k_z \gamma \vec{I}_{N_{y}} \\
    \hat{L}_{p,p} &= -\vec{i} k_x U_0 + \frac{\gamma}{Re Pr}\left\{\eta_0\left[\sv_0'' + 2\sv_0'D_y - \sv_0\left(k_x^2 \vec{I}_{N_{y}} -D_{yy} + k_z^2 \vec{I}_{N_{y}} \right)\right] + \eta_0'(\sv_0' + \sv_0D_y)\right\}
\end{align*}
\newpage
The sub-operators $\hat{\vec{B}}_{i_{\chi}}, i=1,2,3,$ are as follows:
\begin{align*}
\hat{\vec{B}}_{1_{\chi}} &= \begin{bmatrix}
0  &  0  &  \vec{I}_{N_{y}}  & 0 & 0 & 0 & 0 \\
0 &  -\frac{1}{\gamma M_r^2} \vec{I}_{3 N_{y}}  &  0  & \frac{1}{Re} \vec{I}_{3 N_{y}} & 0 & 0 & 0 \\
c_3 \eta_0 &  \begin{bmatrix}
    0 & c_3 \eta_0^\prime & 0
\end{bmatrix}  &  0  & 0 & c_3 \eta_0 & \begin{bmatrix}
    0 & c_3 \eta_0^\prime & 0
\end{bmatrix} & -\gamma \vec{I}_{N_{y}} \\
\end{bmatrix}, \\
\hat{\vec{B}}_{2_{\chi}} &= - \vec{I}_{5N_y}, \notag \\
\hat{\vec{B}}_{3_{\chi}} &= \begin{bmatrix}
0  &  0  &  0  & 0 & 0 & 0 & 0 & 0 \\
0  &  0  &  0  & 0 & 0 & 0 & 0 & 0 \\
c_1 \vec{I}_{N_{y}}  &  c_1 \vec{I}_{N_{y}}  &  c_1 \vec{I}_{N_{y}}  & c_1\vec{I}_{N_{y}} & c_1\vec{I}_{N_{y}} & c_1\vec{I}_{N_{y}} & 2 \eta_0 c_3 & c_2\vec{I}_{N_{y}} \\
\end{bmatrix}.
\end{align*} 
Finally, the sub-operators associated with $\hat{\vec{C}}_\chi (k_x, k_z)$ are given by
\begin{align*}
 \hat{\vec{C}}_{1_{\chi}} &= \begin{bmatrix}
    0 & 0 & 0 & 0 & -k_x^2 \vec{I}_{N_{y}} + D_{yy} -k_z^2 \vec{I}_{N_{y}} \\ 
    0 & 0 & 0 & 0 & \vec{i} k_x \vec{I}_{N_{y}} \\ 
    0 & 0 & 0 & 0 & D_{y}\\
    0 & 0 & 0 & 0 & \vec{i} k_z \vec{I}_{N_{y}} \\
    0 & \vec{i} k_x \vec{I}_{N_{y}} & D_y &\vec{i} k_z \vec{I}_{N_{y}} & 0\\
    0&\hat{C}_{\Pi_{11}}&\hat{C}_{\Pi_{12}}&\hat{C}_{\Pi_{13}}&0\\
    0&\hat{C}_{\Pi_{21}}&\hat{C}_{\Pi_{22}}&\hat{C}_{\Pi_{23}}&0\\
    0&\hat{C}_{\Pi_{31}}&\hat{C}_{\Pi_{32}}&\hat{C}_{\Pi_{33}}&0\\
    -k_x^2 \vec{I}_{N_{y}} + D_{yy} - k_z^2 \vec{I}_{N_{y}} &0&0&0&0\\
    \vec{i} k_x \vec{I}_{N_{y}} & 0 & 0 & 0 & 0\\ 
    D_{y} & 0 & 0 & 0 & 0 \\
    \vec{i} k_z \vec{I}_{N_{y}} & 0 & 0 & 0 & 0 & \\
    0&\vec{i} k_x \vec{I}_{N_{y}} & D_y & \vec{i} k_z \vec{I}_{N_{y}} & 0\\
    \end{bmatrix}, \\
    \hat{\vec{C}}_{2_{\chi}} &= \begin{bmatrix}
    \vec{i} k_x \vec{I}_{N_{y}} &0&0&0&0\\ 
    D_{y}&0&0&0&0\\
    \vec{i} k_z \vec{I}_{N_{y}} &0&0&0&0\\
    0&\vec{i} k_x \vec{I}_{N_{y}} &0&0&0\\ 
    0&D_{y}&0&0&0\\
    0&\vec{i} k_z \vec{I}_{N_{y}}&0&0&0\\
    0&0&\vec{i} k_x \vec{I}_{N_{y}}&0&0\\ 
    0&0&D_{y}&0&0\\
    0&0&\vec{i} k_z \vec{I}_{N_{y}}&0&0\\
    0&0&0&\vec{i} k_x \vec{I}_{N_{y}}&0\\ 
    0&0&0&D_{y}&0\\
    0&0&0&\vec{i} k_z \vec{I}_{N_{y}}&0\\
    0 & 0 & 0 & 0 & \vec{i} k_x \vec{I}_{N_{y}}\\ 
    0 & 0 & 0 & 0 & D_{y}\\
    0 & 0 & 0 & 0 & \vec{i} k_z \vec{I}_{N_{y}}\\
    \end{bmatrix}, \\
   \hat{\vec{C}}_{{3_{2}}_\chi} &= \begin{bmatrix}
     0&\vec{i} k_x \vec{I}_{N_{y}}&0&0&0\\ 
    0&D_{y}&0&0&0\\
    0&\vec{i} k_z \vec{I}_{N_{y}} &0&0&0\\
    0&0&\vec{i} k_x \vec{I}_{N_{y}} &0&0\\ 
    0&0&D_{y}&0&0\\
    0&0&\vec{i} k_z \vec{I}_{N_{y}}&0&0\\
    0&0&0&\vec{i} k_x \vec{I}_{N_{y}}&0\\ 
    0&0&0&D_{y}&0\\
    0&0&0&\vec{i} k_z \vec{I}_{N_{y}}&0\\
    0&\vec{i} k_x \vec{I}_{N_{y}}&0& 0 &0\\
    0& 0 & \vec{i} k_x \vec{I}_{N_{y}}& 0 &0\\
    0& 0 &0 & \vec{i} k_x \vec{I}_{N_{y}} &0\\
    0&D_y &0& 0 &0\\
    0& 0 & D_y & 0 &0\\
    0& 0 &0 & D_y &0\\
    0&\vec{i} k_z \vec{I}_{N_{y}} &0& 0 &0\\
    0& 0 & \vec{i} k_z \vec{I}_{N_{y}} & 0 &0\\
    0& 0 &0 & \vec{i} k_z \vec{I}_{N_{y}} &0\\
    \vec{i} k_x \vec{I}_{N_{y}}&0&0&0&0\\ 
    D_{y}&0&0&0&0\\
    \vec{i} k_z \vec{I}_{N_{y}}&0&0&0&0\\
    \end{bmatrix},
\end{align*}
where
    \begin{align*}
    \hat{C}_{\Pi_{11}} &=\eta_0\left(-\frac43 k_x^2 \vec{I}_{N_{y}} +D_{yy}-k_z^2 \vec{I}_{N_{y}}\right)+\eta_0'D_y, \
    \hat{C}_{\Pi_{12}} =\vec{i} k_x\left(\frac13\eta_0D_y+\eta_0'\right), \\
    \hat{C}_{\Pi_{13}} &=-k_x k_z\frac13\eta_0, \
    \hat{C}_{\Pi_{21}} =\vec{i} k_x\left(\frac13\eta_0D_y-\frac23\eta_0'\right), \\
    \hat{C}_{\Pi_{22}} &=\eta_0\left(-k_x^2 \vec{I}_{N_{y}} +\frac43D_{yy}-k_z^2 \vec{I}_{N_{y}}\right)+\frac43\eta_0'D_y, \
    \hat{C}_{\Pi_{23}} =\vec{i} k_z\left(\frac13\eta_0D_y-\frac23\eta_0'\right), \\
    \hat{C}_{\Pi_{31}} &=-k_x k_z\frac13\eta_0, \
    \hat{C}_{\Pi_{32}} =\vec{i} k_z\left(\frac13\eta_0D_y+\eta_0'\right), \
    \hat{C}_{\Pi_{33}} =\eta_0\left(-k_x^2 \vec{I}_{N_{y}}+D_{yy}-\frac43k_z^2 \vec{I}_{N_{y}} \right)+\eta_0'D_y.
\end{align*}

\section{Structured Input-Output and Resolvent Modes: Wall-Normal Collocation/Grid Points}
\label{app:modes-and-grid_points}

\begin{figure}
\captionsetup[subfigure]{justification=centering}
  \centering
  \begin{subfigure}[b]{0.495\linewidth}
    \includegraphics[width=1\textwidth]{Structured_Forcingmodes_Mach_half_1.pdf}
    \caption{Forcing modes ($N_y=100$)} \label{fig:StrcIO_forcingmodes_N100}
    \end{subfigure}
     \begin{subfigure}[b]{0.495\linewidth}
    \includegraphics[width=1\textwidth]{Structured_Responsemodes_Mach_half_1.pdf}
    \caption{Response modes ($N_y=100$)} \label{fig:StrcIO_responsemodes_N100}
    \end{subfigure}
\begin{subfigure}[b]{0.495\linewidth}
    \includegraphics[width=1\textwidth]{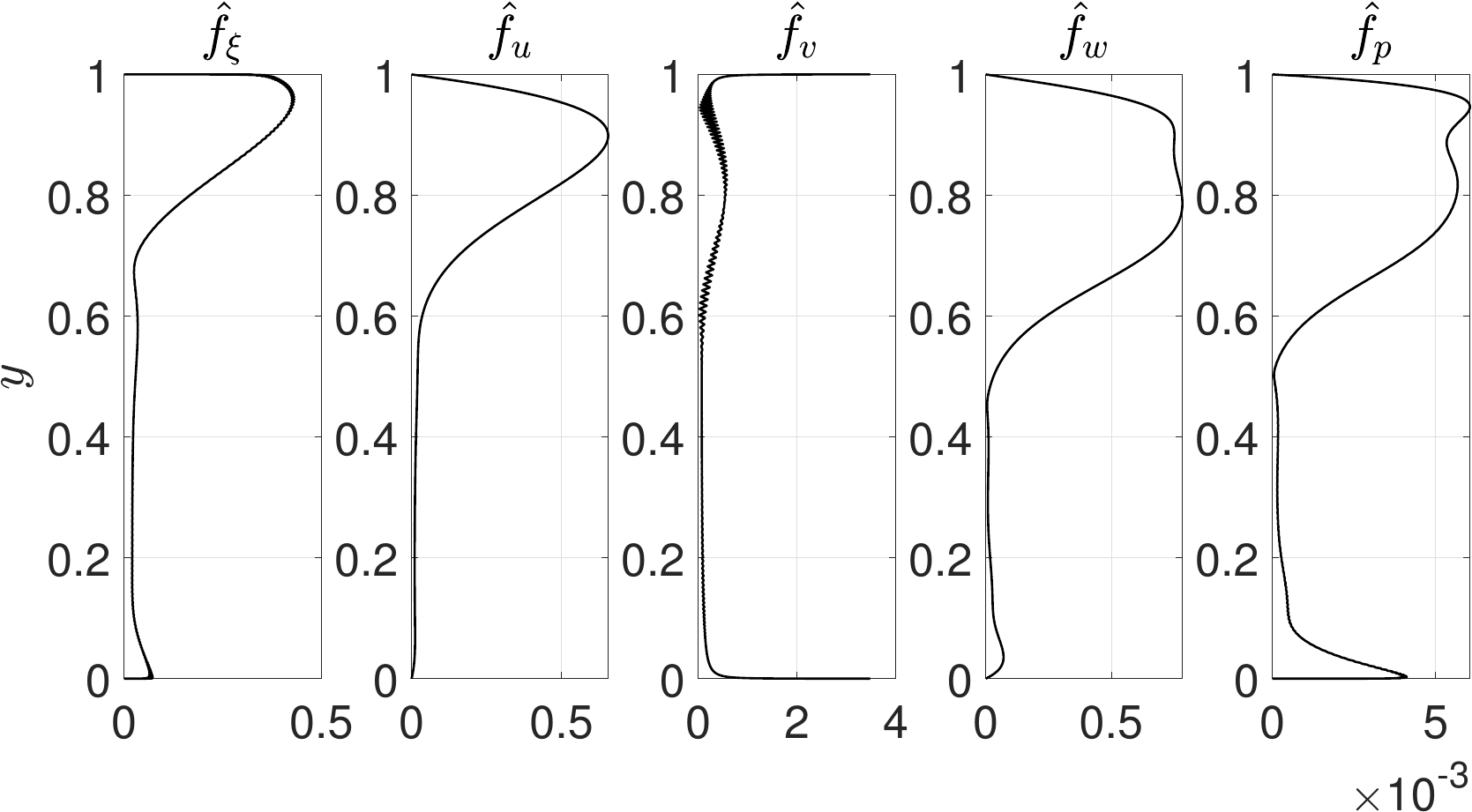}
    \caption{Forcing modes ($N_y=300$)} \label{fig:StrcIO_forcingmodes_N300}
  \end{subfigure}
  \begin{subfigure}[b]{0.495\linewidth}
    \includegraphics[width=1\textwidth]{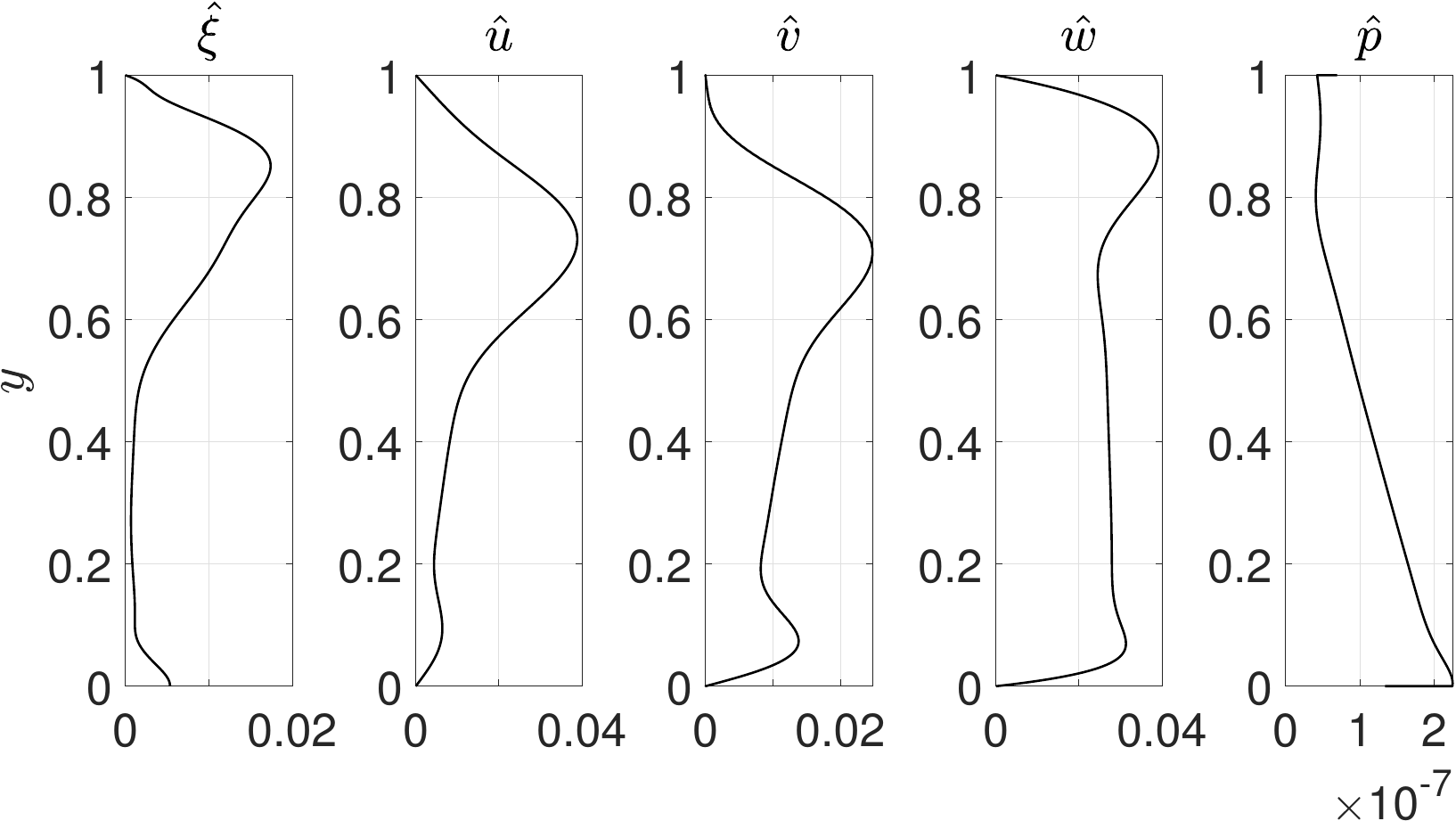}
    \caption{Response modes ($N_y=300$)} \label{fig:StrcIO_responsemodes_N300}
  \end{subfigure}
\caption{{Absolute values of the structured I/O forcing and response modes for $M_r=0.5$ and $(k_x,k_z,\omega)=(0.01,11.24,-0.01)$.
The results here correspond to number of wall-normal collocation/grid points $N_y = 100$ and $N_y=300$.}}
  \label{fig:StrcIO_modes_grid-points}
\end{figure}

\begin{figure}
\captionsetup[subfigure]{justification=centering}
  \centering
  \begin{subfigure}[b]{0.495\linewidth}
    \includegraphics[width=1\textwidth]{Resolvent_Forcingmodes_Mach_half_1.pdf}
    \caption{Forcing modes ($N_y=100$)} \label{fig:Resolvent_forcingmodes_N100}
    \end{subfigure}
     \begin{subfigure}[b]{0.495\linewidth}
    \includegraphics[width=1\textwidth]{Resolvent_Responsemodes_Mach_half_1.pdf}
    \caption{Response modes ($N_y=100$)} \label{fig:Resolvent_responsemodes_N100}
    \end{subfigure}
\begin{subfigure}[b]{0.495\linewidth}
    \includegraphics[width=1\textwidth]{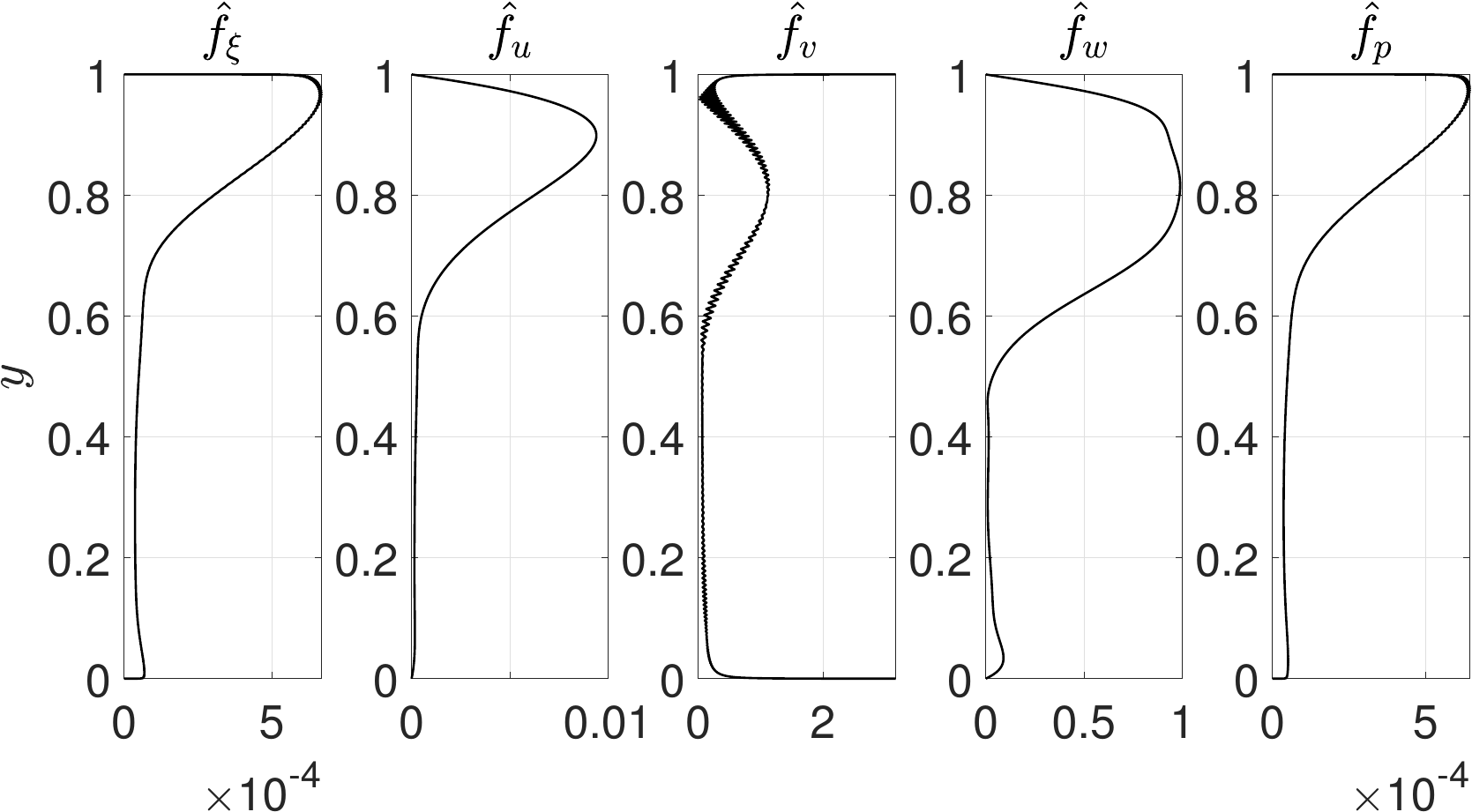}
    \caption{Forcing modes ($N_y=300$)} \label{fig:Resolvent_forcingmodes_N300}
  \end{subfigure}
  \begin{subfigure}[b]{0.495\linewidth}
    \includegraphics[width=1\textwidth]{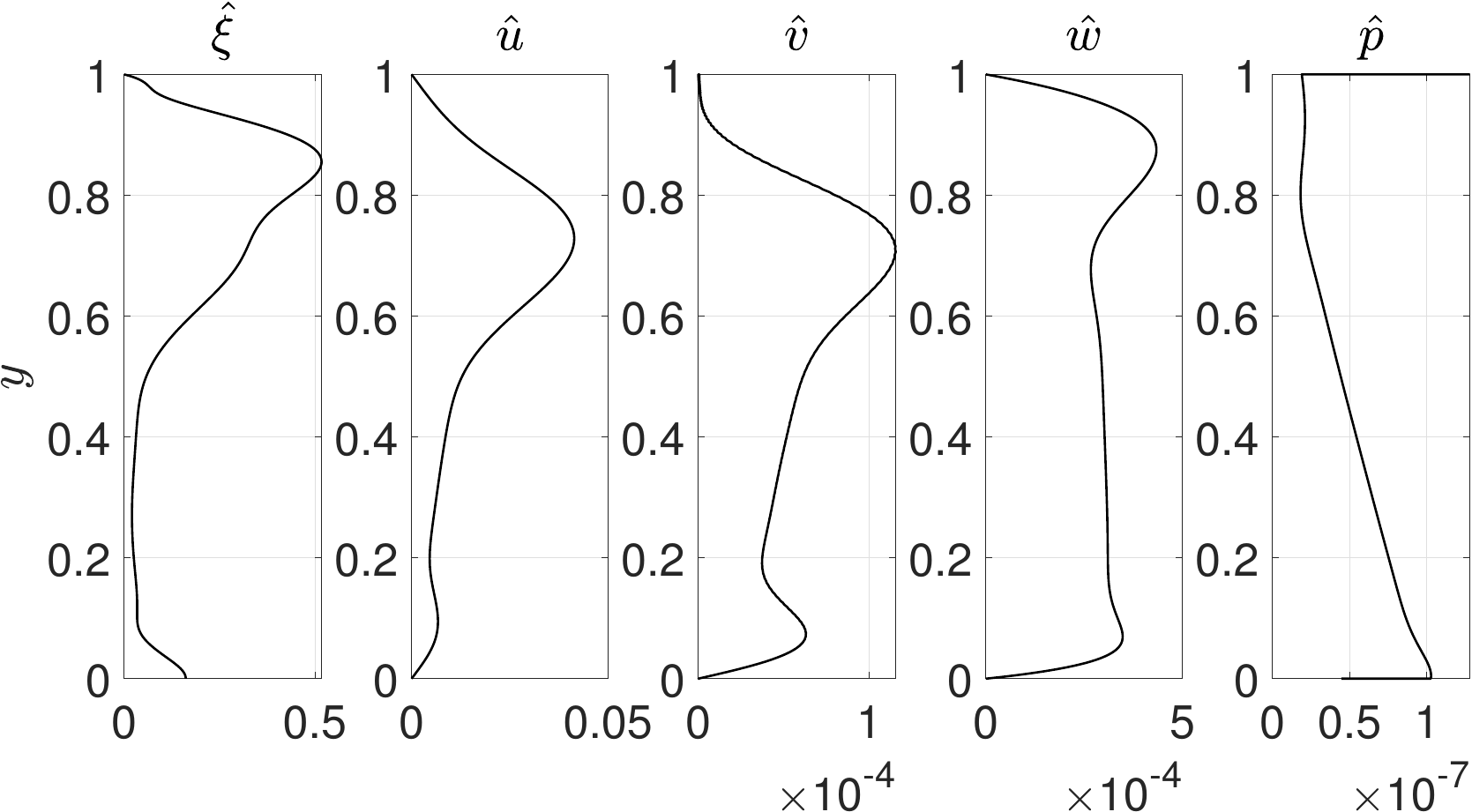}
    \caption{Response modes ($N_y=300$)} \label{fig:Resolvent_responsemodes_N300}
  \end{subfigure}
\caption{{Absolute values of resolvent forcing and response modes for $M_r=0.5$ and $(k_x,k_z,\omega)=(0.01,11.24,-0.01)$.
The results here correspond to number of wall-normal collocation/grid points $N_y = 100$ and $N_y=300$.}}
  \label{fig:Resolvent_modes_grid-points}
\end{figure}

{
Consider the compressible plane Couette flow at $M_r=0.5$ and $(k_x,k_z,\omega) = (0.01, 11.24, -0.01)$, which corresponds to the largest resolvent gain over the wavenumber pair and temporal frequency grid considered for the results shown in Fig. \ref{fig:Resolvent_Mach_half}. 
While there are `oscillations' in both the structured I/O and resolvent $\hat{f}_v$ modes and the resolvent $\hat{v}$ mode at this $(k_x,k_z,\omega)$ tuple and flow parameter setting (see Fig. \ref{fig:forcing_and_response_modes_Mach_half_1}), these are not likely caused by a lack of wall-normal collocation/grid points. 
We investigated the behavior of structured I/O and resolvent modes for higher values of $N_y$ (number of wall-normal grid points) 
and the results are shown in Figs. \ref{fig:StrcIO_modes_grid-points}, \ref{fig:Resolvent_modes_grid-points}.
Going by these results, it is clear that the oscillations persist even at larger $N_y$---while this is obvious in the case of the forcing modes $\hat{f}_v$, a closer inspection reveals attenuated but persistent oscillations in the resolvent response modes $\hat{v}$ as well---thereby indicating that the oscillations are not associated with a lack of grid points in the wall-normal direction.
It is also noteworthy that these oscillations only appear in this one isolated case, which might be indicative of a particular flow behavior. 
%
%
A careful investigation, one which we plan to carry out in our future studies, will be required to find the root cause behind the oscillatory modes and the significance in terms of physical flow behavior (if any).} 

\section{Computation Time Scaling with Number of Wall-Normal Collocation/Grid Points}
\label{app:computation_times_structuredIO_reslvent}

\begin{figure}
\captionsetup[subfigure]{justification=centering}
  \centering
  \begin{subfigure}[b]{0.5\linewidth}
    \includegraphics[width=1\textwidth]{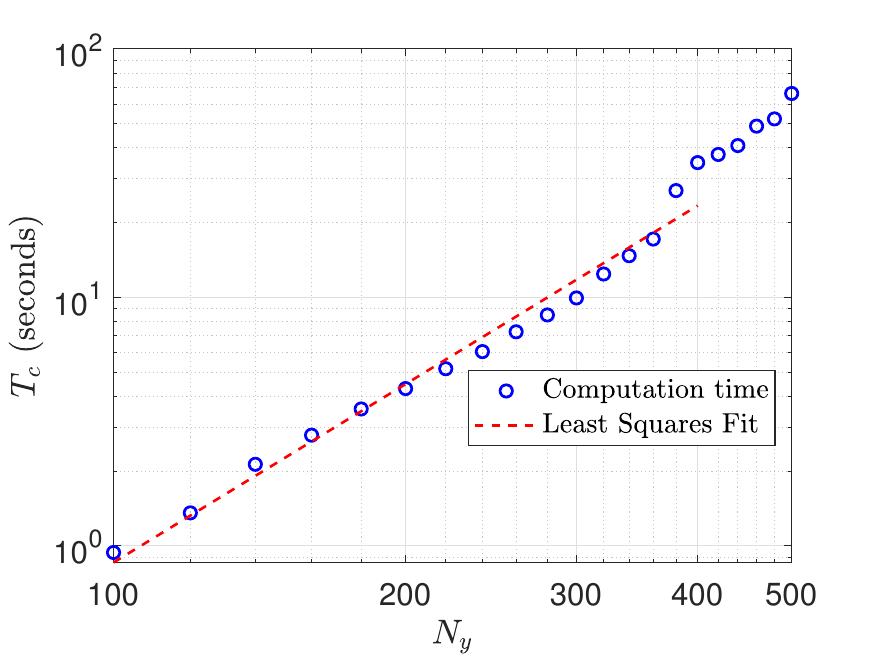}
    \caption{Structured I/O: $\mu$ upper bound} \label{fig:Comp_time_mu_ub_osbal}
  \end{subfigure}%
  \begin{subfigure}[b]{0.5\linewidth}
    \includegraphics[width=1\textwidth]{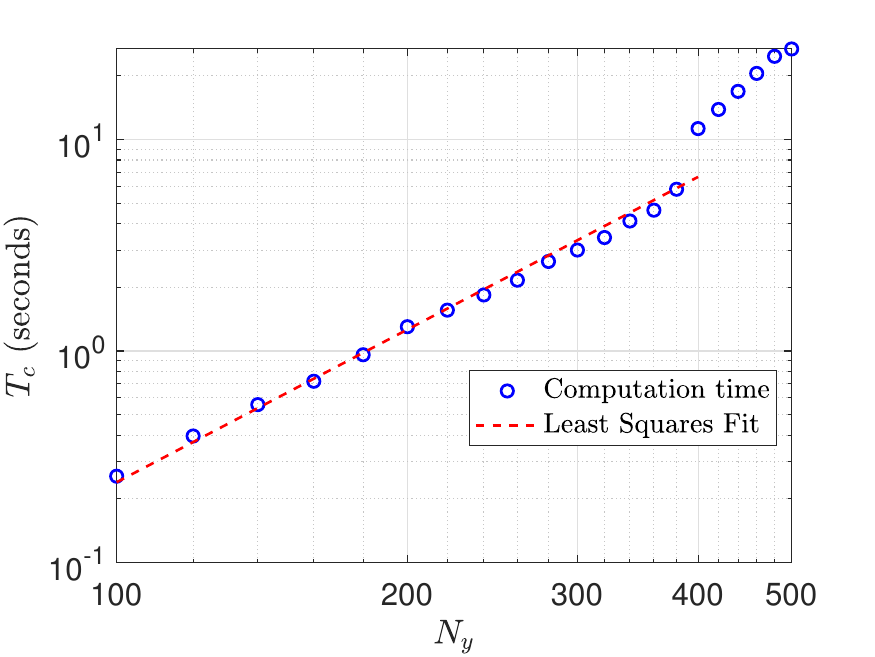}
    \caption{Structured I/O: $\mu$ lower bound} \label{fig:Comp_time_mu_lb_PI}
  \end{subfigure}
   \begin{subfigure}[b]{0.5\linewidth}
    \includegraphics[width=1\textwidth]{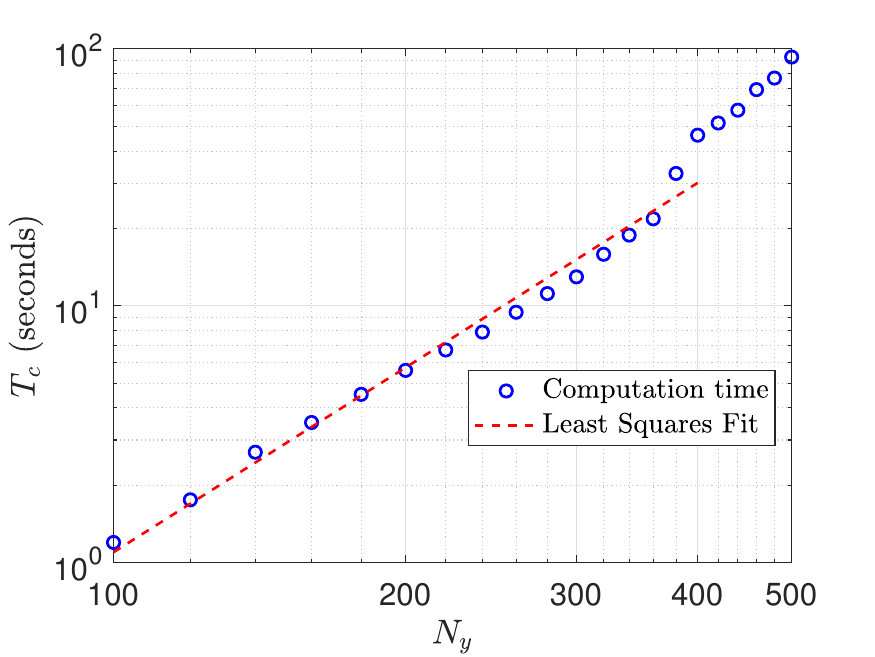}
    \caption{Structured I/O: $\mu$ upper \& lower bounds combined} \label{fig:Comp_time_mu}
  \end{subfigure}%
  \begin{subfigure}[b]{0.5\linewidth}
    \includegraphics[width=1\textwidth]{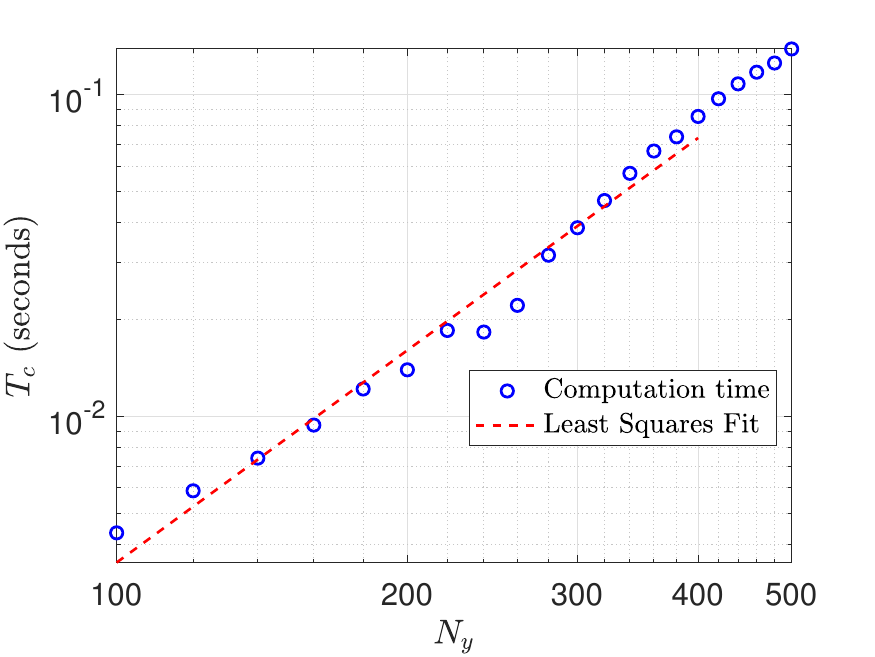}
    \caption{Resolvent} \label{fig:Comp_time_resolvent}
  \end{subfigure}
\caption{{Computation times associated with the proposed structured I/O framework and the resolvent analysis 
for $M_r=0.5$ and $(k_x,k_z,\omega)=(0.01,11.24,-0.01)$. 
The results here are plotted in log-scale and correspond to number of wall-normal collocation points $N_y$ between 100 and 500 (more specifically, we utilize the following set $N_y = \{100,120,140,160,\dots,500\}$).
Note that the least squares fits are done up to $N_y = 400$ 
%
and the structured I/O computation times here includes: (a) the $\mu$ upper bound (computed using `osbal' command in MATLAB's Robust Control Toolbox); (b) the $\mu$ lower bound, structured I/O modes and the structured uncertainty (computed through Power iteration outlined in Algorithm \ref{alg:piter}).
Computation times for the resolvent analysis are associated with the singular value decomposition via MATLAB's `svds' command.}}
  \label{fig:Computation_times}
\end{figure}

{
Consider the compressible plane Couette flow at $M_r=0.5$ and $(k_x,k_z,\omega) = (0.01, 11.24, -0.01)$, which corresponds to the largest resolvent gain over the wavenumber pair and temporal frequency grid considered for the results shown in Fig. \ref{fig:Resolvent_Mach_half}. 
The average computation times associated with the structured I/O and resolvent analysis are provided in Fig. \ref{fig:Computation_times} for 
wall-normal collocation/grid points $N_y = \{100,120,140,160,\dots,500\}$\footnote[4]{
We have utilized a desktop computer with 3.61 GHz 12-th Gen Intel(R) Core(TM) i7-12700K processor containing 12 cores and 16 GB RAM for these computations.}. 
%
%
%
Assuming that the computation times $T_c$ scale polynomially with the wall-normal grid points $N_y$, we have $T_c = a N_y^d$ describing the general trend of how the computation times scale with the problem dimension. 
This gives a straight-line on a log-log plot, given by 
\begin{equation*}
    \log(T_c) = c + d \log(N_y)
\end{equation*}
where $c = \log(a)$.
We can use least squares to fit the constants $(c,d)$, and the fitted lines are shown in Fig. \ref{fig:Computation_times}, for which we have utilized up to $N_y = 400$ to capture the computational cost primarily due to the floating point operations per second (or, FLOPs) as non-FLOPs aspects start to play a major role in the overall computation cost for $N_y > 400$ and lead to the increase in computation times illustrated in Fig. \ref{fig:Computation_times}. 
%
%
Thus, these least squares fits (for up to $N_y=400$, as shown in Fig. \ref{fig:Computation_times}) correspond to the following:
\begin{eqnarray*}
    \text{Structured I/O-$\mu$ upper bound:}~& (c,d) = (-11.1384, 2.3853), \\
    \text{Structured I/O-$\mu$ lower bound:}~& (c,d) = (-12.4706, 2.3977), \\
    \text{Structured I/O-$\mu$ upper \& lower bounds combined:}~& (c,d) = (-10.9106, 2.3892), \\
    \text{Resolvent:}~& (c,d) = (-15.7269, 2.1978).
\end{eqnarray*}
The upper bound calculations, which are done using the `osbal' command available in MATLAB's Robust Control Toolbox, scale roughly as $O(N_y^2)$ since these computations are largely dominated by vector-matrix products and the least squares fit approximately captures this trend, as shown in Fig. \ref{fig:Comp_time_mu_ub_osbal}.
A similar set of comments are applicable to the power iteration employed for lower bound, structured I/O modes and structured uncertainty computations (see Fig. \ref{fig:Comp_time_mu_lb_PI}). 
The resolvent analysis involves taking a singular value decomposition (SVD) of the resolvent operator $\mathcal{R} (k_x,k_z,\omega) = \left(\vec{i} \omega \vec{I}_{n_q} - \hat{\vec{L}}(k_x, k_z) \right)^{-1}$ and we have done that through the `svds' command in MATLAB for our results. 
This particular version of SVD employs Algorithm 3.1 in \cite{baglama2005augmented}, where doing a matrix-vector product carries the largest computational cost.
For a square matrix of dimension $n$, this involves $O(n^2)$ FLOPs with some constant factor depending on the number of iterations. 
Thus, the resolvent computations are also expected to roughly scale as $O(N_y^2)$, which is corroborated by the computation time data and least squares fit shown in Fig. \ref{fig:Comp_time_resolvent}.
Note that as $N_y$ becomes sufficiently large, memory access becomes a key contributor to overall computation time in conjunction with the FLOPs. 
While both structured I/O and resolvent computation times increase due to these non-FLOPs factors, the effects are more prevalent in $\mu$ bound computations and leads to the rapid increase in corresponding computation times for $N_y > 400$ (see Figs. \ref{fig:Comp_time_mu_ub_osbal}, \ref{fig:Comp_time_mu_lb_PI}, \ref{fig:Comp_time_mu}). 
Based on this observation here, we restricted $N_y \leq 400$ to get the least squares fit shown in Fig. \ref{fig:Computation_times} (as mentioned above). 
%
%
However, an alternative set of least squares fits---by accounting for all the points in the $N_y$ set (which are not shown in Fig. \ref{fig:Computation_times})---correspond to the following:
\begin{eqnarray*}
    \text{Structured I/O-$\mu$ upper bound:}~& (c,d) = (-12.3349, 2.6154), \\
    \text{Structured I/O-$\mu$ lower bound:}~& (c,d) = (-14.8076, 2.8472), \\
    \text{Structured I/O- $\mu$ upper \& lower bounds combined:}~& (c,d) = (-12.3972, 2.6752), \\
    \text{Resolvent:}~& (c,d) = (-16.2154, 2.2826),
\end{eqnarray*}
which provide a more practical trend of the computation times encompassing all potential contributing factors.} 

\end{document}